\documentclass[a4paper,onecolumn,11pt,accepted=2022-03-31]{quantumarticle}
\pdfoutput=1
\usepackage[utf8]{inputenc}
\usepackage[english]{babel}
\usepackage[T1]{fontenc}
\usepackage{amsmath}
\usepackage{hyperref}

\usepackage{tikz}
\usepackage{lipsum}

\begin{document}

\title{Extraction of autonomous quantum coherences}

\author{Artur Slobodeniuk}
\email{aslobodeniuk@karlov.mff.cuni.cz}
\affiliation{Department of Optics, Palack\'{y} University, 17. listopadu 12, 77146 Olomouc, Czech Republic}
\affiliation{Department of Condensed Matter Physics, Faculty of Mathematics and Physics,
Charles University, \\ Ke Karlovu 5, CZ-121 16 Prague, Czech Republic}
\orcid{0000-0001-5798-0431}

\author{Tom\'{a}\v{s} Novotn\'{y}}
\email{tno@karlov.mff.cuni.cz}
\affiliation{Department of Condensed Matter Physics, Faculty of Mathematics and Physics,
Charles University, \\ Ke Karlovu 5, CZ-121 16 Prague, Czech Republic}
\orcid{0000-0001-7014-4155}

\author{Radim Filip}
\email{filip@optics.upol.cz}
\affiliation{Department of Optics, Palack\'{y} University, 17. listopadu 12, 77146 Olomouc, Czech Republic}
\orcid{0000-0003-4114-6068}

\maketitle

\begin{abstract}
Quantum coherence is an essential resource to gain advantage over classical physics and technology. Recently, it has been proposed that a low-temperature environment can induce quantum coherence of a spin without an external coherent pump. We address a critical question if such coherence is extractable by a weak coupling to an output system dynamically affecting back the spin-environment coupling. Describing the entire mechanism, we prove that such extraction is generically possible for output spins (also oscillators or fields) and, as well, in a fermionic analogue of such a process. We compare the internal spin coherence and output coherence over temperature and characteristic frequencies. The proposed optimal coherence extraction opens paths for the upcoming experimental tests with atomic and solid-state systems.
\end{abstract}

\section{Introduction}

Quantum coherence of atomic and solid-state spins is at the core of many quantum technology applications 
\cite{Streltsov2017} and necessary to gain quantum advantage for modern applications over classical technology. It is highly required for quantum sensing \cite{Schmitt2017,Zhou2020}, quantum communication 
\cite{Reiserer2016,Awschalom2018}, quantum simulators \cite{Hensgens2017,Drost2017,Slot2017}, energy harvesting 
\cite{Scholes2017,Romero2017}, quantum thermodynamics \cite{Klatzow2019, Ono2020,Latune2021} and, of course, in quantum computing 
\cite{Bradley2019,Stephen2019}. It also stimulates a broad recent theoretical analysis 
\cite{DelSanto2020,DelSanto2018,Tan2017,Micadei2020_1,Alonso2016,Haase2018,Czajkowski2019} and new discussions 
\cite{Novo2016,Micadei2020,Diaz2020,Demkowicz2017,Seah2020,Miller2020,Francica2020,Seah2019,Latune2019,Tupkary2021}. The simplest quantum coherence can appear in a single spin as superposition of free-Hamiltonian eigenstates. A classical coherent external force can be used to induce spin quantum coherence. On the other hand, coupling of spin to the surrounding environment typically destroys quantum coherence \cite{Zurek2003}.

Can such a single spin, and later a larger spin ensemble, receive a steady-state quantum coherence without any external coherent drive, purely from a coherent coupling with its environment? If that is possible, quantum technology can be, in principle, more autonomous. Maybe, as broadly discussed, even Nature can already use such coherence to gain quantum advantages without external drive \cite{Mohseni2014,Duan2017}. It might be hard to imagine that a low-temperature environment coherently builds up quantum coherence. However, the recent theoretical considerations predict such an open door. Indeed, single spin coherence can generically appear by cooling the phononic environment to a low temperature \cite{Guarnieri2018}. Quantum coherence of the spin rises there from the coherence of engineered interaction with the environment.

Subsequent studies confirmed similar phenomena \cite{Guarnieri2020,Reppert2020,Roman2020} and also bring the first experimental proposal in the solid-state systems \cite{Purkayastha2020}. The path to the first experiments seems to be open. However, an essential aspect remains still unsolved. Is such autonomous coherence extractable, i.e. can an output system continuously and even weakly connected to the spin extract the steady-state coherence? Would such weak extraction not disturb the coherence generation? These essential questions are already critical for a fundamental understanding and detectability of quantum coherence as a resource. Simultaneously, to answer them will be necessary for more autonomous quantum devices and their applications. Apparently, our current picture about an extraction of coherence caused by an external drive is not sufficient. It assumes that strong coherent external drive is largely immune to coupling the output system. As the autonomous coherences appear by an interaction with environment, their extraction is still an open issue.  

The existing experimental proposal \cite{Purkayastha2020} has opened this key question of extraction of autonomous coherence into an intracavity field. The extraction assumed a weak linear coupling of the cavity field to the spin in a simple static approximation without any back action on the spin and coupling with the environment. For the cavity-field frequency much below the spin frequency and mean spin population almost unchanged (see equation (70) in Supplementary Notes \cite{Purkayastha2020}), the spin coherence drives the mean coherent displacement of the high-quality cavity field. It scales only by the spin-cavity coupling strength (see equations~(70)-(73) in Supplementary Notes 
\cite{Purkayastha2020}) and is inversely proportional to the field frequency.

Here, we build upon that analysis and propose a minimal extraction mechanism suitable for fields, oscillators, spins, and fermions. We compare that static approximation to our full dynamical description with an essential back action of extraction on spin-environment interaction. In this way, we thoroughly answer the question about autonomous coherence extractability. To provide an in-depth analysis, we adopt the spin as the output system, as illustrated in 
Fig.~\ref{fig:fig_1}~(a). However, as the spin is weakly excited, the results remain valid for the oscillator of field mode as the output system. For the sake of completeness, we present an adequate extension to fermionic systems. Importantly, we confirm that the autonomous coherence is generically extractable to all these output systems. It paves a road to the first experimental tests. In the full dynamical description, extracted coherence nontrivially depends on the environment temperature and characteristic system frequencies. For the spin system, we reach coherence maximum asymptotically for a low temperature. Interestingly, the coherence maximum appears at finite temperature for a fermionic analogue. It is stimulating for further proposals to obtain autonomous quantum coherences.

\section{Spin-boson model}

\begin{figure}
	\centering
	\includegraphics[width=\linewidth]{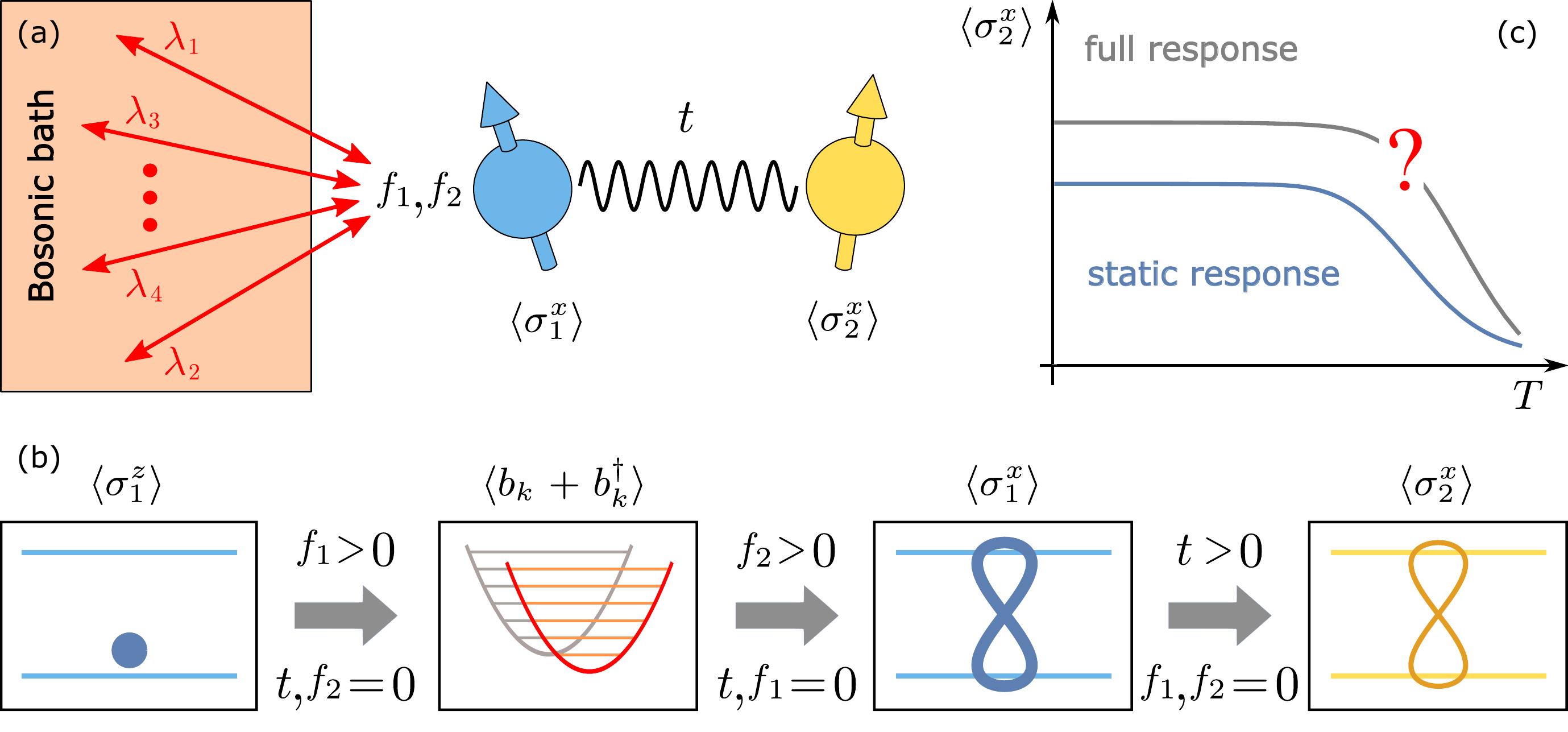}
	\caption{\label{fig:fig_1} {\bf Extraction of autonomous quantum coherence.} {\bf (a)} Model of the system. 
The internal spin $1$ with a coherence $\langle\sigma_1^x\rangle$ interacts with the bosonic degrees of freedom of the environmental bath, represented by the red two-headed arrows. The intensity of interaction with the $k$-th bosonic mode is characterized by the weight parameter $\lambda_k$. The coupling of the internal spin with the bath is defined by the pair of real parameters $f_1,f_2$. The internal spin interacts with the output spin $2$ with a coherence 
$\langle\sigma_2^x\rangle$ via spin-spin coupling term. The strength of coupling is defined by the parameter $t$. The bath temperature is $T$. {\bf (b)} Scheme of derivation the coherence of the output spin 
$\langle\sigma_2^x\rangle$ in the static response approximation. The internal spin with the non-zero thermal average 
$\langle\sigma_1^z\rangle$ but vanishing $\langle\sigma_1^x\rangle$ (first panel) induces the non-zero shift of the boson degrees of freedom 
$\langle b_k+b_k^\dag\rangle$ (second panel) via the $f_1$ term. The bosonic shift acts back on to the output spin via 
the $f_2$ term and generates a non-zero coherence of the internal spin $\langle\sigma_1^x\rangle$ (third panel). The part of this coherence is transferred to the output spin via the $t$ term and generates non-zero coherence $\langle\sigma_2^x\rangle$ (fourth panel). {\bf (c)} The sketch of the temperature dependence of the extracted coherence $\langle\sigma_2^x\rangle$ derived in the static response approximation (blue curve) and in the quantum full dynamical response (gray curve). }
\end{figure}

To analyze the extraction of autonomous quantum coherence, we minimally extend the model 
\cite{Guarnieri2018,Purkayastha2020}. Fig.~\ref{fig:fig_1}~(a) schematically illustrates the model we use to describe the extraction of quantum coherence. The Hamiltonian of the model $H=H_{12}+H_B+H_{B1}$ consists of three parts, described below. We consider a single (internal) spin $1$ coupled weakly with a strength $t$ to an output spin $2$. The output spin $2$ will extract the quantum coherence of the initial spin $1$. The overall two-spin Hamiltonian
\begin{equation}
\label{hoping}
H_{12}=\frac{\hbar\omega_1}{2}\sigma_1^z+\frac{\hbar\omega_2}{2}\sigma_2^z + 
\hbar t (\sigma_1^+\sigma_2^-+\sigma_2^+\sigma_1^-),
\end{equation}
describes the spin-spin coupling, where $\hbar\omega_1$ and $\hbar\omega_2$ are the energies of the first and second spins, respectively. Here the Pauli spin operators $\sigma_1^j$ and $\sigma_2^j$ (with $j=x,y,z$) correspond to the spins $1$ and $2$. The raising $\sigma_n^+$ and lowering $\sigma_n^-$ operators of $n(=1,2)$-th spin are 
$\sigma_n^\pm=(\sigma_n^x\pm i\sigma_n^y)/2$. Note that, the coupling term in Eq.~\eqref{hoping} is not able to generate the output coherence if it is not present in the internal spin. As the coupling is weak $t\ll \omega_1,\omega_2$ and the internal spin $1$ is weakly excited, any excitation of the output spin $2$ will be small too. In this limit, we can equivalently consider the output spin $2$ as the two lowest levels of an oscillator or field of radiation. It generalizes the description and covers detailed analysis also for the proposal \cite{Purkayastha2020}. Numerical simulations check the validity of weak coupling approximation for small $t<\omega_1,\omega_2$, see Appendices~~\ref{sec:coherence_first_spin} and \ref{sec:coherence_second_spin} for details. 

As in the previous works \cite{Guarnieri2018,Purkayastha2020}, the internal spin $1$ is coupled to a multimode phononic bath (generally, bosonic bath) described by the Hamiltonian
\begin{equation}
\label{bath}
H_B=\sum_k \Omega_k b_k^\dag b_k,
\end{equation}
through the engineered spin-boson coupling
\begin{equation}
\label{inter}
H_{B1}=(f_1\sigma_1^z+f_2\sigma_1^x)\sum_k \lambda_k(b_k^\dag+b_k).
\end{equation}
The boson system is described by creation $b_k^\dag$ and annihilation $b_k$ operators and spectrum $\Omega_k>0$. 
Subscript $k$ parameterizes all the boson degrees of freedom. The interaction between the internal spin $1$ and bosons are defined by real-valued coupling constants $f_1$ and $f_2$ \cite{Guarnieri2018,Purkayastha2020}. Steady-state equilibrium quantum coherence autonomously appears at a low temperature if both $f_1,f_2$ are non-vanishing 
\cite{Guarnieri2018,Purkayastha2020}. In Ref.~\cite{Guarnieri2018}, the Hamiltonian 
without the output subsystem was considered, and hence the extraction of autonomous quantum coherence remained an open problem. In Ref.~\cite{Purkayastha2020}, a perturbative {\em static} analysis of the extraction predicted a coherence of the output cavity-field coupled to the spin $1$. As is described in Fig.~\ref{fig:fig_1}~(b), such a static approximation does not consider the crucial back-action of the output system on the spin-environment. Although, the static approach is an inspiring approximation, the complete dynamical description is needed to address extractability convincingly. In such a dynamical theory, extractable quantum coherence can not only change, as depicted schematically at Fig.~\ref{fig:fig_1}~(c), but it could also potentially vanish, which would be critical.

In the full treatment of the perturbative analysis, quantum coherence of the output spin $2$ in our spin model 
(see Appendices~\ref{sec:spin_boson_hamiltonian}, \ref{sec:perturbation_analysis}, \ref{sec:averaging_procedure},
\ref{sec:integral_representation}, and \ref{sec:coherence_detuned_second_spin} for details) can reach
\begin{eqnarray}
\label{coherence_s_2}
\langle\sigma_2^x\rangle_T =
\frac{4f_1 f_2 t}{\omega_2}\tanh\Big(\frac{\beta\hbar\omega_2}{2}\Big)\tanh\Big(\frac{\beta\hbar\omega_1}{2}\Big)
\int_0^\infty d\xi\,\mathcal{I}(\xi)
\frac{\xi \coth\Big(\frac{\beta\xi}{2}\Big)-\hbar\omega_1\coth\Big(\frac{\beta\hbar\omega_1}{2})}{\xi(\xi^2-\hbar^2\omega_1^2)},
\end{eqnarray}
where $\beta\equiv(k_BT)^{-1}$ is the inverse temperature in energy units, $k_B$ is the Boltzmann constant, $T$ is the temperature of the system in Kelvin scale and 
$\mathcal{I}(\xi)\equiv\sum_k\lambda_k^2\delta(\xi-\Omega_k)$ is the bath spectral density function.
Simultaneously, the mean value $\langle \sigma_2^y\rangle_T$ is zero. The extracted quantum coherence is proportional to $f_1$ and $f_2$ as in the previous analysis \cite{Purkayastha2020}. Similarly, it linearly scales with the coupling strength $t$ of the interaction term in Eq.~\eqref{hoping}. The frequency dependence on $\omega_1$ and $\omega_2$ is factorized. Note that the coherence of the second spin scales linearly with the coherence of the first spin 
\begin{equation}
\label{coherence_s_1}
\langle\sigma_1^x\rangle_T=-4f_1f_2 \tanh\Big(\frac{\beta\hbar\omega_1}{2}\Big)\int_0^\infty d\xi\,\mathcal{I}(\xi)
\frac{\xi\coth\Big(\frac{\beta\xi}{2}\Big)-\hbar\omega_1\coth\Big(\frac{\beta\hbar\omega_1}{2}\Big)}{\xi\left(\xi^2-\hbar^2\omega_1^2\right)},
\end{equation} 
derived in Appendix~\ref{sec:coherence_detuned_first_spin} and satisfies the relation  
\begin{equation}
\label{main1}
\langle\sigma_2^x\rangle_T=-\frac{t}{\omega_2}\tanh\Big(\frac{\beta\hbar\omega_2}{2}\Big)\langle\sigma_1^x\rangle_T,
\end{equation}
valid for any spectral density function  $\mathcal{I}(\xi)$. 
This property can be easily observed in the static approximation 
(see details in Appendix~\ref{sec:qualitative_analysis_bosons}), where the coherence of the second spin 
is derived in three steps, depicted in Fig.~\ref{fig:fig_1}~(b). 
In the first step, supposing the smallness of the parameter $t$, one considers the spins as non-interacting and 
estimates the non-zero thermal average of the first spin $\langle\sigma_1^z\rangle_T^{st}=-\tanh(\beta\hbar\omega_1/2)$ 
(first panel in Fig.~\ref{fig:fig_1}~(b)). Here, the superscript ``{\it st}'' means ``static''. Then, replacing the operator $\sigma_1^z$ by its average value 
$\langle\sigma_1^z\rangle_T^{st}$, one gets the linear in $(b_k+b_k^\dag)$ correction to the bosonic Hamiltonian. 
This correction causes the non-zero average value 
$\langle b_k+b_k^\dag\rangle_T^{st}=-2f_1\langle\sigma_1^z\rangle_T^{st}\lambda_k/\Omega_k$ for the $k$-th bosonic mode (second panel in Fig.~\ref{fig:fig_1}~(b)). In the second step, the average $\langle b_k+b_k^\dag\rangle_T^{st}$ generates the non-zero coherence of the first spin $\langle\sigma_1^x\rangle^{st}_T\approx -4f_1f_2[\langle\sigma_1^z\rangle_T^{st}]^2\Omega/(\hbar\omega_1)$  due to the $f_2\sigma_1^x$ term in Hamiltonian \eqref{inter} 
(third panel in Fig.~\ref{fig:fig_1}~(b)). 
Here $\Omega\equiv\sum_k\lambda_k^2/\Omega_k=\int_0^\infty d\xi\,\mathcal{I}(\xi)/\xi$. Finally, $\langle\sigma_1^x\rangle_T^{st}$ induces the non-zero coherence of the second spin 
$\langle\sigma_2^x\rangle_T^{st}\approx -(t/\omega_2)\tanh(\beta\hbar\omega_2/2)\langle\sigma_1^x\rangle_T^{st}$ due to
the spin-spin interaction term in Hamiltonian \eqref{hoping} (fourth panel in Fig.~\ref{fig:fig_1}~(b)). Note that the obtained result is valid only if $t\ll|\omega_1-\omega_2|,\omega_1,\omega_2$, since under these conditions both spins can be considered separately and weakly interacting, while the results \eqref{coherence_s_2} and \eqref{coherence_s_1} are valid also for $\omega_1=\omega_2$ case (see details in Appendices~\ref{sec:coherence_first_spin} and \ref{sec:coherence_second_spin}).

Both temperature and frequency dependency, however, needs more detailed analysis beyond the static approximation.
For the generic case of the ohmic spectral density function $\mathcal{I}(\xi)=A\xi\exp(-\xi/\hbar\omega_c)$ 
\cite{Breuer2007, Weiss2012}, the coherence expression reads
\begin{equation}
\langle\sigma_2^x\rangle_T=\frac{4 A f_1 f_2 t}{\omega_2 } \tanh\left(\frac{\beta\hbar\omega_2}{2}\right)
\tanh\left(\frac{\beta\hbar\omega_1}{2}\right)\mathcal{F}\Big(\frac{\omega_1}{\omega_c},\beta\hbar\omega_1\Big),
\end{equation}
where
\begin{equation}
\mathcal{F}\Big(\frac{\omega_1}{\omega_c},\beta\hbar\omega_1\Big)=\int_0^\infty dx\,
e^{-\frac{\omega_1}{\omega_c}x}\,\frac{x\coth \left(\frac{\beta\hbar\omega_1 x}{2}\right)
	-\coth\left(\frac{\beta\hbar\omega_1}{2}\right)}{x^2-1},
\end{equation}
is a dimensionless function of parameters $\omega_1/\omega_c$ and $\beta\hbar\omega_1$. 
In the low-temperature limit $\beta\hbar\omega_1\gg1$, quantum coherence reaches
\begin{equation}
\langle\sigma_2^x\rangle_T\approx\frac{4 A f_1 f_2 t}{\omega_2} \tanh\left(\frac{\beta\hbar\omega_2}{2}\right)
\Big[-e^{\frac{\omega_1}{\omega_c}}\textrm{Ei}\Big(-\frac{\omega_1}{\omega_c}\Big)
-\frac{\pi^2}{3\beta^2\hbar^2\omega_1^2}\Big],
\end{equation}
where $\textrm{Ei}(x)=-\int_{-x}^\infty dz\, e^{-z}/z$ is the exponential integral function \cite{Bateman1953}.

\begin{figure}
	\centering
	\includegraphics[width=\linewidth]{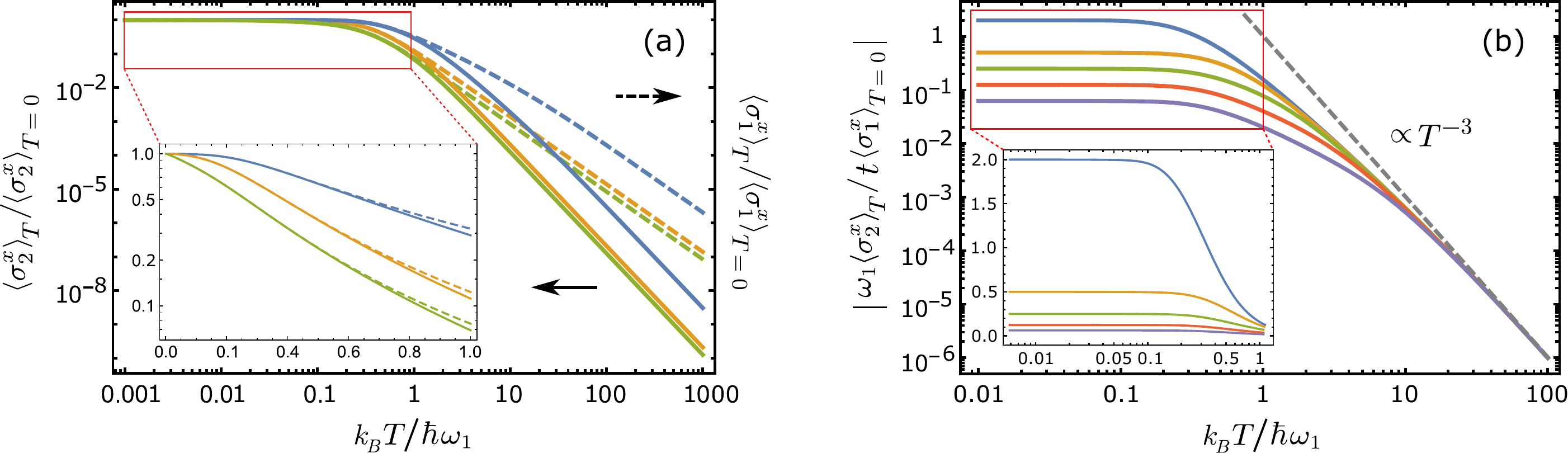}
	\caption{\label{fig:fig_2} {\bf Normalized autonomous coherences in the spin-boson model.} 
	  {\bf (a)} Log-log scale plot of normalized coherences
		$\langle\sigma_2^x\rangle_T/\langle\sigma_2^x\rangle_{T=0}$ (solid lines) and
		$\langle\sigma_1^x\rangle_T/\langle\sigma_1^x\rangle_{T=0}$ (dashed lines) for different values of
		$\omega_1/\omega_c=0.01,1,100$ (blue, yellow and green curves, respectively) for the particular case of
		$\omega_2/\omega_1=3$ as a function of temperature $T$. Here $\omega_c$ is the cut-off energy of the
		bosonic spectral function of the ohmic type. The inset is the log-linear scale plot of the same functions 
		with the same color notation and parameters. It represents the low-temperature behavior of normalized coherences. 
		The extractable coherence of the spin $2$ starts to increase fast for $k_BT<\hbar\omega_1$ and saturates sooner at
		the maximum for smaller $\omega_1/\omega_c$. The backaction of the extraction process does not slow coherence 
		increase in the low-temperature region. {\bf (b)} Log-log scale plot of absolute value of dimensionless ratio
		$\langle\sigma_2^x\rangle_T/\langle\sigma_1^x\rangle_{T=0}$ multiplied by the normalization factor 
    $\omega_1/t$ for $\omega_1/\omega_c=0.01$ and different values of $\omega_2/\omega_1=0.5,2,4,8,16$ 
    (blue, yellow, green, red and purple curves, respectively) as a function of temperature $T$. It represents the 
		dependence of the coherence $\langle\sigma_2^x\rangle_T$ on $\omega_2$. The gray dashed line depicts
		the asymptote of all curves, which scales as $T^{-3}$. The inset shows the linear-log plot of the same functions
		with the same color notation and parameters at small temperatures. It is advantageous to reach 
		$\omega_2/\omega_1<1$, to extract quantum coherence to the spin $2$ more efficiently.}
\end{figure}
Both the internal spin $1$ and output spin $2$ coherences, normalized to their asymptotic values at zero temperature, are compared in Fig.~\ref{fig:fig_2}~(a). Both the coherences increase {\em monotonously} with temperature reduction as in the static approximation. They demonstrate nearly linear behavior at log-log scale for the high temperature region $k_BT/\hbar\omega_1\gg 1$. Therefore, the normalized coherences have a power-law dependence,
\begin{align}
\langle\sigma_1^x\rangle_{T\rightarrow \infty}&\approx -f_1f_2\frac{\Omega}{3\hbar\omega_1}
(\beta\hbar\omega_1)^2=
 -f_1f_2A\frac{\omega_c}{3\omega_1}(\beta\hbar\omega_1)^2,\\ 
\langle\sigma_2^x\rangle_{T\rightarrow \infty}&\approx \frac{f_1f_2t}{\omega_1}\frac{\Omega}{6\hbar\omega_1}
(\beta\hbar\omega_1)^3=\frac{Af_1f_2t}{\omega_1}\frac{\omega_c}{6\omega_1}
(\beta\hbar\omega_1)^3,
\end{align}
which, however, remain valid only for $k_BT\gg\hbar\omega_1$. 
Note that the aforementioned power-law behavior remains valid for the case of an arbitrary spectral density function $\mathcal{I}(\xi)$ if the condition $\int_0^\infty d\xi\,\mathcal{I}(\xi)/\xi<+\infty$ is satisfied 
(see Eq.~(\ref{eq:ht_limit_spin}) in Appendix~\ref{sec:coherence_detuned_second_spin}). 
In particular, it is valid for the generalized Ohmic spectral density function $\mathcal{I}(\xi,s)\sim \xi^s\exp(-\xi/\hbar\omega_c)$ with $s>0$ (sub-Ohmic for $0<s<1$ and super-Ohmic for $s>1$, respectively).
 Extracted spin $2$ coherence increases with a larger slope at the log-log scale. At the high-temperature limit, autonomous coherence is relatively negligible. As the temperature reduces to $k_BT\approx \hbar\omega_1$, both coherences increase almost up to their maximum values and the normalized output coherence gets closer to the internal one. The latter is evident from the inset of 
Fig.~\ref{fig:fig_2}~(a). Note that
a wider band of the spectral function with smaller $\omega_1/\omega_c$ can already stimulate a faster increase of extractable coherence. Reducing temperature further, the coherence saturates when $k_BT/\hbar\omega_1\ll 1$ much earlier for $\omega_1/\omega_c \ll 1$. It reaches maximum asymptotically at zero temperature. It is advantageous to have $\omega_1/\omega_c$ as small as possible to reach a plateau near the maximum of extractable coherence sooner by lowering the temperature.  Indeed, the Eq.~\eqref{main1} proves quantum coherence is extractable, expectably, scaled by the coupling strength $t$. The extraction depends both on temperature $T$ and exclusively on $\omega_2$. Frequency $\omega_1$ does not advantageously affect the extraction process in this approximation. For large temperature, the transfer coefficient in \eqref{main1} turns to $\langle\sigma_2^x\rangle_T=-\frac{\hbar t}{2k_B T}\langle\sigma_1^x\rangle$. High temperatures thus limit the yielding of the extraction process; on the other hand, for small temperatures, the extraction follows 
$\langle\sigma_2^x\rangle_T=-\frac{t}{\omega_2}\langle\sigma_1^x\rangle$ and extraction is not limited by the temperature. Lower $\omega_2$ is preferable as it increases the extraction. It approaches the result for the static response qualitatively \cite{Purkayastha2020}.

To visualize dependence of the output spin $2$ coherence as a function of temperature for different values of 
$\omega_2$ we consider the following dimensionless ratio
\begin{equation}
\frac{\omega_1}{t}\frac{\langle\sigma_2^x\rangle_T}{\langle\sigma_1^x\rangle_{T=0}}=
\frac{\omega_1}{\omega_2}\tanh\left(\frac{\beta\hbar\omega_2}{2}\right)\tanh
\left(\frac{\beta\hbar\omega _1}{2}\right)\frac{\mathcal{F}\Big(\frac{\omega_1}{\omega_c},\beta\hbar\omega_1\Big)}{
	e^{\frac{\omega_1}{\omega_c}}\textrm{Ei}\Big(-\frac{\omega_1}{\omega_c}\Big)}\propto \langle\sigma_2^x\rangle_T
\end{equation}

Note that for a fixed ratio $\omega_1/\omega_c$ the coherence of the second spin at large temperatures becomes independent of $\omega_2$ and demonstrates a universal behavior. In particular, this feature can be observed from the plots of
the function
\begin{equation}
\Big|\frac{\omega_1}{t}\frac{\langle\sigma_2^x\rangle_{T}}{\langle\sigma_1^x\rangle_{T=0}}\Big|_{T\rightarrow\infty}\approx
\frac{\omega_c}{24\omega_1}\frac{(\beta\hbar\omega_1)^3}{e^{\frac{\omega_1}{\omega_c}}
\Big|\textrm{Ei}\Big(-\frac{\omega_1}{\omega_c}\Big)\Big|}
\propto T^{-3},
\end{equation}
for different $\omega_2$. The transfer of coherence from spin $1$ to spin $2$ is visualized in
Fig.~\ref{fig:fig_2}~(b) for an optimal choice of small $\omega_1/\omega_c$ and different $\omega_2/\omega_1$. 
For high temperatures, $k_BT/\hbar\omega_1\gg1$, the increase is linear in the log-log plot with the scaling $T^{-3}$ as predicted above. The extraction is inefficient, and the temperature reduction down to $k_BT/\hbar\omega_1\ll 1$ is needed to reach efficient transfer of the quantum coherence. A smaller frequency $\omega_2$ of the output spin $2$ is very advantageous in this low-temperature regime as longer as the perturbation approach holds. Transfer to the output spin $2$ coherence is substantially more efficient when 
$\omega_2<\omega_1$. Together with the previous condition, the optimal extraction to the output spin 
(and oscillator or field) needs $\omega_2<\omega_1<\omega_c$.

Note that the coherence of the spin 2 in the low-temperature limit $\beta\hbar\omega_1\gg1$ can be approximated as 
\begin{align}
\label{st_dyn_terms}
\langle\sigma_2^x\rangle_{T\rightarrow 0} \approx
\frac{4f_1 f_2 t}{\hbar\omega_1\omega_2}
\int_0^\infty d\xi\,\mathcal{I}(\xi)\Big[
\frac{1}{\xi}-\frac{1}{\xi+\hbar\omega_1}\Big]=\frac{4f_1 f_2 t}{\hbar\omega_1\omega_2}\Omega-
\frac{4f_1 f_2 t}{\hbar\omega_1\omega_2}\int_0^\infty d\xi\,\frac{\mathcal{I}(\xi)}{\xi+\hbar\omega_1}.
\end{align}
The first term of this expression is nothing but the coherence of spin 2 in the static approximation 
$\langle\sigma_2^x\rangle_T^{st}$ at zero temperature. In this term the contribution of the bosonic bath is 
fully factorized and represented only by the parameter $\Omega$. The factorization reflects the fact that 
the back-action effects are not taken into account in the static approximation. The result is not sensitive to the 
shape of the spectral density function $\mathcal{I}(\xi)$, since two different spectral density functions 
with the same $\Omega$ give the same value of the coherence $\langle\sigma_2^x\rangle_T^{st}$.
The second term can be called {\it dynamical} since it provides the back-action correction to the static contribution. 
Note that the frequency parameter $\omega_1$ of the spin 1 and energy dependent spectral density function 
$\mathcal{I}(\xi)$ are included in the expression inseparably in contrast to the static result. Therefore, the dynamical 
term is sensitive to the shape of the spectral density function $\mathcal{I}(\xi)$. Finally, the second term represents the deviation of the full coherence $\langle\sigma_2^x\rangle_T$ form the static result
$\langle\sigma_2^x\rangle_T^{st}$ 
and hence its value as a function of the temperature answers the question depicted in Fig.~\ref{fig:fig_1}~(c).

Since $\mathcal{I}(\xi)>0$, the static approximation provides only the upper bound 
of the extracted coherence of spin 2 at low temperatures. Therefore, the correct estimation of such a 
coherence needs a better understanding of the dynamical term. However, it seems that this term 
doesn't have an intuitive explanation in terms of the spin-boson model. Moreover, the perturbation analysis of the dynamical contribution in the spin language is bulky due to the non-commutativity of spin operators and the absence of the Wick theorem for them. This makes the analysis of the static and dynamical terms in higher orders of perturbation theory hard. 

Therefore, beyond a general interest, it can be also mathematically valuable to rewrite the spin Hamiltonian, for example, in terms of fermions where the Wick theorem is applicable. It allows the use of the Green function technique and, potentially, makes the final result clearer for understanding. This analysis goes beyond the previous theory for the spins \cite{Guarnieri2018,Purkayastha2020}, and helps to compare the spin and fermionic systems coupled to the same bosonic baths. Additionally, such an analysis can stimulate a broader class of experimental tests and discussions about a coherence in fermionic systems.

\section{Fermion-boson model}

In order to implement the aforementioned fermionization program, we apply the Jordan-Wigner transformation (see \cite{Lieb1961} and references therein)  to the Hamiltonians \eqref{hoping} and \eqref{inter}.
First, we introduce auxiliary (zeroth) subsystem, which is described by the operators $\{\sigma_0^x, \sigma_0^y, \sigma_0^z\}$ and has zero energy. Then we replace the spin operators by the fermionic ones  $a_n,n=0,1,2$, with the standard anti-commutation rules 
$\{a_n,a_m^\dag\}=\delta_{nm}$ 
\begin{align}
\sigma_0^-&\rightarrow a_0, \quad \sigma_0^+\rightarrow a_0^\dag, \quad \sigma_0^z\rightarrow 2a_0^\dag a_0-1, \\
\sigma_n^-&\rightarrow \prod_{m=0}^{n-1}(2a_m^\dag a_m-1)a_n, 
\quad \sigma_n^+\rightarrow \prod_{m=0}^{n-1}(2a_m^\dag a_m-1)a^\dag_n, \quad \sigma_n^z\rightarrow 2a_n^\dag a_n-1,
\end{align}
and make the additional substitutions
\begin{align}
\sigma_1^z=&2a_1^\dag a_1-1\rightarrow 2a_1^\dag a_1, \quad
\sigma_2^z=2a_2^\dag a_2 -1\rightarrow 2a_2^\dag a_2, \\
\label{first_fermion_coherence}
\sigma_1^x=&(\sigma_1^++\sigma_1^-)\rightarrow (\sigma_1^+\sigma_0^-+\sigma_0^+\sigma_1^-)
=-(a_1^\dag a_0+a_0^\dag a_1),
\end{align}
to remove the constant terms, which do not affect the quantum dynamics significantly.   
Then the overall fermion-boson Hamiltonian takes the following form
\begin{align}
\label{hamiltonian_F}
H_F =& \hbar\omega_1 a_1^\dag a_1+\hbar\omega_2 a_2^\dag a_2-\hbar t(a_1^\dag a_2+a_2^\dag a_1)+
\sum_k \Omega_k b_k^\dag b_k+ \\ \nonumber &+
[2f_1a_1^\dag a_1-f_2(a_1^\dag a_0+ a_0^\dag a_1)]\sum_k \lambda_k(b_k^\dag+b_k)
\end{align}
from the sum of spin-boson Hamiltonians \eqref{hoping}, \eqref{bath} and \eqref{inter}.
Note that the fermionization procedure changes a sign of the spin-spin $t\rightarrow -t$ and spin-boson 
$f_2\rightarrow -f_2$ interaction terms. In the following, we will use the subscript $F$ to indicate that we derived the coherences from the fermionized Hamiltonian $H_F$.	

Considering same ohmic spectral characteristics of bosonic bath as for the previous two-spin analysis, we can define and derive the {\it fermionic coherence} of the first spin using the Eq.~\eqref{first_fermion_coherence} (see Appendices~\ref{sec:fermionization} and \ref{sec:green_function} for details)
\begin{equation}
\label{fermcoh1}
\langle\sigma_1^x\rangle_{F,T}\equiv-\langle a_1^\dag a_0+a_1^\dag a_0\rangle_F=
-4f_1f_2A \tanh\Big(\frac{\beta\hbar\omega _1}{2}\Big)\Big[-\frac{\omega_c}{2\omega_1}+
\frac12\mathcal{F}\Big(\frac{\omega_1}{\omega_c},\beta\hbar\omega_1\Big)\Big].
\end{equation}
Here  $\langle\dots\rangle_F$ is the thermal average with the Hamiltonian $H_F$. 
The final expression represents the leading contribution to the coherence. 
We define the fermion coherence of the spin 2 by analogy with the expression for the coherence of the spin 1 
and obtain
\begin{equation}
\label{fermcoh2}
\langle\sigma_2^x\rangle_{F,T}\equiv-\langle a_2^\dag a_0+a_0^\dag a_2\rangle_F=
-\frac{4t f_1f_2A}{\hbar \Delta\omega}\mathcal{F}_F
\Big(\frac{\omega_1}{\omega_c},\beta\hbar\omega_1,\beta\hbar\omega_2\Big),
\end{equation}
where  $\Delta\omega=\omega_2-\omega_1$ and 
\begin{align}
\mathcal{F}_F
\Big(\frac{\omega_1}{\omega_c},\beta\hbar\omega_1,\beta\hbar\omega_2\Big) = 
\frac{\omega_c}{\omega_2}n(\hbar\omega_1)[1-2n(\hbar\omega_2)]+
\int_{-\infty}^\infty dx e^{-\frac{\omega_1}{\omega_c}|x|}\mathcal{K}(x),
\end{align}
with frequency-dependent kernel 
\begin{align}
\mathcal{K}(x)=[N(\hbar\omega_1 x)+n(\hbar\omega_1)]
\Big\{\frac{x}{2(x-1)}\frac{\Delta\omega}{\omega_2}-
\frac{n[\hbar\omega_1(1-x)]x}{(x-1)(x+\frac{\Delta\omega}{\omega_1})}\frac{\Delta\omega}{\omega_1}+
\frac{n(\hbar\omega_2)\omega_1x}{\omega_2(x+\frac{\Delta\omega}{\omega_1})}\Big\}.
\end{align}
We introduced the notations for the Bose-Einstein $N(\xi)=(e^{\beta\xi}-1)^{-1}$ and the Fermi-Dirac 
$n(\xi)=(e^{\beta\xi}+1)^{-1}$ distributions for brevity. Note that the coherence of each spin is a sum of the Hartree and Fock contributions (see Appendices~\ref{sec:fock}, \ref{sec:hartree}, \ref{sec:analysis_of_coherences} and 
\ref{sec:omega_2_small}  for details). It turns out that the Hartree contribution of $j(=1,2)$-th spin coincides with the fermionic coherence of the $j$-th spin obtained in the static approximation (see Appendix~\ref{sec:qualitative_fermions})
\begin{equation}
\langle\sigma_1^x\rangle_{F,T}^{st}\approx \frac{4f_1f_2\Omega}{\hbar\omega_1}n(\hbar\omega_1)
\tanh\Big(\frac{\beta\hbar\omega_1}{2}\Big),
\end{equation}
\begin{equation}
\langle\sigma_2^x\rangle_{F,T}^{st}\approx \frac{4f_1f_2\Omega t}{\hbar\Delta\omega}n(\hbar\omega_1)
\left[\frac{\tanh\Big(\frac{\beta\hbar\omega_1}{2}\Big)}{\omega_1}-\frac{\tanh\Big(\frac{\beta\hbar\omega_2}{2}\Big)}{\omega_2}\right].
\end{equation}
Like in the previous spin-boson case, the influence of the boson bath is fully factorized and represented by the same parameter $\Omega$ for these static terms. The factorization has a natural explanation in the fermion picture. 
Namely, it comes from the structure of the expression for the Hartree term (see Eq.~\eqref{eq:hartree} in
Appendix~\ref{sec:green_function}).        
Then, the dynamical contribution to the fermion coherences is defined by the Fock terms. 
Again, similarly to the spin-boson model, the mutual influence of the bosonic bath and the fermion subsystem 
are not separated in such a case. This property can be seen from the expression for the Fock contribution 
(see Eq.~\eqref{eq:fock} in Appendix~\ref{sec:green_function}). The clear physical meaning of the static and dynamical terms in the fermion system 
provides the potential opportunity to understand better the corresponding terms 
in spin systems via their fermionization.       
 
The final relation between the internal and output coherence reads
\begin{align}
\langle\sigma_2^x\rangle_{F,T}=\frac{t}{\omega_2}\langle\sigma_1^x\rangle_{F,T}\left[-\frac{2\omega_2}{\Delta\omega}
\coth\Big(\frac{\beta\hbar\omega_1}{2}\Big)
\frac{\mathcal{F}_F\Big(\frac{\omega_1}{\omega_c},\beta\hbar\omega_1,\beta\hbar\omega_2\Big)}{\frac{\omega_c}{\omega_1}-
	\mathcal{F}\Big(\frac{\omega_1}{\omega_c},\beta\hbar\omega_1\Big)}\right]=\frac{t}{\omega_2}\langle\sigma_1^x\rangle_{F,T}\mathcal{R}(T).
\end{align}
Note that the ratio $\mathcal{R}(T)$ is a function of both  $\omega_1$ and $\omega_2$ parameters, 
while the same ratio for the spin-boson system (see Eq.~\eqref{main1}) 
is only a function of $\omega_2$.  
Similarly as for the spin coherence, we normalize \eqref{fermcoh1} and \eqref{fermcoh2} 
to their values at zero temperature
\begin{align}
\langle\sigma_1^x\rangle_{F,T=0}&=2f_1f_2A
\Big[\frac{\omega_c}{\omega_1}+e^{\frac{\omega_1}{\omega_c}}\textrm{Ei}\Big(-\frac{\omega_1}{\omega_c}\Big)\Big],\quad
\langle\sigma_2^x\rangle_{F,T=0}
=\frac{t}{\omega_2}\langle\sigma_1^x\rangle_{F,T=0},
\end{align}
which simplifies their analysis to
\begin{align}
\frac{\langle\sigma_1^x\rangle_{F,T}}{\langle\sigma_1^x\rangle_{F,T=0}}&=\tanh\left(\frac{\beta\hbar\omega_1}{2}\right)
\frac{\frac{\omega_c}{\omega_1}-\mathcal{F}\Big(\frac{\omega_1}{\omega_c},\beta\hbar\omega_1\Big)}
{\frac{\omega_c}{\omega_1}+e^{\frac{\omega_1}{\omega_c}}\textrm{Ei}\Big(-\frac{\omega_1}{\omega_c}\Big)},\\
\frac{\langle\sigma_2^x\rangle_{F,T}}{\langle\sigma_2^x\rangle_{F,T=0}}&=
-\frac{2\omega_2}{\Delta\omega}\frac{\mathcal{F}_F\Big(\frac{\omega_1}{\omega_c},\beta\hbar\omega_1,\beta\hbar\omega_2\Big)}{\frac{\omega_c}{\omega_1}+e^\frac{\omega_1}{\omega_c} \textrm{Ei}\Big(-\frac{\omega_1}{\omega_c}\Big)}.
\end{align}
The low-temperature limit relation 
$\langle\sigma_2^x\rangle_{F,T=0}=\frac{t}{\omega_2}\langle\sigma_1^x\rangle_{F,T=0}$ replicates up to the sign the previous result for the spins.
\begin{figure}
	\centering
	\includegraphics[width=\linewidth]{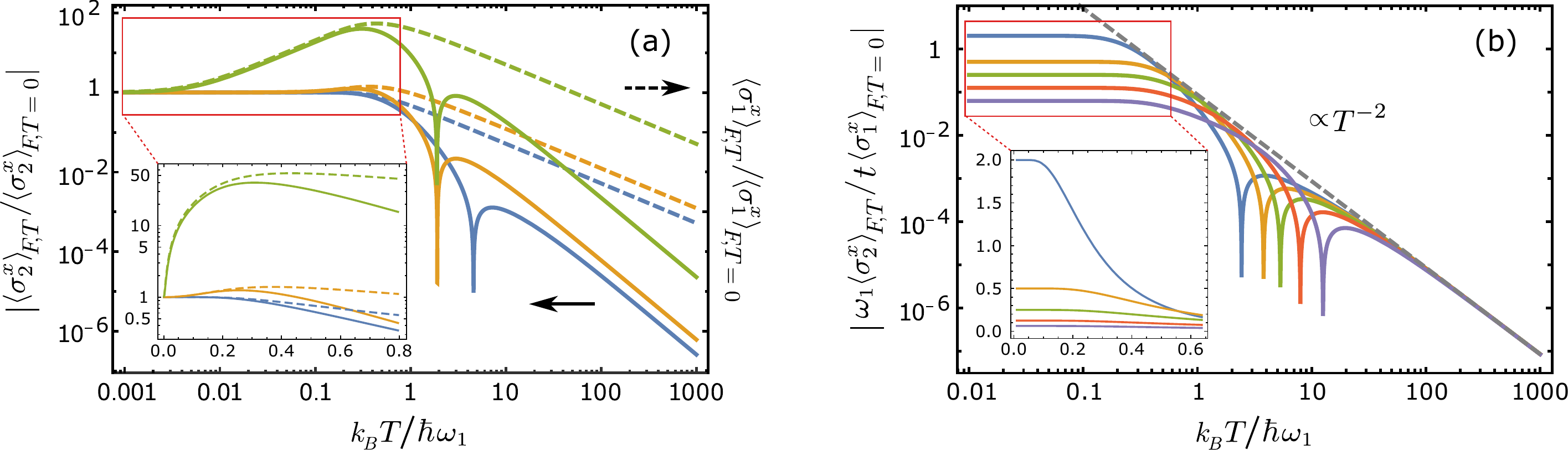}
	\caption{\label{fig:fig_3} {\bf Normalized autonomous coherences in the fermion-boson model.} {\bf (a)}
		Log-log scale plot of the absolute values of the dimensionless coherences for the fermionic system
		$\langle\sigma_2^x\rangle_{F,T}/\langle\sigma_2^x\rangle_{F,T=0}$ (solid lines) and
		$\langle\sigma_1^x\rangle_{F,T}/\langle\sigma_1^x\rangle_{F,T=0}$ (dashed lines) for different values of
		$\omega_1/\omega_c=0.01,1,100$ (blue, yellow and green curves, respectively) in the particular case 
		of $\omega_2/\omega_1=3$ as a function of temperature $T$. The inset shows an advantageous maximum of both
    internal and output coherences at a finite temperature for $\omega_1>\omega_c$. {\bf (b)} Log-log scale plot 
		of the absolute value of the dimensionless ratio
		$\langle\sigma_2^x\rangle_{F,T}/\langle\sigma_1^x\rangle_{F,T=0}$, multiplied by the normalization factor
    $\omega_1/t$, for the fixed value of $\omega_1/\omega_c=0.01$ and different values of 
		$\omega_2/\omega_1=0.5,2,4,8,16$ (blue, yellow, green, red and purple curves, respectively) as a function of 
		temperature $T$. It represents the dependence of the coherence $\langle\sigma_2^x\rangle_{F,T}$ on $\omega_2$. 
		The gray dashed line depicts the asymptote of all curves, which scales as $T^{-2}$. The inset 
		shows the coherence transfer at the low-temperatures similarly to the spin system.}
\end{figure}
The normalized internal and output coherences for the fermionic system are visualized in Fig.~\ref{fig:fig_3}~(a). In the large temperature limit, $k_BT/\hbar\omega_1\gg 1$, the increase of both coherences with decreasing temperature is linear in the log-log plot, similarly as for the spins. It means that also for fermionic systems the increase is only by a power law for large temperatures, and the coherence is negligible. On the other hand, temperature dependence of the internal coherence $\langle\sigma_1^x\rangle_{F,T}/\langle\sigma_1^x\rangle_{F,T=0}$ is {\em non-monotonous} for a large $\omega_1/\omega_c$. It has a maximum near $k_BT/\hbar\omega_1\approx 1$ with a much larger value than at zero temperature. Remarkably, this fermionic model obtains larger internal quantum coherence for a finite temperature away from the zero-temperature limit. Advantageously, the output coherence follows the same non-monotonous tendency for 
$\omega_1>\omega_c$. The origin of such nontrivial behavior results from interesting competition between Hartree and Fock terms of the inner fermionic coherence.  The sudden drop of the output coherence due to zero-crossing is practically irrelevant. The inset shows that the maximum output coherence shifts to the lower temperature with lowering $\omega_c$. However, relative output coherence increase is similar to the internal one. The fermionic system brought new positive phenomena into the autonomous quantum coherence being largest at a finite temperature.

In order to compare the spin and fermion model we consider the high temperature limit
\begin{equation}
-\frac{\omega_1}{t}\frac{\langle\sigma_2^x\rangle_{F,T\rightarrow \infty }}{\langle\sigma_1^x\rangle_{F,T=0}}=
\frac{2\omega_1}{\Delta\omega}\frac{\mathcal{F}_F\Big(\frac{\omega_1}{\omega_c},\beta\hbar\omega_1,\beta\hbar\omega_2\Big)}{\frac{\omega_c}{\omega_1}+e^\frac{\omega_1}{\omega_c} \textrm{Ei}\Big(-\frac{\omega_1}{\omega_c}\Big)}\Big|_{\beta\rightarrow 0}\approx
\frac{\omega_c}{12\omega_1}\frac{(\beta\hbar\omega_1)^2}{\frac{\omega_c}{\omega_1}+e^\frac{\omega_1}{\omega_c}\textrm{Ei}
\Big(-\frac{\omega_1}{\omega_c}\Big)}\propto (\beta\hbar\omega_1)^2,
\end{equation}
where we have taken into account 
\begin{equation}
\mathcal{F}_F\Big(\frac{\omega_1}{\omega_c},\beta\hbar\omega_1,\beta\hbar\omega_2\Big)\Big|_{\beta\rightarrow 0}\approx
\frac{\Delta\omega\omega_c}{24\omega_1^2}(\beta\hbar\omega_1)^2.
\end{equation}

Note that the high temperature behavior of the ratio 
$-\frac{\omega_1}{t}\frac{\langle\sigma_2^x\rangle_{F,T\rightarrow \infty }}{\langle\sigma_1^x\rangle_{F,T=0}}$ is independent of $\omega_2$ and has another power-low dependence
$\propto T^{-2}$ in comparison with the same ratio for the spin system, where it behaves as $\propto T^{-3}$.
Note that the obtained power-law behavior remains the same for the case of 
an arbitrary spectral density function $\mathcal{I}(\xi)$ which satisfies the condition $\int_0^\infty d\xi\,\mathcal{I}(\xi)/\xi<+\infty$ (see Eq.~(\ref{eq:ht_limit_fermion}) in Appendix~\ref{sec:analysis_of_coherences}).  
In particular, it is valid for the case of the generalized Ohmic spectral density function
$\mathcal{I}(\xi,s)\sim \xi^s\exp(-\xi/\hbar\omega_c)$ for $s>0$ (sub-Ohmic for $0<s<1$ and super-Ohmic for $s>1$, respectively).

The transfer of coherence from internal to output part in the fermionic system is analogical to the spin system. Only at high temperatures, it scales with $T^{-2}$ instead of $T^{-3}$ as for the spins. The sudden drops caused by the zero-crossings are practically irrelevant. In the relevant low-temperature regime $k_BT/\hbar\omega_1\leq 1$, the transfer improves for $\omega_2<\omega_1$ as in the previous spin analysis. For the fermions, the optimal regime to obtain and extract autonomous quantum coherence efficiently at a finite temperature needs $\omega_1\gg \omega_c$ and $\omega_2<\omega_1$. It is complementary with the spins, where the large band was optimal. Optimal extraction still profits from the small $\omega_2$ until the approximation fails.

\section{Summary}

We have hereby confirmed the extractibility of autonomous quantum coherence for all basic quantum systems coupled to a generic bosonic environment. The results justified a qualitative prediction from static approximation 
\cite{Purkayastha2020}. They proved that the back action of a weak extraction is fortunately not critical for the coherence generation and extraction in the low-temperature limit. Moreover, we obtained the required frequency and temperature dependences that are not reliably predictable from the static approximation. We optimized the extraction for spins 
(oscillators and fields) and fermions and found different roles of the environment bandwidth to obtain extractable coherence. Importantly, all those effects are numerically confirmed beyond the asymptotic limit for a finite strength of the extraction (detailed analysis is presented in Appendix). Extractability principally opens a way to experimental verifications of proposals with quantum dots \cite{Purkayastha2020} and also to others under development in the superconducting circuits \cite{Touzard2019}, electro-mechanical systems \cite{Ma2021},
and trapped ions \cite{Cai2021}.

We can extend the systems used to generate and extract quantum coherence. The fermionic system generating a maximum of coherence at a finite temperature $k_BT\leq\hbar\omega_1$ inspires further extensions. We can investigate three-level and four-level models to optimize the proposed methods. It is a viable investigation area as such systems appear for both atomic and solid-state systems. A possible maximum of quantum coherence at the finite temperature in the region 
$k_BT>\hbar\omega_1$ remains an open problem. However, further engineering of the systems and interaction may overcome such a limit. Another open problem is an accumulation of coherence in an output system from many such parallel interactions. Recently, it has been demonstrated in a collision model that multiple environments can enhance quantum coherence substantially \cite{Roman2020}. 
Further research can be focused on coherence distillation and use a possible extension of the protocols to more spins or fermions giving an even more considerable amount of the extractable quantum coherence. 
Thermodynamical analysis of autonomous coherence has been already initiated too 
\cite{Guarnieri2020}. These directions for immediate exploration illustrate the novelty of this subject in theory and stimulate current experiments. It is very timely as quantum coherence from a low-temperature environment will directly transform quantum sensors and quantum engines to be more autonomous.

\section{Acknowledgments}

A.S. and R.F. acknowledge the project 20-16577S of the Czech Science
Foundation. R.F. acknowledges also LTAUSA19099 of MEYS CR.

\bibliographystyle{plain}

%\onecolumn\newpage
\appendix

\section{Spin-boson Hamiltonian}
\label{sec:spin_boson_hamiltonian}

We consider the Hamiltonian of two interacting spins coupled to the boson system
\begin{equation}
H=\frac{\omega_1}{2}\sigma_1^z+\frac{\omega_2}{2}\sigma_2^z + t (\sigma_1^+\sigma_2^-+\sigma_2^+\sigma_1^-)+
\sum_k \Omega_k b_k^\dag b_k + (f_1\sigma_1^z+f_2\sigma_1^x)\sum_k \lambda_k(b_k^\dag+b_k).
\end{equation}
Here $\sigma_1^j$ and $\sigma_2^j$ (with $j=x,y,z$) are the Pauli matrices corresponding to the $1$-st and $2$-nd spins with the energies $\omega_1$ and $\omega_2$ respectively.The real valued parameter $t>0$ defines the coupling between the spins. Note that in this section and in further we consider $\omega_1$, $\omega_2$ and $t$ in energy units for brevity. In order to restore the original frequency units, used in the main text, one needs to make the following substitutions $\omega_1\rightarrow \hbar\omega_1$, $\omega_2\rightarrow \hbar\omega_2$ and $t\rightarrow \hbar t$. The raising $\sigma_n^+$ and lowering $\sigma_n^-$ operators of  $n(=1,2)$-th spin are $\sigma_n^\pm=(\sigma_n^x\pm i\sigma_n^y)/2$. The boson system is described by creation $b_k^\dag$ and annihilation $b_k$ operators and spectrum $\Omega_k>0$. Subscript $k$ parameterizes all the boson degrees of freedom. The interaction between spins and bosons are defined by real valued coupling constants $f_1$ and $f_2$.
It is convenient to introduce the following notations for spin Hamiltonian
\begin{equation}
H_{12}=\frac{\omega_1}{2}\sigma_1^z+\frac{\omega_2}{2}\sigma_2^z + t (\sigma_1^+\sigma_2^-+\sigma_2^+\sigma_1^-),
\end{equation}
boson Hamiltonian
\begin{equation}
H_B=\sum_k \Omega_k b_k^\dag b_k,
\end{equation}
and the spin-boson coupling term
\begin{equation}
H_{B1}=(f_1\sigma_1^z+f_2\sigma_1^x)\sum_k \lambda_k(b_k^\dag+b_k).
\end{equation}
As a first step we write the spin Hamiltonian in the basis of triplet and singlet states of two spins
\begin{align}
|1,1\rangle=&|\uparrow_1\rangle|\uparrow_2\rangle, \quad  |1,-1\rangle=|\downarrow_1\rangle|\downarrow_2\rangle, \quad
|1,0\rangle=\frac{1}{\sqrt{2}}(|\uparrow_1\rangle|\downarrow_2\rangle+|\downarrow_1\rangle|\uparrow_2\rangle), 
\\ \nonumber
|0,0\rangle=&\frac{1}{\sqrt{2}}(|\uparrow_1\rangle|\downarrow_2\rangle-|\downarrow_1\rangle|\uparrow_2\rangle),
\end{align}
where $|\uparrow_n\rangle$ and $|\downarrow_n\rangle$ are the eigenstates of the $\sigma_n^z$ operators with eigenvalues $\pm1$ respectively. In this basis the Hamiltonian has the following matrix form
\begin{equation}
H_{12}=\frac12\left[\begin{array}{cccc}
                      \omega_1+\omega_2 & 0 & 0 & 0 \\
                      0 & -\omega_1-\omega_2 & 0 & 0 \\
					  0 & 0 & 2t & \omega_1-\omega_2 \\
					  0 & 0 & \omega_1-\omega_2 & -2t
                  \end{array}\right].
\end{equation}
The non-zero difference $\delta\omega=\omega_1-\omega_2\neq 0$ mixes the spin states with zero angular momentum. We consider their linear combination in order to diagonalize the Hamiltonian matrix
\begin{equation}
|+,0\rangle=\cos(\theta/2) |1,0\rangle + \sin(\theta/2) |0,0\rangle, \quad
|-,0\rangle=-\sin(\theta/2) |1,0\rangle + \cos(\theta/2) |0,0\rangle.
\end{equation}
The  matrix becomes diagonal for the case $\cos\theta=2t/\sqrt{4t^2+\delta\omega^2}$,
$\sin\theta=\delta\omega/\sqrt{4t^2+\delta\omega^2}$. The eigenvalues become $E_1=(\omega_1+\omega_2)/2=\omega$,
$E_2=-(\omega_1+\omega_2)/2=-\omega$, $E_3=\sqrt{t^2+\delta\omega^2/4}=\epsilon$ and $E_4=-\sqrt{t^2+\delta\omega^2/4}=-\epsilon$ for the states $|1,1\rangle$, $|1,-1\rangle$,  $|+,0\rangle$ and $|-,0\rangle$ respectively. The spin operators $\sigma_1^x$ and $\sigma_1^z$ have the following form in the basis
$|1,1\rangle, |1,-1\rangle, |+,0\rangle, |-,0\rangle$
\begin{equation}
\sigma_1^x=\left[\begin{array}{cccc}
                      0 & 0 & \sin\phi & -\cos\phi \\
                      0 & 0 & \cos\phi & \sin\phi \\
											\sin\phi & \cos\phi & 0 & 0 \\
											-\cos\phi & \sin\phi & 0 & 0
                  \end{array}\right],	\quad
\sigma_1^z=\left[\begin{array}{cccc}
                      1 & 0 & 0 & 0 \\
                      0 & -1 & 0 & 0 \\
											0 & 0 & \sin\theta & \cos\theta \\
											0 & 0 & \cos\theta &  -\sin\theta
                  \end{array}\right],                                    							
\end{equation}
where $\phi=\pi/4-\theta/2$.

\section{Perturbation analysis}
\label{sec:perturbation_analysis}

We treat the Hamiltonian $H_0=H_{12}+H_B$ exactly and $H_{B1}$ as a perturbation \cite{Purkayastha2020}.
We present the density matrix operator in the form ($\beta=1/T$, $T$ is the temperature in energetic units)
\begin{equation}
e^{-\beta H}=e^{-\beta(H_{12}+H_B)}X(\beta).
\end{equation}
Here the operator $X(\beta)$ is presented as a $\mathrm{T}$-exponent
\begin{equation}
X(\beta)=\mathrm{T}\Big(-\int_0^\beta d\tau \widetilde{H}_{B1}(\tau)\Big)\approx
1-\int_0^\beta d\tau \widetilde{H}_{B1}(\tau) +
\int_0^\beta d\tau \int_0^\tau d\tau' \widetilde{H}_{B1}(\tau)\widetilde{H}_{B1}(\tau'),
\end{equation}
where
\begin{equation}
\widetilde{H}_{B1}(\tau)=e^{\tau(H_{12}+H_B)}H_{B1}e^{-\tau(H_{12}+H_B)}.
\end{equation}
Then the perturbation term $H_{B1}=SB=[f_1\sigma_1^z+f_2\sigma_1^x][\sum_k \lambda_k (b_k^\dag+b_k)]$ takes the form
\begin{align}
\widetilde{H}_{B1}(\tau)=&\widetilde{S}(\tau)\widetilde{B}(\tau)=
\\ \nonumber=&
[f_1e^{\tau H_{12}}\sigma_1^z e^{-\tau H_{12}}+f_2e^{\tau H_{12}}\sigma_1^xe^{-\tau H_{12}}\Big]
\Big[\sum_k \lambda_k (e^{\tau H_B}b_k^\dag e^{-\tau H_B}+e^{\tau H_B}b_ke^{-\tau H_B})\Big].
\end{align}
Taking into account the structure of the $H_{12}$ and $H_B$ Hamiltonians and using the notation
$\Lambda_1(\tau)=e^{\tau\omega}$, $\Lambda_2(\tau)=e^{\tau\epsilon}$
we obtain
\begin{equation}
e^{\tau H_B}b_k^\dag e^{-\tau H_B}=e^{\tau \Omega_k}b_k^\dag, \quad
e^{\tau H_B}b_ke^{-\tau H_B}=e^{-\tau \Omega_k}b_k,
\end{equation}
\begin{align}
\sigma_1^z(\tau)=e^{\tau H_{12}}\sigma_1^ze^{-\tau H_{12}}=\left[\begin{array}{cccc}
                      1 & 0 & 0 & 0 \\
                      0 & -1 & 0 & 0 \\
											0 & 0 & \sin\theta & \Lambda_2^2(\tau)\cos\theta \\
											0 & 0 & \Lambda_2^{-2}(\tau)\cos\theta &  -\sin\theta
                  \end{array}\right],
\end{align}
\begin{align}									
&\sigma_1^x(\tau)=e^{\tau H_{12}}\sigma_1^xe^{-\tau H_{12}}= \\ \nonumber
&=\left[\begin{array}{cccc}
                      0 & 0 & \Lambda_1(\tau)\Lambda_2^{-1}(\tau)\sin\phi & -\Lambda_1(\tau)\Lambda_2(\tau)\cos\phi \\
                      0 & 0 & \Lambda_1^{-1}(\tau)\Lambda_2^{-1}(\tau)\cos\phi & \Lambda_1^{-1}(\tau)\Lambda_2(\tau)     \sin\phi \\
											\Lambda_1^{-1}(\tau)\Lambda_2(\tau)\sin\phi & \Lambda_1(\tau)\Lambda_2(\tau)\cos\phi & 0 & 0 \\
											-\Lambda_1^{-1}(\tau)\Lambda_2^{-1}(\tau)\cos\phi & \Lambda_1(\tau)\Lambda_2^{-1}(\tau)\sin\phi & 0 & 0
                  \end{array}\right].								
\end{align}
For calculation of the averages of any operator we use the symmetrization of the density operator
\begin{equation}
e^{-\beta H}=\frac12\Big[e^{-\beta(H_{12}+H_B)}X(\beta)+X^\dag(\beta) e^{-\beta(H_{12}+H_B)}\Big].
\end{equation}
Then the partition function $Z$ can be expressed as
\begin{equation}
Z=\mathrm{Tr}[e^{-\beta H}]=\frac{Z_0}{2}\langle X(\beta)+X^\dag(\beta) \rangle_0.
\end{equation}
Here $\langle\star\rangle_0=Z_0^{-1}\mathrm{Tr}[e^{-\beta(H_{12}+H_B)}\star]$ is 
the averaging by the non-interaction density function.
The partition function up to second order correction takes the following form
\begin{equation}
Z=Z_0\Big\{1+\frac12\int_0^\beta d\tau\int_0^\tau d\tau' [\langle\widetilde{H}_{B1}(\tau)\widetilde{H}_{B1}(\tau')\rangle_0+\langle\widetilde{H}_{B1}^\dag(\tau')\widetilde{H}_{B1}^\dag(\tau)\rangle_0]\Big\}.
\end{equation}
Taking the average over the boson degrees of freedom we get
\begin{equation}
Z=Z_0\Big\{1+\int_0^\beta d\tau\int_0^\tau d\tau' \phi(\tau-\tau')\mathrm{Re}\langle\widetilde{S}(\tau)\widetilde{S}(\tau')\rangle_0\Big\},
\end{equation}
where we introduced the bosonic correlation function 
\begin{equation}
\phi(\tau-\tau')=\langle\widetilde{B}(\tau)\widetilde{B}(\tau')\rangle_0=
\langle\widetilde{B}^\dag(\tau')\widetilde{B}^\dag(\tau)\rangle_0=
\sum_k \lambda_k^2 [e^{\Omega_k(\tau-\tau')}\langle b_k^\dag b_k\rangle_0+
e^{-\Omega_k(\tau-\tau')}\langle b_k b_k^\dag\rangle_0].
\end{equation}
The spin dependent part takes the form
\begin{equation}
\widetilde{S}(\tau)\widetilde{S}(\tau')=f_1^2\sigma_1^z(\tau)\sigma_1^z(\tau')+f_2^2\sigma_1^x(\tau)\sigma_1^x(\tau')+
f_1f_2\sigma_1^z(\tau)\sigma_1^x(\tau')+f_1f_2\sigma_1^x(\tau)\sigma_1^z(\tau').
\end{equation}
We are interested in the spin sector only, therefore we integrate out the boson degrees of freedom in total density operator and keep only the spin dependent part $\varrho_S=\mathrm{Tr}_B[e^{-\beta H}]/Z$
\begin{align}
\varrho_S=&\frac{e^{-\beta H_{12}}}{Z_S}\Big[1-\int_0^\beta d\tau\int_0^\tau d\tau' \phi(\tau-\tau')\mathrm{Re}\langle\widetilde{S}(\tau)\widetilde{S}(\tau')\rangle_0\Big]+ \nonumber \\ 
+&\frac12\int_0^\beta d\tau \int_0^\tau d\tau'
\phi(\tau-\tau')\Big[\frac{e^{-\beta H_{12}}}{Z_S}\widetilde{S}(\tau)\widetilde{S}(\tau')+ \textrm{h.c.}\Big].
\end{align}
Here we introduced $Z_S=\mathrm{Tr}[e^{-\beta H_{12}}]=2\cosh(\beta\omega)+2\cosh(\beta\epsilon)$.

\section{Averaging procedure for the second spin operators}
\label{sec:averaging_procedure}

We are interested in the averages  $\langle\sigma_2^j\rangle=\mathrm{Tr}[\varrho_S\sigma_2^j]$,
where $j=x,y,z$. The matrices of corresponding spin operators in the basis  $\{|1,1\rangle, |1,-1\rangle, |+,0\rangle, |-,0\rangle\}$ are
\begin{align}
\sigma_2^z=&\left[\begin{array}{cccc}
                      1 & 0 & 0 & 0 \\
                      0 & -1 & 0 & 0 \\
											0 & 0 & -\sin\theta & -\cos\theta \\
											0 & 0 & -\cos\theta &  \sin\theta
                  \end{array}\right], \quad 
\sigma_2^x=\left[\begin{array}{cccc}
                      0 & 0 & \cos\phi & \sin\phi \\
                      0 & 0 & \sin\phi & -\cos\phi \\
											\cos\phi & \sin\phi & 0 & 0 \\
											\sin\phi & -\cos\phi & 0 & 0
                  \end{array}\right], \nonumber \\	
\sigma_2^y=&\left[\begin{array}{cccc}
                      0 & 0 & -i\cos\phi & -i\sin\phi \\
                      0 & 0 & i\sin\phi & -i\cos\phi \\
											i\cos\phi & -i\sin\phi & 0 & 0 \\
											i\sin\phi & i\cos\phi & 0 & 0
                  \end{array}\right].							
\end{align}
where $\phi=\pi/4-\theta/2$. In this basis the procedure of taking trace becomes matrix trace procedure.
First we calculate the leading terms
\begin{equation}
Z_S\langle\sigma_2^z\rangle_0=\mathrm{Tr}[e^{-\beta H_{12}}\sigma_2^z]=2\sin\theta\sinh(\beta\epsilon)-2\sinh[\beta\omega], \,
Z_S\langle\sigma_2^{x(y)}\rangle_0=\mathrm{Tr}[e^{-\beta H_{12}}\sigma_2^{x(y)}]=0.
\end{equation}
Then we introduce the traces of the product of the spin operators
\begin{align}
\label{eq:F_tau_tau'}
F(\tau,\tau')&=Z_S\mathrm{Re}\langle\widetilde{S}(\tau)\widetilde{S}(\tau')\rangle_0=\nonumber \\
&2f_1^2\Big[\cosh(\beta\omega)+
\sin^2\theta\cosh(\beta\epsilon)+\cos^2\theta\cosh((\beta-2\tau+2\tau')\epsilon)\Big]+\nonumber\\
&2f_2^2\Big[\cos^2\phi \{\cosh(\beta\epsilon-(\epsilon+\omega)(\tau-\tau'))+
\cosh(\beta\omega-(\epsilon+\omega)(\tau-\tau'))\}\Big]+ \nonumber \\
&2f_2^2\Big[\sin^2\phi \{\cosh(\beta\epsilon-(\epsilon-\omega)(\tau-\tau'))+
\cosh(\beta\omega+(\epsilon-\omega)(\tau-\tau'))\}\Big],
\end{align}
\begin{align}
F_1(\tau,\tau')=\mathrm{Tr}&[e^{-\beta H_{12}}\widetilde{S}(\tau)\widetilde{S}(\tau')\sigma_2^z+
\widetilde{S}^\dag(\tau')\widetilde{S}^\dag(\tau)e^{-\beta H_{12}}\sigma_2^z]= \nonumber \\
4f_1^2&\Big[\sin^3\theta\sinh(\beta\epsilon)-\sinh(\beta\omega )+
\sin\theta\cos^2\theta\{\sinh(\epsilon(\beta+2\tau'-2\tau))+ \nonumber \\ 
&+\sinh(\epsilon(\beta-2\tau'))-
\sinh(\epsilon(\beta-2\tau))\}\Big]+ \nonumber \\
4f_2^2&\Big[\sin\theta\sin^2\phi\sinh(\beta\epsilon+(\tau'-\tau)(\epsilon-\omega))+ \nonumber \\ 
&\sin\theta\cos^2\phi\sinh(\beta\epsilon+(\tau'-\tau)(\omega+\epsilon))+ \nonumber \\ &
\cos\theta\sin\phi\cos\phi\{\sinh(\beta\epsilon+\tau'(\omega-\epsilon)-\tau(\omega+\epsilon))-\nonumber \\ 
&\sinh(\beta\epsilon-\tau'(\omega+\epsilon)+\tau (\omega-\epsilon))\}- \nonumber \\ &
\sin^2\phi\sinh(\omega\beta+(\tau'-\tau)(\omega-\epsilon))-
\cos^2\phi\sinh(\beta\omega+(\tau'-\tau)(\omega+\epsilon))\Big],
\end{align}
\begin{align}
\label{eq:F_2_tau_tau'}
F_2(\tau,\tau')&=\mathrm{Tr}[e^{-\beta H_{12}}\widetilde{S}(\tau)\widetilde{S}(\tau')\sigma_2^x+
\widetilde{S}^\dag(\tau')\widetilde{S}^\dag(\tau)e^{-\beta H_{12}}\sigma_2^x]= \nonumber \\ &4f_1 f_2\Big[
-\cos\theta\cos^2\phi\{\cosh(\beta\omega+2\epsilon\tau'-\tau(\omega +\epsilon))+
\cosh(\epsilon(\beta-2\tau)+\tau'(\omega +\epsilon))\}+ \nonumber \\
&+\cos\theta\sin^2\phi\{\cosh(\beta\omega-2\epsilon\tau'-\tau(\omega-\epsilon))+\cosh(\epsilon(\beta-2\tau)+
\tau'(\epsilon -\omega))\}+ \nonumber \\
&+\sin(2\phi)\sin\theta\{(\cosh(\omega\tau')
\cosh(\epsilon(\beta-\tau'))+\cosh(\tau\epsilon)\cosh(\omega(\beta-\tau))\}+ \nonumber \\
&+\sin(2\phi)\{\sinh(\epsilon\tau')
\sinh(\omega(\beta-\tau'))+\sinh(\tau\omega)\sinh(\epsilon(\beta-\tau )))\}\Big],
\end{align}
\begin{equation}
\mathrm{Tr}[e^{-\beta H_{12}}\widetilde{S}(\tau)\widetilde{S}(\tau')\sigma_2^y+
\widetilde{S}^\dag(\tau')\widetilde{S}^\dag(\tau)e^{-\beta H_{12}}\sigma_2^y]=0.
\end{equation}

\section{Integral representation for the spin average values}
\label{sec:integral_representation}

The average values $\langle\sigma_2^x\rangle$ 
and $\langle\sigma_2^z\rangle$ are defined by the convolution integrals of 
$F(\tau,\tau')$, $F_1(\tau,\tau')$ and $F_2(\tau,\tau')$ with $\phi(\tau-\tau')$ in $\tau,\tau'$ domain.
Such the integration can be done in the following way. We introduce the boson spectral function
\begin{equation}
\mathcal{I}(\xi)=\sum_k \lambda_k^2\delta(\xi-\Omega_k)
\end{equation}
and present $\phi(\tau-\tau')$ in the form
\begin{equation}
\phi(\tau-\tau')=\int_0^\infty d\xi\,\mathcal{I}(\xi)
\left\{\frac{e^{\xi(\tau-\tau')}}{e^{\beta\xi}-1}+\frac{e^{-\xi(\tau-\tau')}e^{\beta\xi}}{e^{\beta\xi}-1}\right\},
\end{equation}
taking into account that $\Omega_k>0$.
In this notation we can express the non-zero corrections of average values of spin operators in the form
\begin{equation}
\Delta\langle\sigma_2^z\rangle=\langle\sigma_2^z\rangle-\langle\sigma_2^z\rangle_0=
\frac{1}{2Z_S}\int_0^\beta d\tau \int_0^\tau d\tau'\phi(\tau-\tau')
\Big[F_1(\tau,\tau')-2\langle\sigma_2^z\rangle_0F(\tau,\tau')\Big],
\end{equation}
\begin{equation}
\langle\sigma_2^x\rangle=\langle\sigma_2^x\rangle-\langle\sigma_2^x\rangle_0=
\frac{1}{2Z_S}\int_0^\beta d\tau \int_0^\tau d\tau'\phi(\tau-\tau')F_2(\tau,\tau'), 
\end{equation}
since $\langle\sigma_2^x\rangle_0=0$.
Taking into account the structure of $\phi(\tau-\tau')$ function  and introducing the transformation
\begin{equation}
\mathcal{F}(\xi)=\frac{1}{e^{\beta\xi}-1}\int_0^\beta d\tau\int_0^\tau d\tau' e^{\xi(\tau-\tau')}
\mathcal{F}(\tau,\tau'),
\end{equation}
we can write
\begin{equation}
\Delta\langle\sigma_2^z\rangle=\langle\sigma_2^z\rangle-\langle\sigma_2^z\rangle_0=
\frac{1}{2Z_S}\int_0^\infty d\xi\, \mathcal{I}(\xi)
\Big\{F_1(\xi)-F_1(-\xi)-2\langle\sigma_2^z\rangle_0[F(\xi)-F(-\xi)]\Big\},
\end{equation}
\begin{equation}
\langle\sigma_2^x\rangle=\frac{1}{2Z_S}\int_0^\infty d\xi\,\mathcal{I}(\xi)\Big\{F_2(\xi)-F_2(-\xi)\Big\}.
\end{equation}

\section{Detuned spins: the second spin coherence}
\label{sec:coherence_detuned_second_spin}

Let us consider the case when $|\omega_1-\omega_2|\gg t$. It corresponds to the situation when the energy levels of considered spins are weakly modified by interaction terms. For this case the calculation of the the averages of spin operators can be simplified. Indeed, we consider the leading terms of $\cos(\theta)$ and $\sin(\theta)$ by small parametr 
$t/|\omega_1-\omega_2|$
\begin{align}
\sin\theta=&\frac{(\omega_1-\omega_2)}{\sqrt{4t^2+(\omega_1-\omega_2)^2}}\approx
\text{sgn}(\omega_1-\omega_2)\Big[1-\frac{2t^2}{(\omega_1-\omega_2)^2}\Big],
\nonumber \\
 \cos\theta=&\frac{2t}{\sqrt{4t^2+(\omega_1-\omega_2)^2}}\approx \frac{2t}{|\omega_1-\omega_2|},
\end{align}
substitute them in the expressions for $F(\tau,\tau')$, $F_1(\tau,\tau')$ and $F_2(\tau,\tau')$ and calculate the 
corresponding integrals over $\tau,\tau'$ and $\xi$. We introduce the notations $\eta=\text{sgn}(\omega_1-\omega_2)$ and $\delta=|\omega_1-\omega_2|$ in further calculations for brevity. We focus on the evaluation of the coherence of the second spin operator $\langle\sigma_2^x\rangle$. In order to do it we substitute
\begin{align}
\sin^2\theta\approx& 1-\frac{4t^2}{\delta^2}, \quad \cos^2\theta\approx \frac{4t^2}{\delta^2},\nonumber \\
\cos^2\phi=&\frac{1+\sin\theta}{2}\approx \frac{1+\eta}{2} -\frac{\eta t^2}{\delta^2},
\quad \sin^2\phi=\frac{1-\sin\theta}{2}\approx \frac{1-\eta}{2}+\frac{\eta t^2}{\delta^2},
\end{align}
in Eq.~\eqref{eq:F_2_tau_tau'} and keep the leading term in the series expansion of this expression by the small parameter $t/\delta$. We have
\begin{align}
F_2(\tau,\tau')=\frac{8 f_1 f_2 t}{\omega _1-\omega _2}
&\Big[\sinh \left(\frac{1}{2} \left(\omega _1-\omega _2\right) \tau '\right)
\sinh \left(\frac{1}{2} \left(\omega _1+\omega _2\right) \left(\beta -\tau '\right)\right)+
\nonumber \\&+
\cosh \left(\frac{1}{2} \left(\omega _1+\omega _2\right) \tau '\right)
\cosh \left(\frac{1}{2} \left(\omega _1-\omega _2\right) \left(\beta -\tau '\right)\right)-
\nonumber \\&
-\cosh \left(\frac{1}{2} \left(\omega _1-\omega _2\right) (\beta -2 \tau )+\omega _1 \tau '\right)-
\nonumber \\&-
\cosh \left(-\frac{1}{2}\beta\left(\omega _1+\omega _2\right)+
\left(\omega_2-\omega_1\right) \tau '+\tau\omega _1\right)+
\nonumber \\&+
\sinh \left(\frac{1}{2}\tau\left(\omega_1+\omega_2\right)\right)
\sinh\left(\frac{1}{2}\left(\omega_1-\omega_2\right)(\beta -\tau )\right)+
\nonumber \\&+
\cosh \left(\frac{1}{2} \tau  \left(\omega _1-\omega _2\right)\right)
\cosh \left(\frac{1}{2} \left(\omega_1+\omega_2\right)(\beta -\tau)\right)\Big].
\end{align}
After integration over $\tau,\tau'$ we get
\begin{equation}
F_2(\xi)-F_2(-\xi)=\frac{32 f_1 f_2 t \sinh \left(\frac{\beta  \omega _2}{2}\right)\sinh \left(\frac{\beta  \omega _1}{2}\right) \left(\xi  \coth \left(\frac{\beta  \xi }{2}\right)
-\omega _1 \coth \left(\frac{\beta  \omega _1}{2}\right)\right)}{\xi  \omega _2 \left(\xi ^2-\omega _1^2\right)}
\end{equation}
Therefore the coherence $\langle\sigma_2^x\rangle$ takes the form
\begin{align}
\langle\sigma_2^x\rangle=&\frac{1}{2Z_S}\int_0^\infty d\xi\,\mathcal{I}(\xi)\Big\{F_2(\xi)-F_2(-\xi)\Big\}=\nonumber \\=&
\frac{4f_1 f_2 t}{\omega_2}\tanh\Big(\frac{\beta\omega_2}{2}\Big)\tanh\Big(\frac{\beta\omega_1}{2}\Big)
\int_0^\infty d\xi\,\mathcal{I}(\xi)
\frac{\xi \coth\Big(\frac{\beta\xi}{2}\Big)-\omega _1\coth\Big(\frac{\beta\omega_1}{2}\Big)}{\xi(\xi ^2-\omega _1^2)},
\end{align}
where we use $Z_S$ in the leading order by $t/\delta$: $Z_S=4\cosh(\beta\omega_1/2)\cosh(\beta\omega_2/2)$.

\begin{figure}
	\centering
		\includegraphics{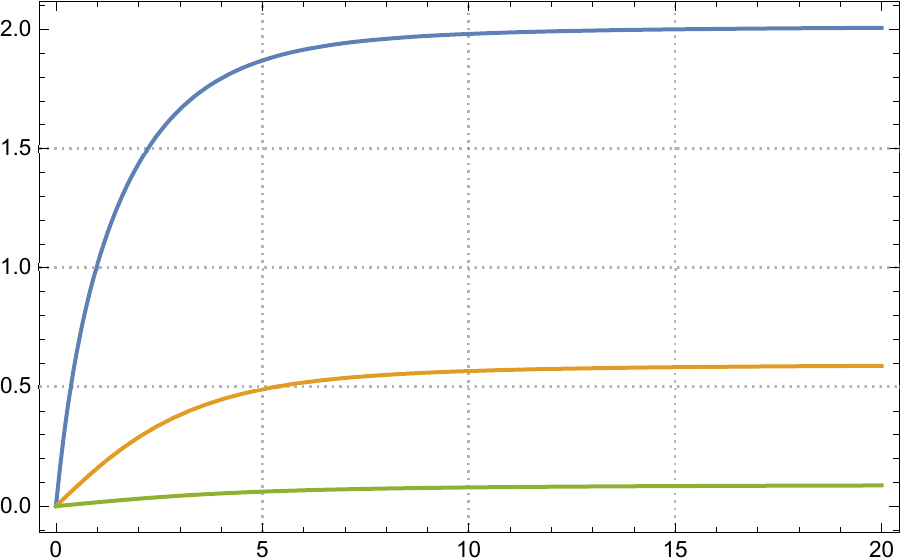}
	\caption{\label{fig:figure_2} Dimensionless function $\mathcal{F}(\omega_1/\omega_c,\beta\omega_1)$ for 
	$\omega_1/\omega_c=0.1$ (blue curve), $\omega_1/\omega_c=1$ (yellow curve), and $\omega_1/\omega_c=10$
	(green curve) as a function of dimensionless parameter $\beta\omega_1 \in[0,20]$. }
\end{figure}

For the particular case of spectral density function $\mathcal{I}(\xi)=A\xi\exp(-\xi/\omega_c)$ this expression reads
\begin{equation}
\langle\sigma_2^x\rangle=\frac{4 A f_1 f_2 t}{\omega _2 } \tanh \left(\frac{\beta  \omega _2}{2}\right)\tanh 
\left(\frac{\beta  \omega _1}{2}\right)\mathcal{F}(\omega_1/\omega_c,\beta\omega_1),
\end{equation}
where
\begin{equation}
\mathcal{F}(\omega_1/\omega_c,\beta\omega_1)=\int_0^\infty dx\,
\exp(-x\omega_1/\omega_c)\frac{x\coth \left(\frac{\beta\omega_1 x}{2}\right)
-\coth \left(\frac{\beta\omega_1}{2}\right)}{x^2-1},
\end{equation}
is a dimensionless function of parameters $\omega_1/\omega_c$ and $\beta\omega_1$. 
Here $\omega_c$ is the cut-off energy of the bosonic spectral function.      
The function $\mathcal{F}(\omega_1/\omega_c,\beta\omega_1)$ for different values of $\omega_1/\omega_c=0.1,1,10$ is presented on the  Fig.~\ref{fig:figure_2}

\begin{figure}
	\centering
		\includegraphics{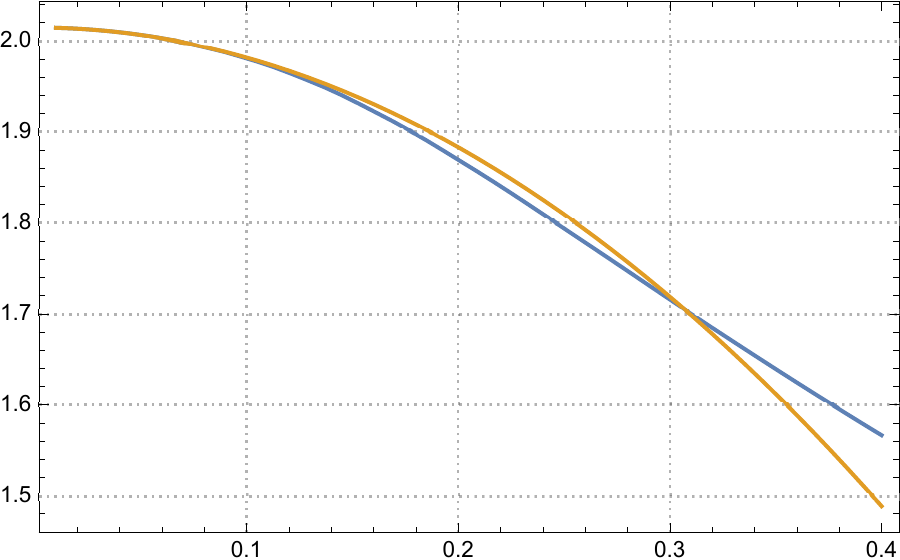}
	\caption{\label{fig:exact_approx_01} Dimensionless function $\mathcal{F}(\omega_1/\omega_c,1/T)$  (blue curve), 
	and its approximation $-e^{\omega_1/\omega_c}\mathrm{Ei}(-\omega_1/\omega_c)-\pi^2T^2/3$ (yellow curve) as a function of 
	dimensionless temperature $T=1/\beta\omega_1\in[0,0.4]$ for $\omega_1/\omega_c=0.1$. The relative deviation of the
	approximation function from the exact one is less than $5\%$ for considered domain.} 
\end{figure}

Let us estimate the value of the integral $\mathcal{F}(\omega_1/\omega_c,\beta\omega_1)$
in the low-temperature limit $\beta\omega_1\gg1$.
In order to do it we present the integral in the form
\begin{align}
\mathcal{F}(\omega_1/\omega_c,\beta\omega_1)=&\int_0^\infty  dx \frac{\exp(-x\omega_1/\omega_c)}{x+1}
-\frac{2}{e^{\beta\omega_1}-1}\text{v.p.}\int_0^\infty  dx \frac{\exp(-x\omega_1/\omega_c)}{x^2-1}+
\nonumber \\+&
\text{v.p.}\int_0^\infty  dx \frac{\exp(-x\omega_1/\omega_c)}{x^2-1}
\frac{2x}{e^{\beta\omega_1 x}-1}.
\end{align}
The first term is
\begin{equation}
\int_0^\infty dx
\frac{\exp(-x\omega_1/\omega_c)}{x+1}=-e^{\omega_1/\omega_c}\mathrm{Ei}(-\omega_1/\omega_c)=
\left\{\begin{array}{c}
\omega_c/\omega_1, \quad \omega_1/\omega_c\gg 1 \\
-\log(\omega_1/\omega_c), \quad \omega_1/\omega_c\ll 1,
\end{array}
\right.
\end{equation}
where $\mathrm{Ei}(x)=-\int_{-x}^\infty dt e^{-t}/t$ is the exponential integral function \cite{Bateman1953}.
This result defines the value of the function $\mathcal{F}(\omega_1/\omega_c,\beta\omega_1)$ 
at zero temperature $\beta\omega_1\rightarrow \infty$. 
Therefore, the second and the third terms describe the temperature dependence 
of $\mathcal{F}(\omega_1/\omega_c,\beta\omega_1)$.
The second term is asymptotically small $\sim e^{-\beta\omega_1}$ at $\beta\omega_1\gg1$, since the integral
\begin{equation}
\text{v.p.}\int_0^\infty  dx \frac{\exp(-x\omega_1/\omega_c)}{x^2-1}=
\frac{1}{2} \Big[e^{\omega_1/\omega_c} \mathrm{Ei}(-\omega_1/\omega_c)-e^{-\omega_1/\omega_c}\mathrm{Ei}(\omega_1/\omega_c)\Big]
\end{equation}
is finite for any $\omega_1/\omega_c$. 
The third term gives the leading correction to $\mathcal{F}(\omega_1/\omega_c,\beta\omega_1)$ at small temperatures 
\begin{align}
\text{v.p.}\int_0^\infty dx \frac{\exp(-x\omega_1/\omega_c)}{x^2-1}\frac{2x}{e^{\beta\omega_1 x}-1}=&
-\sum_{n=1}^\infty\sum_{\zeta=\pm}e^{\zeta(\omega_c^{-1}+\beta n)\omega_1} \mathrm{Ei}(-\zeta(\omega_c^{-1}+\beta n )
\omega_1)=\nonumber \\
=&-\frac{\pi^2}{3}\frac{1}{\beta^2\omega_1^2}+O\Big(\frac{\omega_1/\omega_c}{\beta^3\omega_1^3}\Big).
\end{align}
Therefore, for the case $\omega_c\beta\gg 1$ we have the following approximation formula  
\begin{equation}
\mathcal{F}(\omega_1/\omega_c,\beta\omega_1)\approx -e^{\omega_1/\omega_c}\mathrm{Ei}(-\omega_1/\omega_c)
-\frac{\pi^2}{3\beta^2\omega_1^2}.
\end{equation}
The exact and approximation functions are presented in  Fig.~\ref{fig:exact_approx_01} 
for the case of $\omega_1/\omega_c=0.1$.

Summarizing, one can write the expression for the coherence of the second spin at small temperature 
\begin{equation}
\langle\sigma_2^x\rangle=\frac{4 A f_1 f_2 t}{\omega _2 } \tanh \left(\frac{\beta  \omega _2}{2}\right)
\tanh \left(\frac{\beta  \omega _1}{2}\right)\Big[-e^{\omega_1/\omega_c}\mathrm{Ei}(-\omega_1/\omega_c)
-\frac{\pi^2}{3\beta^2\omega_1^2}\Big].
\end{equation}

Note that there is the linear dependence between the the second $\langle\sigma_2^x\rangle$ and first spin coherences 
$\langle\sigma_1^x\rangle$ in the system, valid for any spectral density function $\mathcal{I}(\xi)$
\begin{equation}
\label{eq:linear}
\langle\sigma_2^x\rangle=-\frac{t}{\omega_2}\tanh
\left(\frac{\beta\omega_2}{2}\right)\langle\sigma_1^x\rangle.
\end{equation}
The derivation of the first spin coherence $\langle\sigma_1^x\rangle$ is presented in the 
Appendix~\ref{sec:coherence_detuned_first_spin}.

One can point out that at small temperatures $\beta\omega_2\gg1$ the coherence $\langle\sigma_2^x\rangle\propto 1/\omega_2$
formally diverges at small $\omega_2$. The qualitative analysis of this situation and the restrictions for values of 
$\omega_1$ and $\omega_2$ in perturbative approach are performed in the 
Appendix~\ref{sec:qualitative_analysis_bosons}.  

The second spin coherence in the high-temperature limit $\beta\rightarrow 0$ 
for an arbitrary spectral density function $\mathcal{I}(\xi)$ is
\begin{align}
\label{eq:ht_limit_spin}
\langle\sigma_2^x\rangle|_{\beta\rightarrow 0}=\frac{f_1f_2t\omega_1\beta^3}{6}
\int_0^\infty d\xi\,\frac{\mathcal{I}(\xi)}{\xi}+O(\beta^5).
\end{align}
 
The expressions for the coherences for arbitrary values of $\omega_1,\omega_2$ and $t$ are presented in the 
Appendices~\ref{sec:coherence_first_spin} and \ref{sec:coherence_second_spin}.    

\section{Detuned spins: the first spin coherence}
\label{sec:coherence_detuned_first_spin}

We use the density operator $\varrho_S$ defined before for calculation of the coherence of the first spin 
$\langle\sigma_1^x\rangle=\langle\sigma_1^x\rangle-\langle\sigma_1^x\rangle_0$.
Repeating all the steps of the calculation performed for the second spin we obtain
that such a coherence is defined by the function
$G_2(\tau,\tau')=\mathrm{Tr}[e^{-\beta H_{12}}\widetilde{S}(\tau)\widetilde{S}(\tau')\sigma_1^x+
\widetilde{S}^\dag(\tau')\widetilde{S}^\dag(\tau)e^{-\beta H_{12}}\sigma_1^x]$, that is an analog of the function
$F_2(\tau,\tau')$ considered before. For arbitrary angle $\theta$ this function is
\begin{align}
G_2(\tau,\tau')=
&4f_1 f_2 \Big[
-\sin (\theta)\sin^2(\phi )\{\sinh \left(\beta\epsilon +\tau '(\omega -\epsilon )\right)+
\sinh (\beta\omega-\tau  (\omega-\epsilon))\} - \nonumber \\
& -\sin (\theta )\cos ^2(\phi )\{\sinh \left(\beta\epsilon-\tau'(\omega +\epsilon )\right)-
\sinh (\beta\omega -\tau(\omega +\epsilon))\}-\nonumber \\
& -\sin ^2(\phi) \{\sinh \left(\beta  \omega +\tau ' (\epsilon -\omega )\right)+\sinh (\beta  \epsilon +\tau  (\omega -\epsilon ))\} + \nonumber \\
& +\cos^2(\phi )\{\sinh (\beta  \epsilon -\tau  (\omega +\epsilon ))-\sinh \left(\beta  \omega -\tau ' (\omega +\epsilon )\right)\} +\nonumber \\
&+\cos (\theta ) \sin (2 \phi)\{\sinh (\omega  (\beta -\tau )) \cosh \left(\epsilon  \left(\tau -2 \tau '\right)\right)
+\nonumber \\ & \qquad \qquad \qquad \quad  +\sinh \left(\omega  \tau '\right) \cosh \left(\epsilon  \left(\beta +\tau '-2 \tau \right)\right)\}\Big].
\end{align}
Substituting the decomposition
\begin{align}
\sin^2\theta\approx& 1-\frac{4t^2}{\delta^2}, \quad \cos^2\theta\approx \frac{4t^2}{\delta^2},\nonumber \\
\cos^2\phi=&\frac{1+\sin\theta}{2}\approx \frac{1+\eta}{2} -\frac{\eta t^2}{\delta^2},
\quad \sin^2\phi=\frac{1-\sin\theta}{2}\approx \frac{1-\eta}{2}+\frac{\eta t^2}{\delta^2},
\end{align}
and keeping the leading order of this function by small parameter $t/\delta$ we get
\begin{align}
G_2(\tau,\tau')=8 f_1 f_2 \sinh \Big(\frac{1}{2} \omega _1 \big(\tau '-\tau \big)\Big) &\Big[\cosh \left(\frac{1}{2} \left(\beta  \left(\omega _2-\omega _1\right)+\omega _1 \left(\tau '+\tau \right)\right)\right)+\nonumber \\ &+\cosh \left(\frac{1}{2} \left(\beta  \left(\omega _1+\omega _2\right)-\omega _1 \left(\tau '+\tau \right)\right)\right)\Big].
\end{align}
Integrating over $\tau, \tau'$ we get
\begin{align}
G_2(\xi)-G_2(-\xi)\approx&\frac{32 f_1 f_2}{\xi  \left(\xi ^2-\omega _1^2\right)} \cosh \left(\frac{\beta  \omega _2}{2}\right) \cosh \left(\frac{\beta  \omega _1}{2}\right) \left[\omega _1-\xi  \coth \left(\frac{\beta  \xi }{2}\right) \tanh \left(\frac{\beta  \omega _1}{2}\right)\right].
\end{align}
Therefore, the coherence of the first spin in the leading order is
\begin{equation}
\langle\sigma_1^x\rangle=-4f_1 f_2\int_0^\infty d\xi\,\mathcal{I}(\xi)
\frac{\xi  \coth(\beta\xi/2)\tanh(\beta\omega_1/2)-\omega _1}{\xi  \left(\xi ^2-\omega _1^2\right)}.
\end{equation}
This is nothing but the coherence of the single spin coupled to the bosonic system, derived in \cite{Purkayastha2020}.

Note that this result is valid only at small $t/\delta\ll1$, since it doesn't contain the contribution 
from the second spin. In order to check the validity of the obtained formula for the case when the ratio
$t/\delta$ is not small we analyze the first spin coherence as well as the second spin coherence 
for arbitrary $t,\omega_1,\omega_2$ in the Appendices~\ref{sec:coherence_first_spin} and 
 \ref{sec:coherence_second_spin}. 

\section{Qualitative analysis of the spin-boson system coherences}
\label{sec:qualitative_analysis_bosons}

In order to understand the proportionality between the coherences of the first and second spin 
\begin{equation}
\langle\sigma_2^x\rangle=-\frac{t}{\omega_2}\tanh
\left(\frac{\beta\omega_2}{2}\right)\langle\sigma_1^x\rangle
\end{equation} 
and physical meaning of such proportionality we provide the following qualitative analysis.  
Let us consider the bath with the Hamiltonian
\begin{equation}
H_B=\sum_k \Omega_k b_k^\dag b_k,
\end{equation}
and system of two weakly interacting spins 
\begin{equation}
H_{12}=\frac{\omega_1}{2}\sigma_1^z+\frac{\omega_2}{2}\sigma_2^z + t(\sigma_1^+\sigma_2^-+\sigma_2^+\sigma_1^-),
\end{equation}
at temperature $T=1/\beta$. 
When two systems are far from each other, both of them have corresponding thermal distributions. 
Taking into account that the interaction between the spins are week ($t\ll \omega_1, \omega_2$), we can 
estimate their averages of $\langle\sigma_j^x\rangle=\langle\sigma_j^y\rangle=0$ and 
$\langle\sigma_j^z\rangle=-\tanh(\beta\omega_j/2)$, for $j=1,2$.  
When system of two spins is close to the bath, the interaction term start to play an important role 
\begin{equation}
H_{B1}=(f_1\sigma_1^z+f_2\sigma_1^x)\sum_k \lambda_k(b_k+b_k^\dag).
\end{equation}
Due to this term the first spin start to ``polarize'' the bath system. This process can be observed from
the Hamiltonian
\begin{equation}
H_{BS_1}=\sum_k \Omega_k b_k^\dag b_k + f_1\langle\sigma_1^z\rangle\sum_k \lambda_k(b_k+b_k^\dag) + 
\frac{\omega_1}{2}\langle\sigma_1^z\rangle,
\end{equation}
where we replace the operator $\sigma_1^z$ by its average value $\langle\sigma_1^z\rangle$. 
Due to the interaction of the first spin with the bath the boson operators get the non-zero average values
$\langle b_k+b_k^\dag\rangle\neq0$. This values can be calculated by diagonalization of the   
boson part of the obtained Hamiltonian. It can be done by substitution  
$b_k=\beta_k-f_1\langle\sigma_1^z\rangle\lambda_k/\Omega_k$, where $\beta_k$ is the set of 
boson annihilation operators with the same commutation rules as $b_k$ operators, and zero average values.
Hence, we have
\begin{equation}
\langle b_k+b_k^\dag\rangle=\langle \beta_k+\beta_k^\dag\rangle-2f_1\langle\sigma_1^z\rangle\lambda_k/\Omega_k=
-2f_1\langle\sigma_1^z\rangle\lambda_k/\Omega_k.
\end{equation}   
The non-zero $\langle b_k+b_k^\dag\rangle$ affects the term 
\begin{equation}
f_2\sigma_1^x\sum_k \lambda_k(b_k+b_k^\dag)\rightarrow -2f_1f_2\langle\sigma_1^z\rangle\sigma_1^x
\sum_k \lambda_k^2/\Omega_k=-2f_1f_2\langle\sigma_1^z\rangle\Omega\sigma_1^x.
\end{equation}
Therefore, the back-action of the bath modifies the Hamiltonian of the first spin
\begin{equation}
H_{S_1}^{mod}=\frac{\omega_1}{2}\sigma_1^z-2f_1f_2\langle\sigma_1^z\rangle\Omega\sigma_1^x.
\end{equation}
Due to non-zero $\sigma_1^x$ term, the first spin gets non-zero coherence 
\begin{align}
\langle\sigma_1^x\rangle&=Tr[e^{-\beta H_{S_1}^{mod}}\sigma_1^x]=\nonumber \\ 
-&\tanh\Big(\beta\sqrt{(2f_1f_2\langle\sigma_1^z\rangle\Omega)^2+(\omega_1/2)^2}\Big)
\Big[\frac{-2f_1f_2\langle\sigma_1^z\rangle\Omega}
{\sqrt{(2f_1f_2\langle\sigma_1^z\rangle\Omega)^2+(\omega_1/2)^2}}\Big]\approx
 -\frac{4f_1f_2\langle\sigma_1^z\rangle^2\Omega}{\omega_1}.
\end{align}
The final approximate result is obtained for the case $4f_1f_2\langle\sigma_1^z\rangle\Omega\ll \omega_1$.
This means that the interaction with the bath gives negligibly small correction to the energy and coherence 
of the first spin. In other words the latter result corresponds to weak interaction limit.  

The non-zero $\langle\sigma_1^x\rangle$ induces the effective $\sigma_2^x$ 
term in the second spin system
\begin{equation}
t(\sigma_1^+\sigma_2^-+\sigma_2^+\sigma_1^-)\rightarrow 
t(\langle\sigma_1^+\rangle\sigma_2^-+\sigma_2^+\langle\sigma_1^-\rangle)=
t\frac{\langle\sigma_1^x\rangle}{2}\sigma_2^x.
\end{equation}  
Therefore the Hamiltonian of the second spin is modified
\begin{equation}
H_{S_2}^{mod}=\frac{\omega_2}{2}\sigma_2^z+t\frac{\langle\sigma_1^x\rangle}{2}\sigma_2^x,
\end{equation}
which leads to the non-zero coherence 
\begin{align}
\langle\sigma_2^x\rangle=&Tr[e^{-\beta H_{S_2}^{mod}}\sigma_2^x]= \nonumber \\ 
-&\tanh\Big(\beta\sqrt{(t\langle\sigma_1^x\rangle/2)^2+(\omega_2/2)^2}\Big)
\frac{t\langle\sigma_1^x\rangle}{2\sqrt{(t\langle\sigma_1^x\rangle/2)^2+(\omega_2/2)^2}}\approx 
-\frac{t}{\omega_2}\tanh\Big(\frac{\beta\omega_2}{2}\Big)\langle\sigma_1^x\rangle.
\end{align}
The latter approximate expression is valid only for the case $t\langle\sigma_1^x\rangle\ll\omega_2$. 
Therefore the linear dependence between the coherences of the second and first spins appears if two 
additional restrictions are satisfied a) the weak coupling regime $4f_1f_2\langle\sigma_1^z\rangle\Omega\ll \omega_1$ 
and b) the weak coupling between spins $t\langle\sigma_1^x\rangle\ll\omega_2$. Note that both inequalities are natural 
for the derivations based on the thermodynamical perturbation approach presented before. The first inequality allows 
to take only the leading term in perturbation series to evaluate the result. If this inequality is not satisfied one needs to use the next terms of perturbation series to get the correct answer. The second condition defines the domain of the parameters where the linear dependence between both coherences exists. 
     
\section{General case: the first spin coherence}
\label{sec:coherence_first_spin}

Let us write the expression for $G_2(\tau,\tau')$ in terms of $\theta$ variable only
\begin{align}
G_2(\tau,\tau')=4f_1f_2&\Big[G^{(0)}_2(\tau,\tau')+\sin\theta G^{(1)}_2(\tau,\tau')+
\cos(2\theta)2G^{(2)}_2(\tau,\tau')\Big]=\nonumber \\=
4f_1f_2&\Big[\frac{\cos(2\theta)}{2}\Big\{\sinh(\tau'\omega)(\cosh(\epsilon(\beta-2\tau+\tau'))-
\cosh(\epsilon(\tau'-\beta )))+\nonumber \\+&\sinh(\omega(\beta-\tau))(\cosh(\epsilon(\tau-2\tau'))-\cosh(\tau\epsilon))\Big\}+
\nonumber \\
+\sin\theta&\Big\{\cosh(\tau\omega)\sinh(\epsilon(\beta-\tau))-\sinh(\tau\epsilon)\cosh(\omega(\beta-\tau))+
\nonumber \\+&
\sinh (\tau'\epsilon)\cosh(\omega(\tau'-\beta))+\cosh(\tau'\omega)\sinh(\epsilon(\tau'-\beta))\Big\}+
\nonumber \\+
\frac12\Big\{&\sinh(\tau'\omega)(\cosh(\epsilon(\beta-2\tau+\tau'))+
\cosh(\epsilon(\tau'-\beta)))+
\nonumber \\ +&\sinh(\omega(\beta-\tau))(\cosh(\epsilon(\tau-2\tau'))+\cosh(\tau\epsilon))+
\nonumber \\ +&
2\cosh(\tau'\epsilon)\sinh(\omega(\tau'-\beta))-2\sinh(\tau\omega)\cosh(\epsilon(\beta-\tau))\Big\}\Big].
\end{align}
After integration over $\tau$ and $\tau'$ we get the following expression
\begin{equation}
G_2(\xi)-G_2(-\xi)=4f_1f_2\Big[G^{(0)}_2(\xi)+\sin\theta G^{(1)}_2(\xi)+
\cos(2\theta)G^{(2)}_2(\xi)\Big],
\end{equation}
where 
\begin{align}
G^{(0)}_2(\xi)=&\frac{-1}{(\omega^2-\epsilon^2)(\xi^2-[\omega+\epsilon]^2)(\xi^2-[\omega -\epsilon]^2)}
\Big[\frac{4 \omega  \cosh (\beta  \omega ) \left(\left(\epsilon ^2-\omega ^2\right)^2-\xi ^2 \omega ^2\right)}{\xi }-
\nonumber \\
-&\frac{4 \omega  \cosh (\beta  \epsilon ) \left(\xi ^2 \omega ^2 \left(\xi ^2-\omega ^2\right)+\epsilon ^4 \left(3 \xi ^2-4 \omega ^2\right)+2 \epsilon ^2 \left(\xi ^2 \omega ^2-\xi ^4+\omega ^4\right)+2 \epsilon ^6\right)}{\xi  \left(\xi ^2-4 \epsilon ^2\right)}-\nonumber \\
-&2 \coth \left(\frac{\beta  \xi }{2}\right) \Big(\frac{\omega  \epsilon  \sinh (\beta  \epsilon ) \left(-7 \xi ^2 \omega ^2+\xi ^4+2 \omega ^4+\epsilon ^2 \left(3 \xi ^2+8 \omega ^2\right)-10 \epsilon ^4\right)}{\xi ^2-4 \epsilon ^2}-
\nonumber \\ 
-&\sinh (\beta  \omega ) \left(2 \omega ^2 \left(\xi ^2-\omega ^2\right)+\epsilon ^2 \left(\omega ^2-\xi ^2\right)+\epsilon ^4\right)\Big)\Big]
\end{align}

\begin{align}
G^{(1)}_2(\xi)=&\frac{4\epsilon\left(\xi^2+\omega^2-\epsilon^2\right)(\cosh(\beta\omega)+\cosh(\beta\epsilon ))}
{\xi(\xi^2-[\omega+\epsilon]^2)(\xi^2-[\omega-\epsilon]^2)}-\nonumber \\
-&\frac{4\xi\coth \left(\frac{\beta  \xi }{2}\right) \left(2 \omega  \epsilon  \sinh (\beta  \omega )+
\sinh (\beta  \epsilon ) \left(\xi^2-\omega^2-\epsilon^2\right)\right)}{\xi(\xi^2-[\omega+\epsilon]^2)(\xi^2-[\omega-\epsilon]^2)}
\end{align}

\begin{align}
G^{(2)}_2(\xi)=&\frac{-2\epsilon}{(\omega^2-\epsilon^2)(\xi^2-[\omega+\epsilon]^2)(\xi^2-[\omega-\epsilon]^2)}
\Big[-2\xi\omega\epsilon\cosh(\beta\omega)+
\nonumber \\+&
\frac{2\cosh(\beta\epsilon)\omega\epsilon\left(\xi^4+2\left(\omega^4-2 \omega^2\left(\xi^2+\epsilon^2\right)+\epsilon^4\right)\right)}{\xi\left(\xi^2-4\epsilon^2\right)}+
\coth \left(\frac{\beta  \xi }{2}\right)
\Big(\epsilon  \sinh (\beta  \omega )\times \nonumber \\ 
 \times&\left(\xi ^2+\omega ^2-\epsilon ^2\right)
+\frac{\omega  \sinh (\beta  \epsilon ) \left(3 \xi ^2 \omega ^2-\xi ^4-2 \omega ^4+\epsilon ^2 \left(\xi ^2+8 \omega ^2\right)-6 \epsilon ^4\right)}{\xi ^2-4 \epsilon ^2}\Big)\Big]
\end{align}

In order to check this result we consider the limit $t\rightarrow 0$ for finite value of $|\omega_1-\omega_2|>0$.
In this limit $\sin\theta\rightarrow\text{sign}(\omega_1-\omega_2)=\eta$ and $\cos\theta\sim t\rightarrow 0$, 
$\cos(2\theta)=2\cos^2(\theta)-1\rightarrow -1$. The leading order of this limit is 
\begin{equation}
G_2(\xi)-G_2(-\xi)=4f_1f_2\Big[G^{(0)}_2(\xi)+\eta G^{(1)}_2(\xi)-G^{(2)}_2(\xi)\Big]_{t=0}
\end{equation}
The $\eta$-independent terms can be simplified 
\begin{align}
G^{(0)}_2(\xi)-G^{(2)}_2(\xi)=&\frac{4\omega  \left(\xi ^2-\omega ^2+\epsilon ^2\right) (\cosh (\beta  \omega )+\cosh (\beta  \epsilon ))}{\xi  \left(\xi ^2-(\omega +\epsilon )^2\right) \left(\xi ^2-(\omega -\epsilon )^2\right)} +
\nonumber \\
+&\frac{4\xi  \coth \left(\frac{\beta  \xi }{2}\right) \left(\sinh (\beta  \omega ) \left(-\xi ^2+\omega ^2+\epsilon ^2\right)-2 \omega  \epsilon  \sinh (\beta  \epsilon )\right)}{\xi  \left(\xi ^2-(\omega +\epsilon )^2\right) \left(\xi ^2-(\omega -\epsilon )^2\right)}.
\end{align}
Then taking into account that $\eta\epsilon\rightarrow (\omega_1-\omega_2)/2$, we get 
\begin{equation}
[G^{(0)}_2(\xi)+\eta G^{(1)}_2(\xi)-G^{(2)}_2(\xi)]_{t=0}=\frac{8 \cosh \left(\frac{\beta  \omega _2}{2}\right) \cosh \left(\frac{\beta  \omega _1}{2}\right) \left(\omega _1-\xi  \coth \left(\frac{\beta  \xi }{2}\right) \tanh \left(\frac{\beta  \omega _1}{2}\right)\right)}{\xi  \left(\xi ^2-\omega _1^2\right)}
\end{equation}

Therefore the coherence of the first spin in the leading order  by $t$ is 
\begin{align}
\langle \sigma_1^x\rangle_{t=0}=&\frac{4f_1f_2}{2Z_S}\int_0^\infty d\xi\, \mathcal{I}(\xi)
[G^{(0)}_2(\xi)+\eta G^{(1)}_2(\xi)-G^{(2)}_2(\xi)]\Big|_{t=0}=\nonumber \\
=&-4 f_1 f_2\int_0^\infty d\xi\, \mathcal{I}(\xi)
\frac{\xi\coth\left(\frac{\beta\xi}{2}\right)\tanh\left(\frac{\beta\omega_1}{2}\right)-\omega_1}
{\xi(\xi^2-\omega_1^2)},
\end{align} 
as it was obtained in the previous section. Note that this expression is valid only in the case $t\ll |\omega_1-\omega_2|$. 
Let us consider the opposite limit. To this end we put $\omega_1=\omega_2=\omega$ in general expression, find the  
$G^{(0)}_2(\xi)-G^{(0)}_2(-\xi)$ as a function of $t$, and then consider the limit $t\rightarrow 0$. 
In this limit $\cos\theta=1$ and $\sin\theta=0$. Therefore we get 
\begin{equation}
G_2(\xi)-G_2(-\xi)=4f_1f_2\Big[G^{(0)}_2(\xi)+G^{(2)}_2(\xi)\Big]_{\omega_1=\omega_2}.
\end{equation}
Then 
\begin{align}
\Big[G^{(0)}_2(\xi)+G^{(2)}_2(\xi)\Big]_{\omega_1=\omega_2}=&
\frac{4\omega}{\left(\omega^2-t^2\right)\left(\xi^2-(t+\omega)^2\right)\left(\xi^2-(\omega-t)^2\right)}\times  
\nonumber \\ \times& 
\Big[\frac{\cosh (\beta  \omega ) \left(\omega ^2 \left(\xi ^2-\omega ^2\right)+t^2 \left(\xi ^2+2 \omega ^2\right)-t^4\right)}{\xi}+ \nonumber \\ 
+&\frac{\xi  \cosh (\beta  t) \left(\omega ^2 \left(\xi ^2+6 t^2\right)-3 t^2 \left(\xi ^2-t^2\right)-\omega ^4\right)}{\xi ^2-4 t^2}+
\nonumber \\ 
+&\coth \left(\frac{\beta  \xi }{2}\right) \Big(\frac{t \sinh (\beta  t) \left(-5 \xi ^2 \omega ^2+\xi ^4+\xi ^2 t^2-2 t^4+2 \omega ^4\right)}{\xi ^2-4 t^2}-\nonumber \\
-&
\omega  \sinh (\beta  \omega ) \left(\xi ^2+t^2-\omega ^2\right)\Big)\Big]
\end{align}
This expression in the limit $t\rightarrow 0$ is
\begin{align}
\Big[G^{(0)}_2(\xi)+G^{(2)}_2(\xi)\Big]_{\omega_1=\omega_2,t=0}=(1+\cosh(\beta\omega))\frac{4 \left(\omega -\xi  \coth \left(\frac{\beta  \xi }{2}\right) \tanh \left(\frac{\beta  \omega }{2}\right)\right)}{\xi  \left(\xi ^2-\omega ^2\right)}.
\end{align}
Taking into account that $Z_S=2(1+\cosh(\beta\omega))$ we get the expression for the coherence
\begin{align}
\langle \sigma_1^x\rangle_{t=0}=&\frac{4f_1f_2}{2Z_S}\int_0^\infty d\xi\, \mathcal{I}(\xi)
[G^{(0)}_2(\xi)+G^{(2)}_2(\xi)]_{\omega_1=\omega_2,t=0}=\nonumber \\
&=-4 f_1 f_2\int_0^\infty d\xi\, \mathcal{I}(\xi)
\frac{\xi\coth\left(\frac{\beta\xi}{2}\right)\tanh\left(\frac{\beta\omega}{2}\right)-\omega}
{\xi(\xi^2-\omega^2)},
\end{align} 
which coincides with the one we got previously. Therefore this result is relevant for the case 
$t\ll |\omega_1-\omega_2|$ as well as for the case $t\gg |\omega_1-\omega_2|$, for relatively 
small value of parameter $t$. 

\begin{figure}
	\centering
		\includegraphics[width=\linewidth]{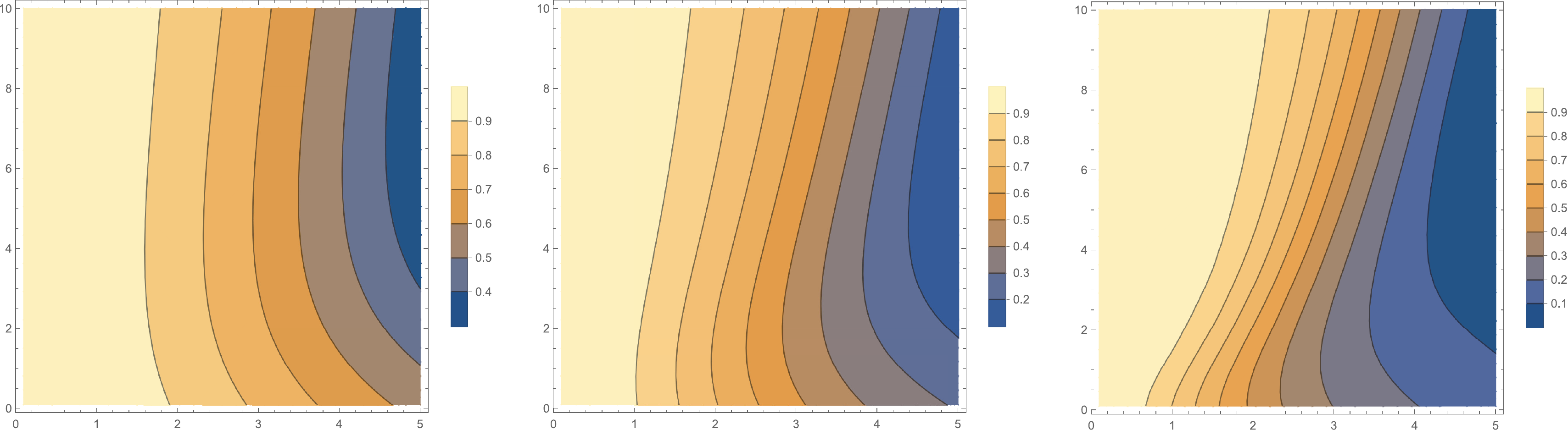}
\caption{\label{fig:sigma1_001} Contour plot of $R_1(t,\omega_1,\omega_2)$
for $\omega_1/\omega_c=0.01$ and different temperatures $\beta\omega_1=0.5,1,2$ 
(left, middle and right figures respectively) for $\omega_2/\omega_1\in[0,10]$ and $t/\omega_1\in[0,5]$. }
\end{figure}

\begin{figure}
	\centering
		\includegraphics[width=\linewidth]{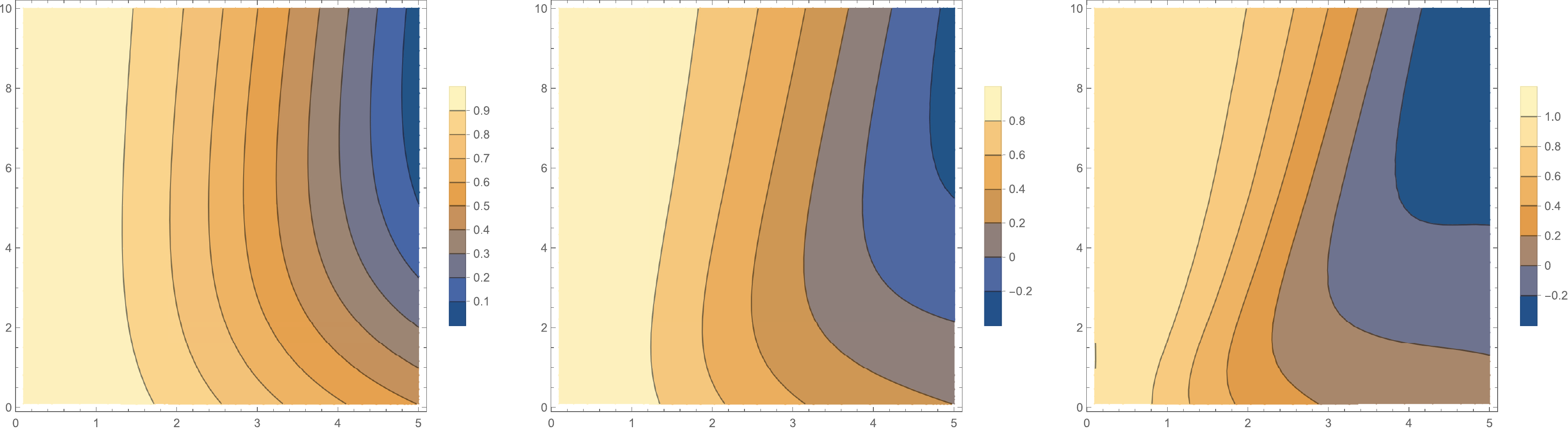}
	\caption{\label{fig:sigma1_1} Contour plot of $R_1(t,\omega_1,\omega_2)$
for $\omega_1/\omega_c=1$ and different temperatures $\beta\omega_1=0.5,1,2$ 
(left, middle and right figures respectively) for $\omega_2/\omega_1\in[0,10]$ and $t/\omega_1\in[0,5]$. }
\end{figure}

\begin{figure}
	\centering
		\includegraphics[width=\linewidth]{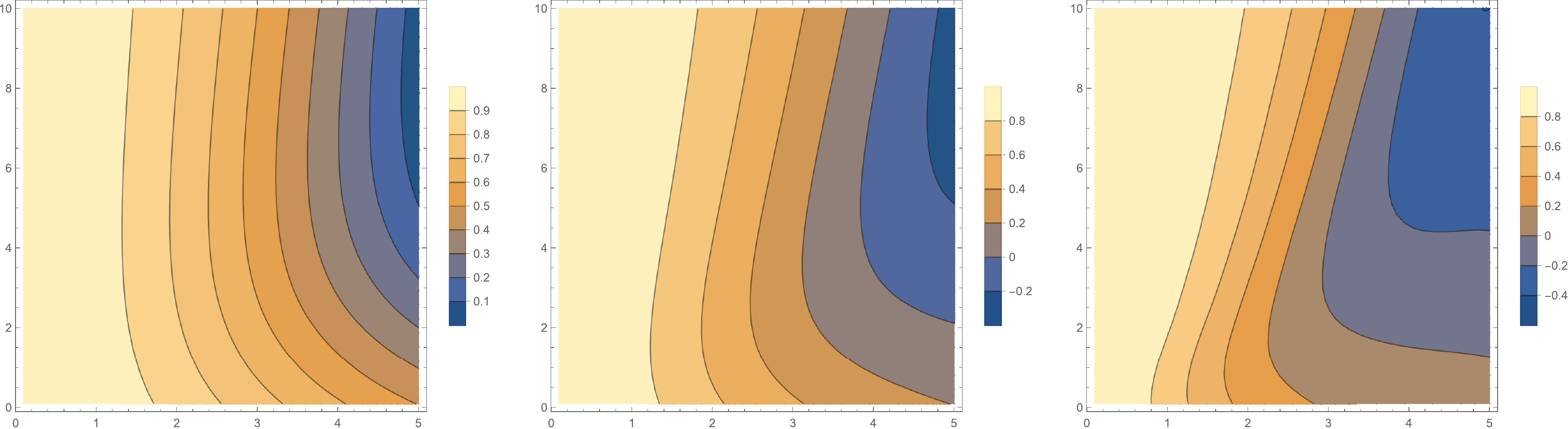}
	\caption{\label{fig:sigma1_100} Contour plot of $R_1(t,\omega_1,\omega_2)$
for $\omega_1/\omega_c=100$ and different temperatures $\beta\omega_1=0.5,1,2$ 
(left, middle and right figures respectively) for $\omega_2/\omega_1\in[0,10]$ and $t/\omega_1\in[0,5]$. }
\end{figure}

In order to check the applicability of the obtained expression for for the domains of parameters where $t\sim|\omega_1-\omega_2|$, we calculate the ratio 
$R_1(t,\omega_1,\omega_2)=\langle \sigma_1^x\rangle/\langle \sigma_1^x\rangle_{t=0}$ as a function of pair of dimensionless parameters 
$(\omega_2/\omega_1,t/\omega_1)$ and find the region where this ratio is close to unity. 
\begin{equation}
R_1(t,\omega_1,\omega_2)=-\frac{\int_0^\infty d\xi\,
\mathcal{I}(\xi)\Big\{G^{(0)}_2(\xi)+\sin\theta G^{(1)}_2(\xi)+
\cos(2\theta)G^{(2)}_2(\xi)\Big\}}{4[\cosh(\beta\omega)+\cosh(\beta\epsilon)]
\tanh\left(\frac{\beta\omega_1}{2}\right)\int_0^\infty d\xi\, \mathcal{I}(\xi)
\frac{\xi\coth\left(\frac{\beta\xi}{2}\right)-\omega_1\coth\left(\frac{\beta\omega_1}{2}\right)}
{\xi(\xi^2-\omega_1^2)}}
\end{equation}
Then taking the particular case of spectral function $\mathcal{I}(\xi)=A\xi\exp(-\xi/\omega_c)$ and introducing the
dimensionless variable $x=\xi/\omega_1$ together with the function 
\begin{equation}
\mathcal{F}\Big(\frac{\omega_1}{\omega_c},\beta\omega_1\Big)=\int_0^\infty dx e^{-\frac{\omega_1}{\omega_c}x}
\frac{x\coth\left(\frac{\beta\omega_1x}{2}\right)-\coth\left(\frac{\beta\omega_1}{2}\right)}
{x^2-1}
\end{equation} 
we get 
\begin{equation}
R_1(t,\omega_1,\omega_2)=-\frac{\omega_1^2\int_0^\infty dx 
xe^{-\frac{\omega_1}{\omega_c}x}\Big\{G^{(0)}_2(\omega_1x)+\sin\theta G^{(1)}_2(\omega_1x)+
\cos(2\theta)G^{(2)}_2(\omega_1x)\Big\}}{4[\cosh(\beta\omega)+\cosh(\beta\epsilon)]
\tanh\left(\frac{\beta\omega_1}{2}\right)\mathcal{F}\Big(\frac{\omega_1}{\omega_c},\beta\omega_1\Big)}.
\end{equation} 
The plots of function $R_1(t,\omega_1,\omega_2)$ for different temperatures and ratio $\omega_1/\omega_c$ are presented on Figs.~\ref{fig:sigma1_001},\ref{fig:sigma1_1} and \ref{fig:sigma1_100}.

\section{General case: the second spin coherence}
\label{sec:coherence_second_spin}

We rewrite the expression for $F_2(\tau,\tau')$ in terms of $\theta$ variable only  
\begin{align}
F_2(\tau,\tau')=&2f_1f_2[\cos\theta F^{(1)}_2(\tau,\tau') - \cos\theta\sin\theta F^{(2)}_2(\tau,\tau')]= \nonumber \\=
&2f_1 f_2\Big[
\cos\theta\Big\{\cosh(\epsilon(\tau'+\beta-2\tau)-\tau'\omega)-
\cosh(\epsilon(\tau'+\beta-2\tau)+\tau'\omega) +\nonumber \\ 
&+\cosh(2\epsilon\tau'-\epsilon\tau-\beta\omega+\tau\omega)-
\cosh(2\epsilon\tau'-\epsilon\tau+\beta\omega-\tau\omega)+ \nonumber \\
&+2\sinh(\epsilon\tau')
\sinh(\omega(\beta-\tau'))+2\sinh(\tau\omega)\sinh(\epsilon(\beta-\tau ))\Big\}- \nonumber \\
&-\cos\theta\sin\theta\Big\{\cosh(\epsilon(\tau'+\beta-2\tau)-\tau'\omega)+
\cosh(\epsilon(\tau'+\beta-2\tau)+\tau'\omega) +\nonumber \\ 
&+\cosh(2\epsilon\tau'-\epsilon\tau-\beta\omega+\tau\omega)+
\cosh(2\epsilon\tau'-\epsilon\tau+\beta\omega-\tau\omega)- \nonumber \\
&-2\cosh(\omega\tau')
\cosh(\epsilon(\beta-\tau'))-2\cosh(\tau\epsilon)\cosh(\omega(\beta-\tau ))
 \Big\}\Big].
\end{align}

After integration over $\tau$ and $\tau'$ we get the following expression

\begin{equation}
F_2(\xi)-F_2(-\xi)=2f_1f_2\cos\theta[F^{(1)}_2(\xi)-\sin\theta F^{(2)}_2(\xi)],
\end{equation}
where 
\begin{align}
F^{(1)}_2(\xi)=&\frac{8\epsilon\frac{\sinh(\beta\omega)}{\xi}\left[\left(\omega^2-\epsilon^2\right)^2-
\xi^2\left(\omega^2+\epsilon^2\right)\right]+16\xi\omega\epsilon^2\sinh(\beta\epsilon)}{(\omega^2-\epsilon^2)(\xi^2-[\omega -\epsilon]^2)(\xi^2-[\omega+\epsilon]^2)}+ 
\nonumber \\
&+\frac{8\epsilon
\omega\coth\left(\frac{\beta\xi}{2}\right)\left(\xi^2-\omega^2+\epsilon^2\right)(\cosh(\beta\omega)-
\cosh(\beta\epsilon))}{(\omega^2-\epsilon^2)(\xi^2-[\omega -\epsilon]^2)(\xi^2-[\omega+\epsilon]^2)} ,
\end {align}
\begin{align}
F^{(2)}_2(\xi)=&\frac{8\epsilon\frac{\sinh(\beta\epsilon)}{\xi}\left[\left(\omega^2-\epsilon^2\right)^2
-\xi^2\left(\omega^2+\epsilon^2\right)\right]+16\xi\omega\epsilon^2\sinh(\beta\omega)}{(\omega^2-\epsilon^2)(\xi^2-[\omega -\epsilon]^2)(\xi^2-[\omega+\epsilon]^2)}+\nonumber \\
+&\frac{8\epsilon^2\coth\left(\frac{\beta\xi}{2}\right)\left(\xi^2+\omega^2-\epsilon^2\right)(\cosh(\beta\epsilon)-
\cosh(\beta\omega))}{(\omega^2-\epsilon^2)(\xi^2-[\omega -\epsilon]^2)(\xi^2-[\omega+\epsilon]^2)}.
\end{align}

Then taking into account that $\epsilon\cos\theta=t$ and $\eta\epsilon\rightarrow (\omega_1-\omega_2)/2$ 
in $t\rightarrow0$ limit and finite value of $|\omega_1-\omega_2|>0$
we get the answer for the coherence of the second spin  
\begin{align}
\langle \sigma_2^x\rangle=&\frac{2f_1f_2\cos(\theta)}{2Z_S}\int_0^\infty d\xi\, \mathcal{I}(\xi)
[F^{(1)}_2(\xi)-\eta F^{(2)}_2(\xi)]\Big|_{t=0}=
\nonumber \\=&
\frac{4f_1f_2t}{\omega_2}\tanh\left(\frac{\beta\omega _2}{2}\right)
\int_0^\infty d\xi\, \mathcal{I}(\xi)
\frac{\xi\coth\left(\frac{\beta\xi}{2}\right)\tanh\left(\frac{\beta\omega_1}{2}\right)-\omega_1}
{\xi(\xi^2-\omega_1^2)},
\end{align}
as it was obtained before. 

Let us consider the opposite limit $\omega_1\rightarrow \omega_2=\omega$, for which $\cos\theta=1$ and $\sin\theta=0$. 
Then 
\begin{equation}
[F_2(\xi)-F_2(-\xi)]_{\omega_1=\omega_2}=2f_1f_2F^{(1)}_2(\xi)|_{\omega_1=\omega_2}. 
\end{equation}
Taking the series by $t$ and keeping the leading order of the expression we get 
\begin{equation}
F^{(1)}_2(\xi)|_{\omega_1=\omega_2}=\frac{8t}{\xi\omega(\xi^2-\omega^2)}
\Big[\xi\coth\left(\frac{\beta\xi}{2}\right)(\cosh(\beta\omega)-1)-\omega\sinh(\beta\omega)\Big],
\end{equation} 
which gives us 
\begin{equation}
\langle \sigma_2^x\rangle_{\omega_{1,2}=\omega}=\frac{4f_1f_2t}{\omega}\tanh\left(\frac{\beta\omega}{2}\right)
\int_0^\infty d\xi\, \mathcal{I}(\xi)
\frac{\xi\coth\left(\frac{\beta\xi}{2}\right)\tanh\left(\frac{\beta\omega}{2}\right)-\omega}
{\xi(\xi^2-\omega^2)}.
\end{equation}
This expression coincides with the previous expression, obtained for the case of $t\ll|\omega_1-\omega_2|$. 

\begin{figure}
	\centering
		\includegraphics[width=\linewidth]{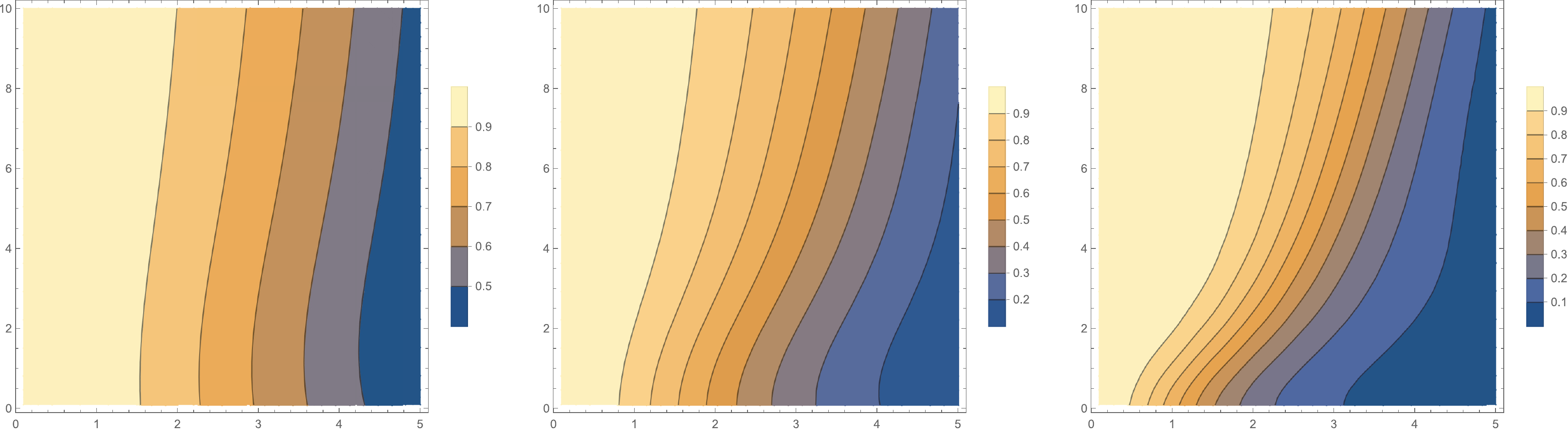}
	\caption{\label{fig:sigma2_001} Contour plot of $R_2(t,\omega_1,\omega_2)$
for $\omega_1/\omega_c=0.01$ and different temperatures $\beta\omega_1=0.5,1,2$ 
(left, middle and right figures respectively) for $\omega_2/\omega_1\in[0,10]$ and $t/\omega_1\in[0,5]$. }
\end{figure}

\begin{figure}
	\centering
		\includegraphics[width=\linewidth]{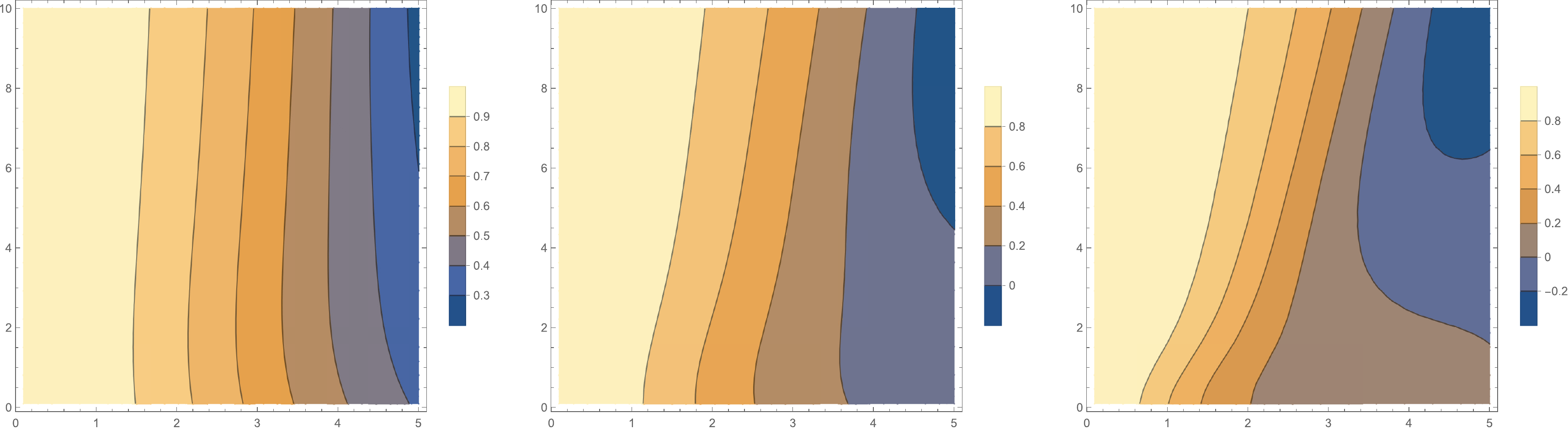}
	\caption{\label{fig:sigma2_1} Contour plot of $R_2(t,\omega_1,\omega_2)$
for $\omega_1/\omega_c=1$ and different temperatures $\beta\omega_1=0.5,1,2$ 
(left, middle and right figures respectively) for $\omega_2/\omega_1\in[0,10]$ and $t/\omega_1\in[0,5]$. }
\end{figure}

\begin{figure}
	\centering
		\includegraphics[width=\linewidth]{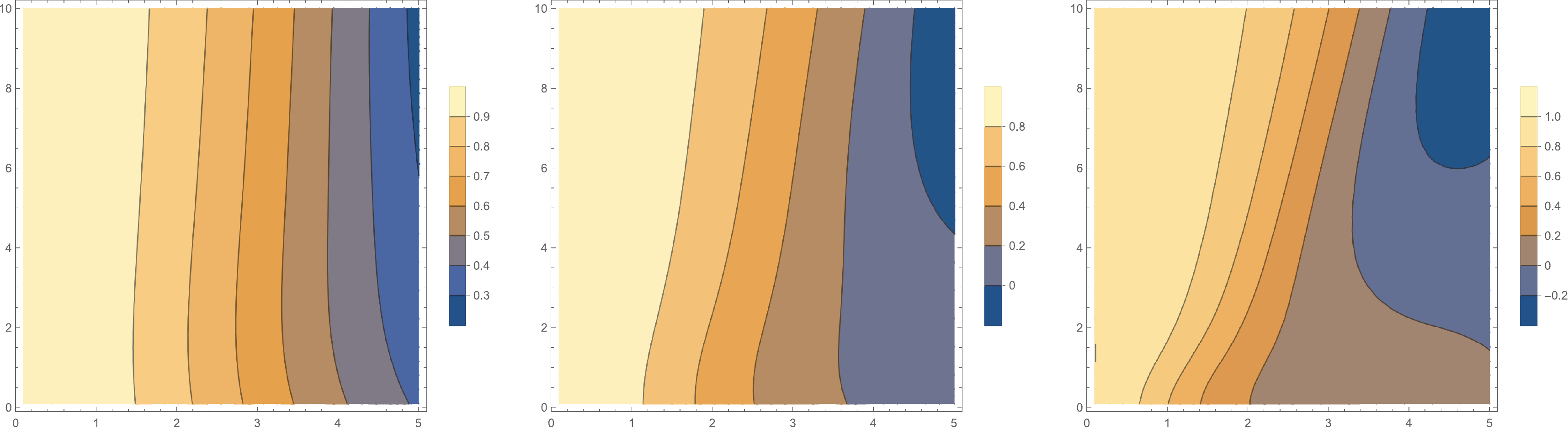}
	\caption{\label{fig:sigma2_100} Contour plot of $R_2(t,\omega_1,\omega_2)$
for $\omega_1/\omega_c=100$ and different temperatures $\beta\omega_1=0.5,1,2$ 
(left, middle and right figures respectively) for $\omega_2/\omega_1\in[0,10]$ and $t/\omega_1\in[0,5]$. }
\end{figure}

In order to verify the universality of the latter result we estimate the ratio of the coherence for arbitrary values 
$t/\omega_1\in[0,10]$ and $\omega_2/\omega_1\in[0,9]$ to the the leading term of the coherence in $t\rightarrow 0$ limit  
$R_2(t,\omega_1,\omega_2)=\langle\sigma_2^x\rangle/\langle\sigma_2^x\rangle_{t\rightarrow 0}$
\begin{equation}
R_2(t,\omega_1,\omega_2)=\frac{\big(\omega_2\cos\theta/[\cosh(\beta\omega)+\cosh(\beta\epsilon)]\big)\int_0^\infty d\xi\,
\mathcal{I}(\xi)\Big\{F^{(1)}_2(\xi)-\sin\theta F^{(2)}_2(\xi)\Big\}}{8t
\tanh\left(\frac{\beta\omega_2}{2}\right)\tanh\left(\frac{\beta\omega_1}{2}\right)\int_0^\infty d\xi\, \mathcal{I}(\xi)
\frac{\xi\coth\left(\frac{\beta\xi}{2}\right)-\omega_1\coth\left(\frac{\beta\omega_1}{2}\right)}
{\xi(\xi^2-\omega_1^2)}},
\end{equation}
which for the spectral density function $\mathcal{I}(\xi)=A\xi\exp(-\xi/\omega_c)$ takes the form 
\begin{equation}
R_2(t,\omega_1,\omega_2)=\frac{\omega_1^2\omega_2\cos\theta\int_0^\infty dx\,x e^{-\frac{\omega_1}{\omega_c}x}
\Big\{F^{(1)}_2(\omega_1x)-\sin\theta F^{(2)}_2(\omega_2x)\Big\}}{8t[\cosh(\beta\omega)+\cosh(\beta\epsilon)]
\tanh\left(\frac{\beta\omega_2}{2}\right)\tanh\left(\frac{\beta\omega_1}{2}\right)
\mathcal{F}\Big(\frac{\omega_1}{\omega_c},\beta\omega_1\Big)},
\end{equation}
where we introduced dimensionless parameter $x=\xi/\omega_1$. The plots of function $R_2(t,\omega_1,\omega_2)$ for different temperatures and ratio $\omega_1/\omega_c$ are presented on Figs.~\ref{fig:sigma2_001},
\ref{fig:sigma2_1} and \ref{fig:sigma2_100}.

\section{Fermionization of the spin-boson model}
\label{sec:fermionization}

Let us consider the Hamiltonian of two spins interacting with the bosons 
\begin{equation}
H=\frac{\omega_2}{2}\sigma_2^z+\frac{\omega_1}{2}\sigma_1^z -
t(\sigma_2^+ \sigma_1^-+\sigma_1^+ \sigma_2^-)+
\sum_k \Omega_k b_k^\dag b_k+[f_1\sigma_1^z-f_2\sigma_1^x]\sum_k \lambda_k(b_k^\dag+b_k).
\end{equation}
Note that we change the sign in spin-spin interaction $t\rightarrow -t$ and spin-boson interaction 
$f_2\rightarrow -f_2$ parameters for simplicity in further calculations. 
It was shown before that the perturbation theory calculations with this Hamiltonian are bulky due to non-commutativity of spin operators and absence of the Wick theorem for them. Therefore, it looks
interesting to rewrite the spin Hamiltonian in terms of bosons or fermions where the Wick theorem is applicable.
This allows us to use the Green's function technique and, potentially, make the final result more clear for understanding. As an example we fermionize the Hamiltonian with the help of the Jordan-Wigner transformation \cite{Lieb1961}.
In order to do it we introduce the additional (``zeroth'') spin subsystem, which is described by the operators $\{\sigma_0^x, \sigma_0^y, \sigma_0^z\}$ and has zero energy. Then we replace the spin operators by the fermion ones
\begin{align}
\sigma_0^-&=a_0, \quad \sigma_0^+=a_0^\dag, \quad \sigma_0^z=2a_0^\dag a_0-1, \\
\quad \sigma_1^-&=(2a_0^\dag a_0-1)a_1, \quad \sigma_1^+=(2a_0^\dag a_0-1)a_1^\dag,
\quad \sigma_1^z=2a_1^\dag a_1-1, \\
\sigma_2^-&=(2a_1^\dag a_1-1)(2a_0^\dag a_0-1)a_2,
\quad \sigma_2^+=(2a_1^\dag a_1-1)(2a_0^\dag a_0-1)a_2^\dag, \quad \sigma_2^z=2a_2^\dag a_2-1,
\end{align}
and make the additional substitution
\begin{align}
\frac{\omega_1}{2}\sigma_1^z=&\omega_1(a_1^\dag a_1 -\frac12)\rightarrow \omega_1 a_1^\dag a_1, \quad
\frac{\omega_2}{2}\sigma_2^z=\omega_2(a_2^\dag a_2 -\frac12)\rightarrow \omega_2 a_2^\dag a_2,
\end{align}
\begin{align}
\sigma_1^x=&(\sigma_1^++\sigma_1^-)\rightarrow (\sigma_1^+\sigma_0^-+\sigma_0^+\sigma_1^-)=\nonumber \\
=&(2a_0^\dag a_0-1)a_1^\dag a_0 + a_0^\dag(2a_0^\dag a_0-1)a_1=-(a_1^\dag a_0+a_0^\dag a_1).
\end{align}
Then the Hamiltonian takes the following form
\begin{equation}
H \rightarrow \omega_2 a_2^\dag a_2 +\omega_1 a_1^\dag a_1 + t(a_2^\dag a_1+a_1^\dag a_2)+
\sum_k \Omega_k b_k^\dag b_k+
[2f_1a_1^\dag a_1 +f_2(a_1^\dag a_0+ a_0^\dag a_1)]\sum_k \lambda_k(b_k^\dag+b_k).
\end{equation}
Note that the fermionization procedure changes the sign of the spin exchange terms.
The coherence of the first spin can be written in terms of $ \mathrm{T}_\tau$-ordered fermion Green's function 
$G_{\alpha\beta}(\tau)=-\langle \mathrm{T}_\tau a_\alpha(\tau)a_\beta^\dag(0)\rangle$ as
\begin{equation}
\langle\sigma_1^x\rangle|_{t,f_2}\rightarrow -\langle a_1^\dag a_0+a_0^\dag a_1\rangle|_{-t,-f_2}=-[G_{01}(0_-)+G_{10}(0_-)]|_{-t,-f_2}.
\end{equation}
Here subscripts define the correspondence rule between spin and fermion models ---
the coherence in the spin system with parameters $t,f_2$ corresponds to the coherence
in the fermion system with parameters $-t,-f_2$ and vice versa.
We introduce the second spin coherence by analogy
\begin{equation}
\langle\sigma_2^x\rangle|_{t,f_2}\rightarrow -\langle a_2^\dag a_0+a_0^\dag a_2\rangle|_{-t,-f_2}=
-[G_{02}(0_-)+G_{20}(0_-)]|_{-t,-f_2}.
\end{equation}
Therefore we need to calculate the fermion Green's function for the latter (fermionized) Hamiltonian.

\section{Green's function method}
\label{sec:green_function}

It is convenient to present the Hamiltonian 
\begin{equation}
\label{eq:fermion_hamiltonian}
H=\omega_1 a_1^\dag a_1 + \omega_2 a_2^\dag a_2 + t(a_1^\dag a_2 + a_2^\dag a_1) +\sum_k \Omega_k b_k^\dag b_k+
[2f_1a_1^\dag a_1 +f_2(a_1^\dag a_0+ a_0^\dag a_1)]\sum_k \lambda_k(b_k^\dag+b_k)
\end{equation}
as a sum of three parts: fermion Hamiltonian
\begin{equation}
H_F=\left[\begin{array}{ccc}
                     a_2^\dag & a_1^\dag & a_0^\dag
										\end{array}\right]\left[\begin{array}{ccc}
                     \omega_2 & t & 0 \\
										 t & \omega_1 & 0 \\
										 0 & 0 & 0
										\end{array}\right]
\left[\begin{array}{ccc}
                     a_2 \\
                     a_1 \\
                     a_0 \\
										\end{array}\right]=
a_\mu^\dag[\mathcal{H}_F]_{\mu\nu}a_\nu,
\end{equation}
where we introduce $3\times3$ matrix $\mathcal{H}_F$ and a summation over repeated (greek) indices;
boson Hamiltonian
\begin{equation}
H_B=\sum_k \Omega_k b_k^\dag b_k;
\end{equation}
and fermion-boson interaction 
\begin{equation}
H_{BF}=[2f_1a_1^\dag a_1 +f_2(a_1^\dag a_0+ a_0^\dag a_1)]\sum_k \lambda_k(b_k^\dag+b_k).
\end{equation}
The latter term is considered as a perturbation. It gives the contribution to the full 
fermion Green's function $G_{\alpha\beta}(\tau-\tau')=-\langle \mathrm{T}_\tau a_\alpha(\tau)a_\beta^\dag(\tau')\rangle$, where the averaging is defined as 
$\langle\star\rangle= \mathrm{Tr}[e^{-\beta H}\star]/\mathrm{Tr}[e^{-\beta H}]$. We evaluate the contribution to the full Green's function in terms of the non-interaction fermion $G^{(0)}_{\alpha\beta}(\tau-\tau')$ and boson 
$D^{(0)}(\tau-\tau')$ Greens' functions.

Let us consider the fermion subsystem first. The non-interacting Green's function reads
\begin{equation}
G^{(0)}_{\alpha\beta}(\tau-\tau')=-\langle \mathrm{T}_\tau a_\alpha(\tau)a_\beta^\dag(\tau')\rangle_0,
\end{equation}
where the averaging is defined as $\langle\star\rangle_0=\mathrm{Tr}[e^{-\beta H_F}\star]/\mathrm{Tr}[e^{-\beta H_F}]$. This Green's function has a compact form in the Matsubara representation
\begin{equation}
G^{(0)}_{\alpha\beta}(ip_n)=\Big[\frac{1}{ip_n-\mathcal{H}_F}\Big]_{\alpha\beta}=
U_{\alpha\mu}\frac{\delta_{\mu\nu}}{ip_n-\epsilon_\mu}[U^\dag]_{\nu\beta}=
U_{\alpha\mu}\widetilde{G}^{(0)}_{\alpha\beta}(ip_n)[U^\dag]_{\nu\beta}.
\end{equation}
Here $p_n=(2n+1)\pi/\beta$ with integer $n$, $\epsilon_2=(\omega_1+\omega_2)/2+\sqrt{(\omega_1-\omega_2)^2/4+t^2}$, $\epsilon_1=(\omega_1+\omega_2)/2-\sqrt{(\omega_1-\omega_2)^2/4+t^2}$, $\epsilon_0=0$, $\sin\theta=2t/\sqrt{(\omega_1-\omega_2)^2+4t^2}$ and $\cos\theta=(\omega_2-\omega_1)/\sqrt{(\omega_1-\omega_2)^2+4t^2}$.
The $3\times3$ matrix $U$ diagonalizes the fermion Hamiltonian matrix $\mathcal{H}_F$
\begin{equation}
\mathcal{H}_F=U \left[\begin{array}{ccc}
           \epsilon_2 & 0 & 0 \\
					 0 & \epsilon_1 & 0 \\
					 0 & 0 & \epsilon_0
\end{array}\right] U^\dag, \quad
U=\left[\begin{array}{ccc}
           \cos(\theta/2) & -\sin(\theta/2) & 0 \\
					 \sin(\theta/2) &  \cos(\theta/2) & 0 \\
					  0 & 0 & 1
\end{array}\right].
\end{equation}
The non-interacting boson Green's function is
\begin{equation}
D^{(0)}(\tau-\tau')=-\langle \mathrm{T}_\tau \phi(\tau)\phi(\tau')\rangle_0,
\end{equation}
where we introduced the boson field $\phi=\sum_k \lambda_k(b_k^\dag+b_k)$ and statistical averaging $\langle\star\rangle_0= \mathrm{Tr}[e^{-\beta H_B}\star]/\mathrm{Tr}[e^{-\beta H_B}]$.
Then
\begin{align}
D^{(0)}(\tau)=&-\sum_{k,l} \lambda_k\lambda_l\langle \mathrm{T}_\tau (b_k e^{-\tau\Omega_k}+b_k^\dag e^{\tau\Omega_k})
(b_l+b_l^\dag)\rangle_0=\sum_k \lambda_k^2 D^{(0)}(k, \tau)= \nonumber \\ =&-\theta(\tau)\sum_k \lambda_k^2 \Big[N_ke^{\tau\Omega_k}+(N_k+1)e^{-\tau\Omega_k}\Big]-\theta(-\tau)\sum_k \lambda_k^2 \Big[N_ke^{-\tau\Omega_k}+
(N_k+1)e^{\tau\Omega_k}\Big],
\end{align}
where $N_k=N(\Omega_k)=(e^{\beta\Omega_k}-1)^{-1}$. The corresponding Matsubara's Green's function with $\omega_m=2\pi m/\beta$ and integer $m$ reads
\begin{equation}
D^{(0)}(k, i\omega_m)=\int_0^\beta d\tau e^{i\omega_m\tau} D^{(0)}(k, \tau)=-\Big[\frac{1}{i\omega_m+\Omega_k}-\frac{1}{i\omega_m-\Omega_k}\Big]=
-\frac{2\Omega_k}{\omega_m^2+\Omega_k^2}.
\end{equation}
The leading non-zero correction to the fermion Green's function is
\begin{equation}
G_{\alpha\beta}(\tau)\approx G^{(0)}_{\alpha\beta}(\tau)+\Delta G^F_{\alpha\beta}(\tau)+\Delta G^H_{\alpha\beta}(\tau).
\end{equation}
It consists of Fock
\begin{equation}
\label{eq:fock}
\Delta G^F_{\alpha\beta}(\tau)=-\int_0^\beta d\tau_1\int_0^\beta d\tau_2 D^{(0)}(\tau_1-\tau_2) G^{(0)}_{\alpha\alpha'}(\tau-\tau_1)M_{\alpha'\beta'}G^{(0)}_{\beta'\alpha''}(\tau_1-\tau_2)
M_{\alpha''\beta''}G^{(0)}_{\beta''\beta}(\tau_2).
\end{equation}
and Hartree terms
\begin{equation}
\label{eq:hartree}
\Delta G_{\alpha\beta}^{H}(\tau)=M_{\alpha''\beta''}G^{(0)}_{\beta''\alpha''}(0)\int_0^\beta d\tau_1  G^{(0)}_{\alpha\alpha'}(\tau-\tau_1)M_{\alpha'\beta'}G^{(0)}_{\beta'\beta}(\tau_1)\int_0^\beta d\tau_2 D^{(0)}(\tau_1-\tau_2).
\end{equation}
Note that there is a summation over the repeated indices. The matrix $M_{\alpha\beta}$ is
\begin{equation}
M=\left[\begin{array}{ccc}
           0 & 0 & 0 \\
					 0 & 2f_1 & f_2 \\
					 0 & f_2 & 0
\end{array}\right].
\end{equation}

\section{Fock contribution}
\label{sec:fock}

We transform the correction into the Matsubara representation using the substitutions
\begin{align}
G^{(0)}_{\alpha\beta}(\tau)=\frac{1}{\beta}\sum_{p_n}e^{-ip_n\tau}G^{(0)}_{\alpha\beta}(ip_n),\quad
D^{(0)}(\tau)=\frac{1}{\beta}\sum_k \lambda_k^2 \sum_{\omega_m}e^{-i\omega_m\tau}D^{(0)}(k,i\omega_m),
\end{align}
where $G^{(0)}_{\alpha\beta}(ip_n)$ and $D^{(0)}(k,i\omega_m)$ have been derived before.
For the Fock term we have
\begin{equation}
\Delta G_{\alpha\beta}^F(ip_n) = G^{(0)}_{\alpha\alpha'}(ip_n)M_{\alpha'\beta'}
\sum_k\lambda_k^2 \Big[-\frac{1}{\beta}\sum_{\omega_m} D^{(0)}(k,i\omega_m)
G^{(0)}_{\beta'\alpha''}(ip_n-i\omega_m)\Big] M_{\alpha''\beta''}G^{(0)}_{\beta''\beta}(ip_n).
\end{equation}
We evaluate the sum in the square brackets
\begin{align}
-\frac{1}{\beta}\sum_{\omega_m} D^{(0)}(k,i\omega_m) &G^{(0)}_{\beta'\alpha''}(ip_n-i\omega_m)=\nonumber \\
=&U_{\beta'\mu}\Big\{-\frac{1}{\beta}\sum_{\omega_m} D^{(0)}(k,i\omega_m) \widetilde{G}^{(0)}_{\mu\nu}(ip_n-i\omega_m)\Big\}[U^\dag]_{\nu\alpha''}= \nonumber \\=&
U_{\beta'\mu} \Big\{\frac{\delta_{\mu\nu}}{\beta}\sum_{\omega_m}\frac{2\Omega_k}{\omega_m^2+\Omega_k^2}
\frac{1}{ip_n-i\omega_m-\epsilon_\mu}\Big\}[U^\dag]_{\nu\alpha''}= \nonumber \\=&
U_{\beta'\mu} \delta_{\mu\nu}\Big[\frac{N_k+n(\epsilon_\mu)}{ip_n-\epsilon_\mu+\Omega_k}+
\frac{N_k+1-n(\epsilon_\mu)}{ip_n-\epsilon_\mu-\Omega_k}\Big][U^\dag]_{\nu\alpha''}.
\end{align}
and introduce the function
\begin{align}
F(\epsilon_\mu)=&\sum_k \lambda_k^2 \Big[\frac{N_k+n(\epsilon_\mu)}{ip_n-\epsilon_\mu+\Omega_k}+
\frac{N_k+1-n(\epsilon_\mu)}{ip_n-\epsilon_\mu-\Omega_k}\Big]=\nonumber \\=&
 \int_0^\infty d\xi\, \mathcal{I}(\xi)
\Big[\frac{N(\xi)+n(\epsilon_\mu)}{ip_n-\epsilon_\mu+\xi}+
\frac{N(\xi)+1-n(\epsilon_\mu)}{ip_n-\epsilon_\mu-\xi}\Big],
\end{align}
with $n(\epsilon_\mu)=(e^{\beta\epsilon_\mu}+1)^{-1}$ and $\mathcal{I}(\xi)=\sum_k\lambda_k^2\delta(\xi-\Omega_k)$. Then the Fock correction takes the form
\begin{equation}
\Delta G^F_{\alpha\beta}(ip_n) = U_{\alpha\mu'}\widetilde{G}^{(0)}_{\mu'\nu'}(ip_n)
[U^\dag MU]_{\nu'\mu} \delta_{\mu\nu}F(\epsilon_\mu)[U^\dag MU]_{\nu\mu''}
\widetilde{G}^{(0)}_{\mu''\nu''}(ip_n)[U^\dag]_{\nu''\beta}.
\end{equation}
Then we evaluate the matrix sum
\begin{equation}
[\widetilde{M}]_{\alpha\beta}=[U^\dag M U]_{\alpha\beta}=
[U^\dag]_{\alpha\mu} M_{\mu\nu} U_{\nu\beta}=
\left[\begin{array}{ccc}
                     2f_1\sin^2(\theta/2) & f_1\sin\theta & f_2\sin(\theta/2) \\
					 f_1\sin\theta & 2f_1\cos^2(\theta/2) & f_2\cos(\theta/2) \\
					 f_2\sin(\theta/2) & f_2\cos(\theta/2) & 0
\end{array}\right],
\end{equation}
and calculate the matrix elements
\begin{equation}
F_{\nu'\mu''}=[\widetilde{M}]_{\nu'\mu}\delta_{\mu\nu}F(\epsilon_\mu)[\widetilde{M}]_{\nu \mu''}.
\end{equation}
The matrix $F_{\nu'\mu''}$ is real valued $\text{Im}[F_{\nu'\mu''}]=0$ and symmetric $F_{\nu'\mu''}=F_{\mu''\nu'}$. Therefore we have
\begin{align}
F_{22}=& f_1^2 \Big[4F(\epsilon_2) \sin^4(\theta/2)+F(\epsilon_1)\sin^2(\theta )\Big]+f_2^2 F(\epsilon_0) \sin ^2(\theta/2), \\
F_{11}=&f_1^2 \Big[F(\epsilon_2)\sin^2(\theta )+
4F(\epsilon_1) \cos^4(\theta/2)\Big]+f_2^2 F(\epsilon_0)\cos^2(\theta/2),\\
F_{00}=& f_2^2 \Big[F(\epsilon_2) \sin ^2(\theta/2)+F(\epsilon_1) \cos^2(\theta/2)\Big],\\
F_{21}=&2f_1^2\sin\theta\Big[F(\epsilon_2)\sin^2(\theta/2)+F(\epsilon_1)\cos^2(\theta/2)\Big]+
f_2^2F(\epsilon_0)\sin (\theta)/2,\\
F_{20}=&2f_1 f_2\sin(\theta/2)\Big[F(\epsilon_2)\sin^2(\theta/2)+F(\epsilon_1)\cos^2(\theta/2)\Big],\\
F_{10}=&2f_1f_2\cos(\theta/2)\Big[F(\epsilon_2)\sin^2(\theta/2)+F(\epsilon_1)\cos^2(\theta/2)\Big],
\end{align}
Finally the Fock correction takes the simple form
\begin{equation}
\Delta G^F_{\alpha\beta}(ip_n) = U_{\alpha\mu'}\frac{1}{ip_n-\epsilon_{\mu'}}
F_{\mu'\mu''}\frac{1}{ip_n-\epsilon_{\mu''}}[U^\dag]_{\mu''\beta}.
\end{equation}
We are interested in the matrix elements $\Delta G^F_{01}(\tau=0_-)$ and $\Delta G^F_{02}(\tau=0_-)$ which correspond to Fock contribution in $\langle a_1^\dag a_0\rangle$ and $\langle a_2^\dag a_0\rangle$ respectively. To do it we evaluate the sum
\begin{equation}
\Delta G^F_{\alpha\beta}(\tau)=\frac{1}{\beta}\sum_{p_n}e^{-ip_n\tau}\Delta G^F_{\alpha\beta}(ip_n),
\end{equation}
and take the limit $\tau\rightarrow 0_-$ in the final result. We have
\begin{align}
\Delta G^F_{01}(ip_n)=2f_1f_2\frac{F(\epsilon_2)\sin^2(\theta/2)+F(\epsilon_1)\cos^2(\theta/2)}{ip_n-\epsilon_0}
\Big[\frac{\sin^2(\theta/2)}{(ip_n-\epsilon_2)}+\frac{\cos^2(\theta/2)}{(ip_n-\epsilon_1)}\Big],
\end{align}
\begin{align}
\Delta G^F_{02}(ip_n) =f_1f_2\sin\theta\frac{F(\epsilon_2)\sin^2(\theta/2)+F(\epsilon_1)\cos^2(\theta/2)}
{ip_n-\epsilon_0}
\Big[\frac{1}{(ip_n-\epsilon_2)}-\frac{1}{(ip_n-\epsilon_1)}\Big].
\end{align}
These expressions are simplified the limit $t\rightarrow 0$, which equals to the case $\sin\theta\rightarrow 0$ and $\cos\theta\rightarrow 1$. Here and further in the text we consider the case $\omega_2>\omega_1$ for brevity.
Then in the leading order by small parameter $\sin\theta$ we have 
\begin{align}
\Delta G^F_{01}(ip_n) \approx \frac{2f_1f_2 F(\epsilon_1)}{(ip_n-\epsilon_0)(ip_n-\epsilon_1)},\quad
\Delta G^F_{02}(ip_n) \approx \sin\theta \frac{f_1f_2F(\epsilon_1)}{(ip_n-\epsilon_0)}
\Big[\frac{1}{(ip_n-\epsilon_2)}-\frac{1}{(ip_n-\epsilon_1)}\Big].
\end{align}
One can note the correspondence between the obtained corrections
\begin{equation}
\Delta G^F_{02}(ip_n)=\frac{t}{ip_n-\epsilon_2}\Delta G^F_{01}(ip_n).
\end{equation}
Let us evaluate now $\Delta G^F_{01}(0_-)$ and  $\Delta G^F_{02}(0_-)$ in the limit $t\rightarrow 0$
\begin{align}
\Delta G^F_{01}(0_-)=2f_1f_2 \int_0^\infty d\xi\, \mathcal{I}(\xi)\Big[&-\frac{(2N(\xi)+1)\epsilon_1+(2n(\epsilon_1)-1)\xi}{2\epsilon_1(\xi^2-\epsilon_1^2)}+
\frac{(2n(\epsilon_1)-1)n(\epsilon_1)}{\epsilon_1\xi}+
\nonumber \\ &+
\frac{[N(\xi)+1]n(\epsilon_1)}{\xi(\xi-\epsilon_1)}+
\frac{N(\xi)n(\epsilon_1)}{\xi(\xi+\epsilon_1)}\Big],
\end{align}
\begin{align}
\Delta G^F_{02}(0_-)=\sin\theta f_1f_2\int_0^\infty d\xi\, \mathcal{I}(\xi)
\Big[W(\epsilon_2)-W(\epsilon_1)\Big],
\end{align}
where we introduced the function
\begin{align}
W(w)=&\frac{1}{2w}\Big[\frac{N(\xi)+1-n(\epsilon_1)}{\epsilon_1+\xi}-\frac{N(\xi)+n(\epsilon_1)}{\xi-\epsilon_1}\Big]
+\frac{N(\xi)+1-n(\epsilon_1)}{(\epsilon_1+\xi)(\epsilon_1+\xi-w)}
n(\epsilon_1+\xi)
\nonumber \\
+&\frac{n(w)}{w}\Big[\frac{N(\xi)+n(\epsilon_1)}{w-\epsilon_1+\xi}+\frac{N(\xi)+1-n(\epsilon_1)}{w-\epsilon_1-\xi}\Big]+
\frac{N(\xi)+n(\epsilon_1)}{(\epsilon_1-\xi)(\epsilon_1-\xi-w)}n(\epsilon_1-\xi).
\end{align}

\section{Hartree contribution}
\label{sec:hartree}

Let us consider the correction which appears due to the Hartree term
\begin{equation}
\Delta G_{\alpha\beta}^{H}(\tau)=M_{\alpha''\beta''}G^{(0)}_{\beta''\alpha''}(0)\int_0^\beta d\tau_1  G^{(0)}_{\alpha\alpha'}(\tau-\tau_1)M_{\alpha'\beta'}G^{(0)}_{\beta'\beta}(\tau_1)\int_0^\beta d\tau_2 D^{(0)}(\tau_1-\tau_2).
\end{equation}
The first multiplier can be evaluated in the following way. We calculate this term in Matsubara representation
\begin{equation}
M_{\alpha''\beta''}G^{(0)}_{\beta''\alpha''}(ip_n)=M_{\alpha''\beta''}U_{\beta''\mu}
\frac{\delta_{\mu\nu}}{ip_n-\epsilon_\mu}[U^\dag]_{\nu\alpha''}=
\sum_{\mu=0}^2\frac{[U^\dag MU]_{\mu\mu}}{ip_n-\epsilon_\mu}
\end{equation}
We evaluated the matrix $\widetilde{M}=U^\dag MU$ before. Its diagonal matrix elements are
\begin{equation}
\widetilde{M}_{22}=2f_1\sin^2(\theta/2), \quad \widetilde{M}_{11}=2f_1\cos^2(\theta/2), \quad \widetilde{M}_{00}=0.
\end{equation}
It results in the following
\begin{align}
M_{\alpha''\beta''}G^{(0)}_{\beta''\alpha''}(0)=&\frac{1}{\beta}
\sum_{p_n}e^{-ip_n\tau}M_{\alpha''\beta''}G^{(0)}_{\beta''\alpha''}(ip_n)\Big|_{\tau=0_-}=\nonumber \\
=&2f_1\sin^2(\theta/2)n(\epsilon_2)+2f_1\cos^2(\theta/2)n(\epsilon_1).
\end{align}
Then we estimate
\begin{align}
\int_0^\beta d\tau_1  G^{(0)}_{\alpha\alpha'}(\tau-\tau_1)M_{\alpha'\beta'}G^{(0)}_{\beta'\beta}(\tau_1)=&
\frac{1}{\beta}\sum_{p_n}e^{-ip_n\tau}U_{\alpha\mu}\frac{\delta_{\mu\nu}}{ip_n-\epsilon_\nu}\widetilde{M}_{\nu\mu'}
\frac{\delta_{\mu'\nu'}}{ip_n-\epsilon_{\mu'}}[U^\dag]_{\nu'\beta}=\nonumber \\
=&\frac{1}{\beta}\sum_{p_n}e^{-ip_n\tau}R_{\alpha\beta}.
\end{align}
We are interested in $\Delta G_{01}^{H}(\tau)$ and $\Delta G_{02}^{H}(\tau)$.
Therefore we calculate the corresponding matrix elements
\begin{equation}
R_{01}=\frac{f_2}{(ip_n-\epsilon_0)}\Big[\frac{\sin^2(\theta/2)}{ip_n-\epsilon_2}+
\frac{\cos^2(\theta/2)}{ip_n-\epsilon_1}\Big], \quad R_{02}=\frac{f_2\sin\theta}{2(ip_n-\epsilon_0)}
\Big[\frac{1}{ip_n-\epsilon_2}-\frac{1}{ip_n-\epsilon_1}\Big].
\end{equation}
The latter term reads
\begin{align}
\int_0^\beta d\tau_2 D^{(0)}(\tau_1-\tau_2)=&\frac{1}{\beta}\sum_k \lambda_k^2 
\sum_{\omega_m} D^{(0)}(k,i\omega_m)e^{-i\omega_m \tau_1} \int_0^\beta d\tau_2  
e^{i\omega_m \tau_2}=\nonumber \\=
&-2\int_0^\infty d\xi\frac{\mathcal{I}(\xi)}{\xi}=-2\Omega.
\end{align}
As a result we have
\begin{equation}
\Delta G^H_{01}(ip_n)=-\frac{4f_1f_2\Omega}{(ip_n-\epsilon_0)}
\Big[\sin^2(\theta/2)n(\epsilon_2)+\cos^2(\theta/2)n(\epsilon_1)\Big]\Big[\frac{\sin^2(\theta/2)}{ip_n-\epsilon_2}+\frac{\cos^2(\theta/2)}{ip_n-\epsilon_1}\Big].
\end{equation}
\begin{equation}
\Delta G^H_{02}(ip_n)=-\frac{2f_1f_2\Omega\sin\theta}{ip_n-\epsilon_0}
\Big[\sin^2(\theta/2)n(\epsilon_2)+\cos^2(\theta/2)n(\epsilon_1)\Big]
\Big[\frac{1}{ip_n-\epsilon_2}-\frac{1}{ip_n-\epsilon_1}\Big].
\end{equation}
Considering $\epsilon_0=0$ and taking the limit $\sin\theta\rightarrow 0, \cos\theta\rightarrow 1$ we get
\begin{equation}
\Delta G^H_{01}(0_-)=-\frac{4f_1f_2\Omega n(\epsilon_1)}{\beta}\sum_{p_n}\frac{e^{-ip_n\tau}}{ip_n(ip_n-\epsilon_1)}\Big|_{\tau=0_-}\!\!\!=
\frac{2f_1f_2}{\epsilon_1}\Omega n(\epsilon_1)\tanh\Big(\frac{\beta\epsilon_1}{2}\Big),
\end{equation}
\begin{align}
\Delta G^H_{02}(0_-)=&-\frac{2f_1f_2\sin\theta \Omega n(\epsilon_1)}{\beta}\sum_{p_n}
\frac{e^{-ip_n\tau}}{ip_n}\Big[\frac{1}{ip_n-\epsilon_2}-\frac{1}{ip_n-\epsilon_1}\Big]\Big|_{\tau=0_-}\!\!\!=
\nonumber \\=&
f_1f_2\sin\theta n(\epsilon_1)\Omega
\Big[\frac{\tanh\Big(\frac{\beta\epsilon_2}{2}\Big)}{\epsilon_2}-
\frac{\tanh\Big(\frac{\beta\epsilon_1}{2}\Big)}{\epsilon_1}\Big],
\end{align}
where we took into account the sum
\begin{equation}
\frac{1}{\beta}\sum_{p_n}
\frac{e^{-ip_n\tau}}{ip_n(ip_n-w)}=-\frac{1}{2w} + \frac{e^{-\tau w}}{w}n(w).
\end{equation}

\section{Analysis of the coherences}
\label{sec:analysis_of_coherences}

In this section we analyze the coherence of the first $\langle\sigma_1^x\rangle_F$ and second $\langle\sigma_1^x\rangle_F$ spins. Here the subscript $F$ indicates that we derived the coherences from the fermionized Hamiltonian. In accordance with the aforementioned fermion-boson correspondence we have
\begin{align}
\langle\sigma_1^x\rangle_F=-\langle a_1^\dag a_0+a_0^\dag a_1\rangle|_{-t,-f_2}=
&-[G_{01}(0_-)+G_{10}(0_-)]|_{-t,-f_2}=\nonumber \\=&-2[\Delta G^F_{01}(0_-)+\Delta G^H_{01}(0_-)]|_{-t,-f_2},
\end{align}
\begin{align}
\langle\sigma_2^x\rangle_F=-\langle a_2^\dag a_0+a_0^\dag a_2\rangle|_{-t,-f_2}=
&-[G_{02}(0_-)+G_{20}(0_-)]|_{-t,-f_2}=\nonumber \\=&-2[\Delta G^F_{02}(0_-)+\Delta G^H_{02}(0_-)]|_{-t,-f_2}.
\end{align}
Here we took into account that the non-interaction fermion Green's function has zero off-diagonal 
matrix elements and hence $G_{01}(0_-)=\Delta G^F_{01}(0_-)+\Delta G^H_{01}(0_-)$, 
$G_{02}(0_-)=\Delta G^F_{02}(0_-)+\Delta G^H_{02}(0_-)$. 

Taking the sum of Fock $\Delta G^F_{01}(0_-)$ and Hartree $\Delta G^F_{01}(0_-)$ contributions and we get 
\begin{align}
\langle\sigma_1^x\rangle_F=&-\langle a_1^\dag a_0+a_1^\dag a_0\rangle|_{-t,-f_2}=\nonumber \\ 
=&-4f_1f_2 \int_0^\infty d\xi\, \mathcal{I}(\xi)\Big[-\frac12\frac{\tanh(\beta\omega_1/2)}{\xi\omega_1}+
\frac{\xi\coth(\beta\xi/2)\tanh(\beta\omega_1/2)-\omega_1}{2\xi(\xi^2-\omega_1^2)}\Big],
\end{align}
where we took into account that $\epsilon_1\approx \omega_1$ in the limit $t\rightarrow 0$.
This expression has a similar integrand as the coherence derived in the spin representation 
\begin{equation}
\langle\sigma_1^x\rangle=-4f_1f_2\int_0^\infty d\xi\,
\mathcal{I}(\xi)\frac{\xi\coth(\beta\xi/2) \tanh(\beta\omega_1/2)-\omega_1}{\xi(\xi^2-\omega_1^2)}.
\end{equation}
One can observe that the coherences satisfy the equation
\begin{equation}
\langle\sigma_1^x\rangle=2\langle\sigma_1^x\rangle_F-4f_1f_2\frac{\tanh(\beta\omega_1/2)}{\omega_1}\int_0^\infty d\xi\frac{\mathcal{I}(\xi)}{\xi}=2\langle\sigma_1^x\rangle_F-4f_1f_2\Omega\frac{\tanh(\beta\omega_1/2)}{\omega_1},
\end{equation}
which is valid for any spectral density function $\mathcal{I}(\xi)$. Note that 
the $\langle\sigma_1^x\rangle$ and $\langle\sigma_1^x\rangle_F$ have the opposite signs.

\begin{figure}
	\centering
		\includegraphics{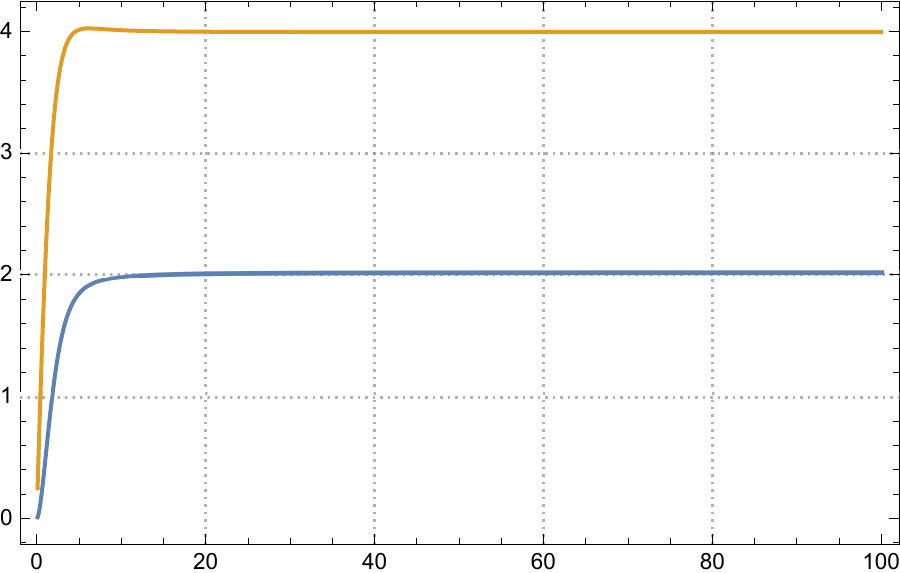}
	\caption{\label{fig:spin_fermion_01} Dimensionless functions $Y_S(\beta\omega_1,\omega_1/\omega_c)$ (blue curve) and
	$-Y_F(\beta\omega_1,\omega_1/\omega_c)$ (yellow curve) for $\omega_1/\omega_c=0.1$, as a function of dimensionless parameter $\beta\omega_1 \in[0,100]$.}
\end{figure}

\begin{figure}
	\centering
		\includegraphics{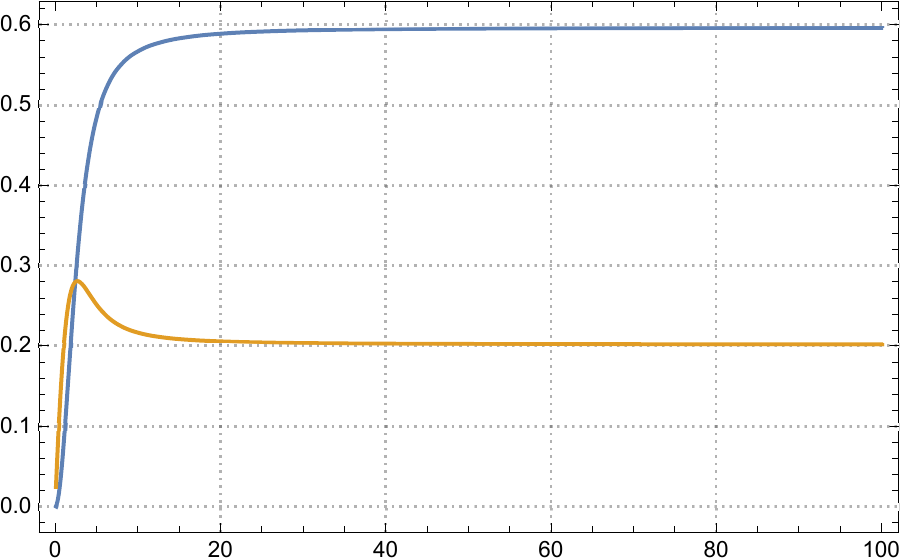}
	\caption{\label{fig:spin_fermion_1} Dimensionless functions $Y_S(\beta\omega_1,\omega_1/\omega_c)$ (blue curve) and
	$-Y_F(\beta\omega_1,\omega_1/\omega_c)$ (yellow curve) for $\omega_1/\omega_c=1$, as a function of dimensionless parameter $\beta\omega_1 \in[0,100]$.}
\end{figure}

\begin{figure}
	\centering
		\includegraphics{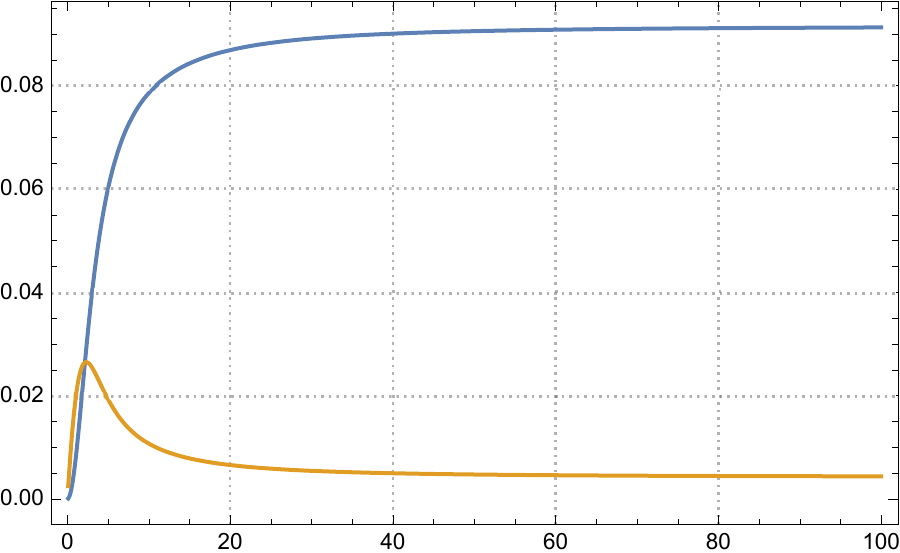}
	\caption{\label{fig:spin_fermion_10} Dimensionless functions $Y_S(\beta\omega_1,\omega_1/\omega_c)$ (blue curve) and
	$-Y_F(\beta\omega_1,\omega_1/\omega_c)$ (yellow curve) for $\omega_1/\omega_c=10$, as a function of dimensionless parameter $\beta\omega_1 \in[0,100]$.}
\end{figure}

Let us calculate and plot both functions as a function of normalized temperature for the particular case of the spectral density function $\mathcal{I}(\xi)=A\xi\exp(-\xi/\omega_c)$. To do it we introduce the new dimensionless parameter 
$x=\xi/\omega_1$ 
\begin{equation}
\langle\sigma_1^x\rangle=-4f_1f_2A\int_0^\infty dx
e^{-x\omega_1/\omega_c}\frac{x\coth(\beta\omega_1 x/2) \tanh(\beta\omega_1/2)-1}{x^2-1}.
\end{equation}
\begin{equation}
\langle\sigma_1^x\rangle_F= -4f_1f_2 A\int_0^\infty dx  e^{-x\omega_1/\omega_c}
\Big[-\frac12\tanh(\beta\omega_1/2)+
\frac{x\coth(\beta\omega_1 x/2)\tanh(\beta\omega_1/2)-1}{2(x^2-1)}\Big].
\end{equation}
 In order to compare the coherence in spin and fermion systems we evaluate numerically and plot the corresponding functions
\begin{equation}
Y_S(\omega_1/\omega_c, \beta\omega_1)=\int_0^\infty dx
e^{-x\omega_1/\omega_c}\frac{x\coth(\beta\omega_1 x/2) \tanh(\beta\omega_1/2)-1}{(x^2-1)},
\end{equation}
and
\begin{equation}
Y_F(\omega_1/\omega_c, \beta\omega_1)=\int_0^\infty dx e^{-x\omega_1/\omega_c}
\Big[-\frac12\tanh(\beta\omega_1/2)+
\frac{x\coth(\beta\omega_1 x/2)\tanh(\beta\omega_1/2)-1}{2(x^2-1)}\Big]
\end{equation}
for different values $\omega_1/\omega_c=0.1,1,10$ as a function of $\beta\omega_1$ and plot these results.
For $\omega_1/\omega_c=0.1$ we have Fig.~\ref{fig:spin_fermion_01}, for $\omega_1/\omega_c=1$ we have 
Fig.~\ref{fig:spin_fermion_1} and for $\omega_1/\omega_c=10$ we have
Fig.~\ref{fig:spin_fermion_10} respectively.

\begin{figure}
	\centering
		\includegraphics{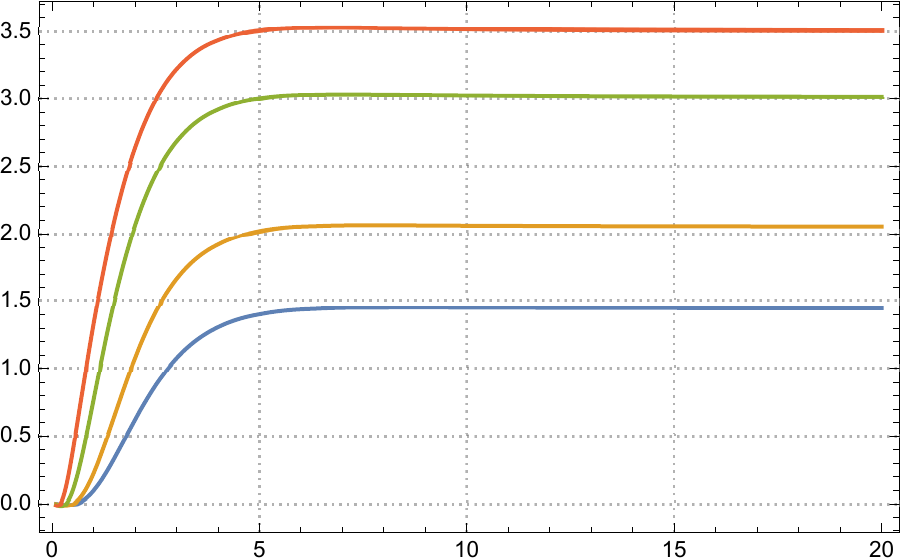}
	\caption{\label{fig:double_spin_fermion_01} Dimensionless function $\mathcal{G}(\beta\omega_1,\omega_1/\omega_c)$ for $\omega_2/\omega_1=1.5$ (blue curve), $\omega_2/\omega_1=2$ (yellow curve), $\omega_2/\omega_1=4$ (green curve) and $\omega_2/\omega_1=8$ (red curve) for the case of $\omega_1/\omega_c=0.1$, as a function of dimensionless parameter $\beta\omega_1 \in[0,20]$.}
\end{figure}

\begin{figure}
	\centering
		\includegraphics{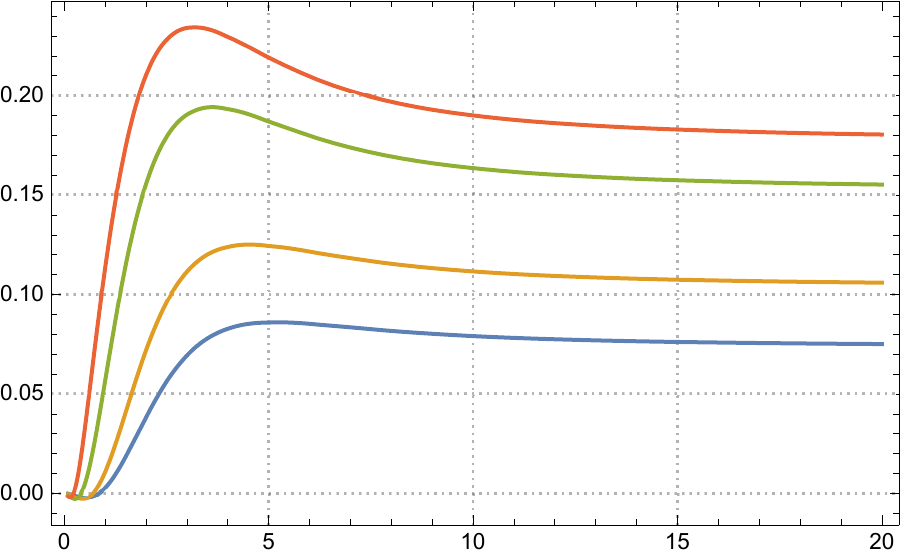}
	\caption{\label{fig:double_spin_fermion_1} Dimensionless function $\mathcal{G}(\beta\omega_1,\omega_1/\omega_c)$ for $\omega_2/\omega_1=1.5$ (blue curve), $\omega_2/\omega_1=2$ (yellow curve), $\omega_2/\omega_1=4$ (green curve) and $\omega_2/\omega_1=8$ (red curve) for the case of $\omega_1/\omega_c=1$, as a function of dimensionless parameter $\beta\omega_1 \in[0,20]$.}
\end{figure}

\begin{figure}
	\centering
		\includegraphics{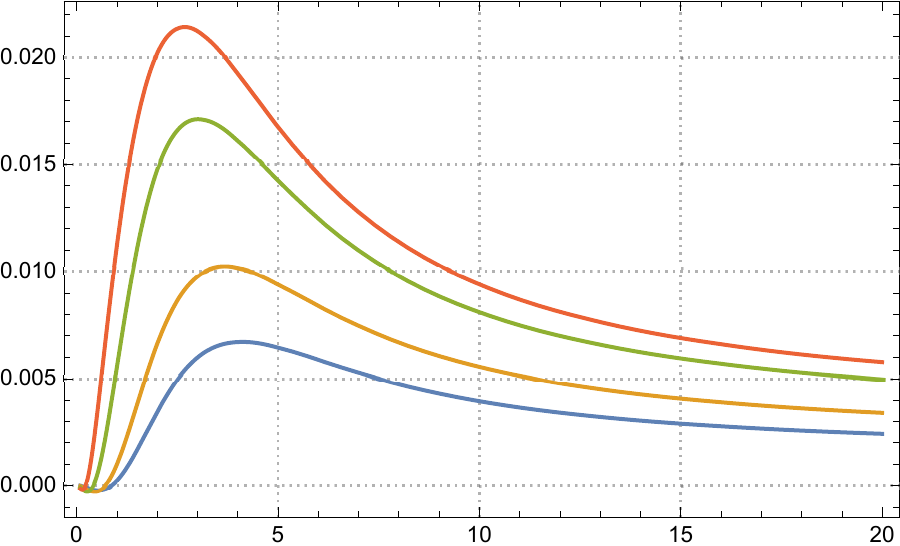}
	\caption{\label{fig:double_spin_fermion_10} Dimensionless function $\mathcal{G}(\beta\omega_1,\omega_1/\omega_c)$ for $\omega_2/\omega_1=1.5$ (blue curve), $\omega_2/\omega_1=2$ (yellow curve), $\omega_2/\omega_1=4$ (green curve) and $\omega_2/\omega_1=8$ (red curve) for the case of $\omega_1/\omega_c=10$, as a function of dimensionless parameter $\beta\omega_1 \in[0,20]$.}
\end{figure}

Combining together $\Delta G^F_{02}(0_-)$ and $\Delta G^H_{02}(0_-)$ we get the result 
\begin{align}
\Delta& G_{02}(0_-)=\sin\theta f_1f_2\int_0^\infty d\xi\, \mathcal{I}(\xi)
\Big[W(\epsilon_2)-W(\epsilon_1) + n(\epsilon_1)\Big\{\frac{\tanh(\beta\epsilon_2/2)}{\xi\epsilon_2}-
\frac{\tanh(\beta\epsilon_1/2)}{\xi\epsilon_1}\Big\}\Big]=\nonumber \\=&
\sin\theta f_1f_2\int_0^\infty d\xi\, \mathcal{I}(\xi) [N(\xi)+n(\epsilon_1)]
\Big\{\frac{1}{2\epsilon_1(\xi-\epsilon_1)}-\frac{1}{2\epsilon_2(\xi-\epsilon_1)}-
\frac{n(\epsilon_1-\xi)}{\xi(\xi-\epsilon_1)}+ \nonumber \\
& \qquad \qquad \qquad \qquad \qquad \qquad \qquad \qquad +\frac{n(\epsilon_2)}{\epsilon_2(\xi+\Delta\epsilon)}+
\frac{n(\epsilon_1-\xi)}{(\xi-\epsilon_1)(\xi+\Delta\epsilon)}\Big\}+
\nonumber \\+&
\sin\theta f_1f_2\int_0^\infty d\xi\, \mathcal{I}(\xi)[N(\xi)+1-n(\epsilon_1)]\Big\{
\frac{n(\xi+\epsilon_1)}{(\xi+\epsilon_1)(\xi-\Delta\epsilon)}-\frac{n(\epsilon_2)}{\epsilon_2(\xi-\Delta\epsilon)}-
\frac{n(\epsilon_1+\xi)}{\xi(\xi+\epsilon_1)} 
+ \nonumber \\
& \qquad \qquad \qquad \qquad \qquad \qquad \qquad \qquad +
\frac{1}{2(\xi+\epsilon_1)\epsilon_2}-
\frac{1}{2(\xi+\epsilon_1)\epsilon_1}\Big\}+
\nonumber \\+&
\sin\theta f_1f_2\int_0^\infty d\xi\, \mathcal{I}(\xi) \frac{n(\epsilon_1)}{\xi\epsilon_2}[1-2n(\epsilon_2)],
\end{align}
where $\Delta\epsilon=\epsilon_2-\epsilon_1>0$. In order to understand the behavior of this function with the parameters of the system we consider the particular case of $t=0.1\omega_1$, introduce the spectral density function $\mathcal{I}(\xi)=A\xi\exp(-\xi/\omega_c)$ and plot the dimensionless function $\mathcal{G}(\beta\omega_1,\omega_1/\omega_c)=
-\Delta G_{02}(0_-)/(A\sin\theta f_1f_2)$ for the parameters $\omega_1/\omega_c=0.1,1,10$  at different ratio $\omega_2/\omega_1=1.5,2,4,8$. The corresponding plots are presented on Figs.~\ref{fig:double_spin_fermion_01}, 
\ref{fig:double_spin_fermion_1} and \ref{fig:double_spin_fermion_10}.

In the limit $t\rightarrow 0$ the coherence of the second spin takes the form 
\begin{align}
\langle\sigma_2^x\rangle_F&=-2\Delta G_{02}(0_-)|_{-t,-f_2}=-\frac{4t f_1f_2}{\omega_2-\omega_1} 
\times \nonumber \\ \times\Big[&
\int_0^\infty d\xi\, \mathcal{I}(\xi) [N(\xi)+n(\omega_1)]
\Big\{\frac{1}{2\omega_1(\xi-\omega_1)}-\frac{1}{2\omega_2(\xi-\omega_1)}-
\frac{n(\omega_1-\xi)}{\xi(\xi-\omega_1)}+ 
\nonumber \\ &\qquad \qquad \qquad \qquad \qquad  \qquad \qquad +\frac{n(\omega_2)}{\omega_2(\xi+\Delta\omega)}+
\frac{n(\omega_1-\xi)}{(\xi-\omega_1)(\xi+\Delta\omega)}\Big\}+
\nonumber \\+&
\int_0^\infty d\xi\, \mathcal{I}(\xi)[N(\xi)+1-n(\omega_1)]\Big\{
\frac{n(\xi+\omega_1)}{(\xi+\omega_1)(\xi-\Delta\omega)}-
\frac{n(\omega_2)}{\omega_2(\xi-\Delta\omega)}-
\frac{n(\omega_1+\xi)}{\xi(\xi+\omega_1)}+
\nonumber \\ &\qquad \qquad \qquad \qquad \qquad \qquad \qquad +
\frac{1}{2(\xi+\omega_1)\omega_2}-
\frac{1}{2(\xi+\omega_1)\omega_1}\Big\}+
\nonumber \\+&
\int_0^\infty d\xi\, \mathcal{I}(\xi) \frac{n(\omega_1)}{\xi\omega_2}[1-2n(\omega_2)]\Big],
\end{align}
where we took into account $\sin\theta=2t/\sqrt{(\omega_1-\omega_2)^2+4t^2}\approx 
2t/(\omega_2-\omega_1)$ for considered $\omega_2>\omega_1$ case and introduced the short notation
$\Delta\omega=\omega_2-\omega_1>0$.

At the zero-temperature limit the expressions for fermion coherences take the form
\begin{align}
\langle\sigma_1^x\rangle_F=\frac{2f_1f_2}{\omega_1}\int_0^\infty d\xi\frac{\mathcal{I}(\xi)}{\xi+\omega_1}, 
\quad
\langle\sigma_2^x\rangle_F=\frac{2tf_1f_2}{\omega_1\omega_2} \int_0^\infty d\xi \frac{\mathcal{I}(\xi)}{\xi+\omega_1}.
\end{align}
Hence we have the relation between fermion coherences  
\begin{align}
\langle\sigma_2^x\rangle_F=\frac{t}{\omega_2} \langle\sigma_1^x\rangle_F.
\end{align}	
Note that, the expressions for the coherences $\langle\sigma_1^x\rangle_F$ and $\langle\sigma_2^x\rangle_F$ have been  evaluated for the case $\omega_2>\omega_1$. However, it turns out that they have the same functional dependence also in the opposite case $\omega_2<\omega_1$. The proof of this statement is presented in Appendix~\ref{sec:omega_2_small}.

The second fermion coherence inthe high-temperature limit $\beta\rightarrow 0$ for an arbitrary spectral density function $\mathcal{I}(\xi)$ is
\begin{align}
\label{eq:ht_limit_fermion}
\langle\sigma_2^x\rangle_F|_{\beta\rightarrow 0}=-\frac{tf_1f_2\beta^2}{6} 
\int_0^\infty d\xi\, \frac{\mathcal{I}(\xi)}{\xi}+O(\beta^3).
\end{align}

\section{Fermion coherences for $\omega_2<\omega_1$ case}
\label{sec:omega_2_small}

Here we derive the expression for the coherence of the first and the second spins in the fermionized spin-boson model, 
for the case $\omega_1>\omega_2$ and $t/(\omega_1-\omega_2)\ll1$. 

First we consider the general expression for the Hartree contribution 
\begin{equation}
\Delta G^H_{01}(ip_n)=-\frac{4f_1f_2\Omega}{ip_n-\epsilon_0}
\Big[\sin^2(\theta/2)n(\epsilon_2)+\cos^2(\theta/2)n(\epsilon_1)\Big]\Big[\frac{\sin^2(\theta/2)}{ip_n-\epsilon_2}+\frac{\cos^2(\theta/2)}{ip_n-\epsilon_1}\Big].
\end{equation}
\begin{equation}
\Delta G^H_{02}(ip_n)=-\frac{2f_1f_2\Omega\sin\theta}{ip_n-\epsilon_0}
\Big[\sin^2(\theta/2)n(\epsilon_2)+\cos^2(\theta/2)n(\epsilon_1)\Big]
\Big[\frac{1}{ip_n-\epsilon_2}-\frac{1}{ip_n-\epsilon_1}\Big].
\end{equation}
Here $\sin\theta=2t/\sqrt{(\omega_1-\omega_2)^2+4t^2}$, $\cos\theta=(\omega_2-\omega_1)/\sqrt{(\omega_1-\omega_2)^2+4t^2}$.
In considering case these expressions are $\sin\theta\approx 2t/|\omega_1-\omega_2|$, 
$\cos\theta\approx \textrm{sign}(\omega_2-\omega_1)=-1$. Hence, the angle 
$\theta\approx \pi-2t/|\omega_1-\omega_2|$ and  $\sin(\theta/2)\approx 1$,
$\cos(\theta/2)\approx t/|\omega_1-\omega_2|$. Then we have
\begin{equation}
\Delta G^H_{01}(ip_n)\approx -\frac{4f_1f_2}{ip_n-\epsilon_0}\int_0^\infty d\xi \frac{\mathcal{I}(\xi)}{\xi}
n(\epsilon_2)\frac{1}{ip_n-\epsilon_2},
\end{equation}
\begin{equation}
\Delta G^H_{02}(ip_n)\approx -\frac{2f_1f_2\sin\theta}{ip_n-\epsilon_0}\int_0^\infty d\xi \frac{\mathcal{I}(\xi)}{\xi}
n(\epsilon_2)\Big[\frac{1}{ip_n-\epsilon_2}-\frac{1}{ip_n-\epsilon_1}\Big].
\end{equation}
 
The general expressions for the Fock terms are
\begin{align}
\Delta G^F_{01}(ip_n)=2f_1f_2\frac{F(\epsilon_2)\sin^2(\theta/2)+F(\epsilon_1)\cos^2(\theta/2)}{(ip_n-\epsilon_0)}
\Big[\frac{\sin^2(\theta/2)}{(ip_n-\epsilon_2)}+\frac{\cos^2(\theta/2)}{(ip_n-\epsilon_1)}\Big],
\end{align}
\begin{align}
\Delta G^F_{02}(ip_n) =f_1f_2\sin\theta\frac{F(\epsilon_2)\sin^2(\theta/2)+F(\epsilon_1)\cos^2(\theta/2)}
{(ip_n-\epsilon_0)}
\Big[\frac{1}{(ip_n-\epsilon_2)}-\frac{1}{(ip_n-\epsilon_1)}\Big].
\end{align}
In the considering case they read  
 \begin{align}
\Delta G^F_{01}(ip_n)=2f_1f_2F(\epsilon_2)
\frac{1}{(ip_n-\epsilon_0)}\frac{1}{(ip_n-\epsilon_2)},
\end{align}
\begin{align}
\Delta G^F_{02}(ip_n)=f_1f_2\sin\theta F(\epsilon_2)
\frac{1}{(ip_n-\epsilon_0)}
\Big[\frac{1}{(ip_n-\epsilon_2)}-\frac{1}{(ip_n-\epsilon_1)}\Big].
\end{align}

Note that for $\omega_2>\omega_1$ the terms $\Delta G^F_{01}(ip_n)$, $\Delta G^F_{01}(ip_n)$ are
functions of $\epsilon_1$, while for $\omega_1>\omega_2$ they are the same functions of
$\epsilon_2$. Taking into account that in the latter case  $\epsilon_2\approx \omega_1$, $\epsilon_1\approx \omega_2$
and $\epsilon_0=0$ we get the following  
\begin{equation}
\Delta G^H_{01}(ip_n)\approx -\frac{4f_1f_2}{ip_n}\int_0^\infty d\xi \frac{\mathcal{I}(\xi)}{\xi}
n(\omega_1)\frac{1}{ip_n-\omega_1},
\end{equation}
\begin{align}
\Delta G^F_{01}(ip_n)=2f_1f_2F(\omega_1)
\frac{1}{ip_n}\frac{1}{(ip_n-\omega_1)}.
\end{align}
Both terms are independent on the $\omega_2$ and coincide with the analogous expressions for the case $\omega_2>\omega_1$ written in terms of $\omega_1,\omega_2$ and $t$. Therefore, the expression for the coherence of the first spin, 
obtained before for the case of $\omega_2>\omega_1$ remains valid also for the opposite one $\omega_1>\omega_2$. 

Substituting $\epsilon_2\approx \omega_1$, $\epsilon_1\approx \omega_2$ into the 
$\Delta G^F_{02}(ip_n)$ and $\Delta G^F_{02}(ip_n)$ we get 
\begin{equation}
\Delta G^H_{02}(ip_n)\approx (-1)\Big\{-\frac{2f_1f_2}{ip_n}\frac{2t}{|\omega_1-\omega_2|}
\int_0^\infty d\xi \frac{\mathcal{I}(\xi)}{\xi}
n(\omega_1)\Big[\frac{1}{ip_n-\omega_2}-\frac{1}{ip_n-\omega_1}\Big]\Big\}.
\end{equation}
\begin{align}
\Delta G^F_{02}(ip_n)=(-1)\Big\{f_1f_2\frac{2t}{|\omega_1-\omega_2|} F(\omega_1)
\frac{1}{ip_n}
\Big[\frac{1}{(ip_n-\omega_2)}-\frac{1}{(ip_n-\omega_1)}\Big]\Big\}.
\end{align}
One can see that the expressions in parenthesis coincide with the expressions for  
$\Delta G^F_{02}(ip_n)$ and $\Delta G^F_{02}(ip_n)$ in the case $\omega_2>\omega_1$. 
Therefore, the final expression for the coherence of the second spin in the case $\omega_1>\omega_2$ 
can be obtained  from the former one by multiplying it on $-1$. Therefore, we get 
\begin{align}
\langle\sigma_2^x\rangle_F&=-\frac{4t f_1f_2}{\omega_2-\omega_1} 
\times \nonumber \\ \times\Big[&
\int_0^\infty d\xi\, \mathcal{I}(\xi) [N(\xi)+n(\omega_1)]
\Big\{\frac{1}{2\omega_1(\xi-\omega_1)}-\frac{1}{2\omega_2(\xi-\omega_1)}-
\frac{n(\omega_1-\xi)}{\xi(\xi-\omega_1)}+ \nonumber \\ 
& \qquad \qquad \qquad \qquad \qquad \qquad +\frac{n(\omega_2)}{\omega_2(\xi+\Delta\omega)}+
\frac{n(\omega_1-\xi)}{(\xi-\omega_1)(\xi+\Delta\omega)}\Big\}+
\nonumber \\+&
\int_0^\infty d\xi\, \mathcal{I}(\xi)[N(\xi)+1-n(\omega_1)]\Big\{
\frac{n(\xi+\omega_1)}{(\xi+\omega_1)(\xi-\Delta\omega)}-
\frac{n(\omega_2)}{\omega_2(\xi-\Delta\omega)}-
\frac{n(\omega_1+\xi)}{\xi(\xi+\omega_1)}+ 
\nonumber \\ 
& \qquad \qquad \qquad \qquad \qquad \qquad +
\frac{1}{2(\xi+\omega_1)\omega_2}-
\frac{1}{2(\xi+\omega_1)\omega_1}\Big\}+
\nonumber \\+&
\int_0^\infty d\xi\, \mathcal{I}(\xi) \frac{n(\omega_1)}{\xi\omega_2}[1-2n(\omega_2)]\Big],
\end{align}    
where we took into account that $-1/|\omega_1-\omega_2|=1/(\omega_2-\omega_1)$ and introduced the notation 
$\Delta\omega=\omega_2-\omega_1<0$. 
The obtained coherence, as a function of the parameters $\omega_1,\omega_2$ coincides with 
the previously obtained coherence for the case $\omega_2>\omega_1$. 

\section{Qualitative analysis of the fermionized spin-boson system coherences}
\label{sec:qualitative_fermions}

Here we provide qualitative static response calculation in order to compare it with the solution obtained before. 
We consider the bosonic bath with the Hamiltonian
\begin{equation}
H_B=\sum_k \Omega_k b_k^\dag b_k,
\end{equation}
and system of three weakly interacting fermions 
\begin{equation}
H_F=\omega_1a_1^\dag a_1+\omega_2a_2^\dag a_2+t(a_1^\dag a_2+a_2^\dag a_1)+0*a_0^\dag a_0,
\end{equation}
at temperature $T=1/\beta$. 
When two systems are far from each other, both of them have corresponding thermal distributions. 
Taking into account that the interaction between the fermions are week ($t\ll \omega_1, \omega_2$), we can 
estimate their occupation numbers of $\langle a^\dag_j a_j\rangle\approx n(\omega_j)$ 
for $j=1,2$, where $n(\omega)=[\exp(\beta \omega)+1]^{-1}$ is Fermi occupation number   
When system of two fermions is close to the bath, the interaction term start to play an important role 
\begin{equation}
H_{BF}=(2f_1 a_1^\dag a_1+f_2 [a_1^\dag a_0+a_0^\dag a_1])\sum_k \lambda_k(b_k+b_k^\dag).
\end{equation}
Due to this term, the non-zero occupation number $\langle a_1^\dag a_1\rangle$ ``polarizes'' the bath system. 
This process can be observed from the Hamiltonian
\begin{equation}
H_{BF_1}=\sum_k \Omega_k b_k^\dag b_k + 2f_1\langle a_1^\dag a_1\rangle\sum_k \lambda_k(b_k+b_k^\dag) + 
\omega_1\langle a_1^\dag a_1\rangle,
\end{equation}
where we replace the operator $a_1^\dag a_1$ by its average value $\langle a_1^\dag a_1\rangle$. 
The interaction of the first spin with the bath induces the non-zero average values
$\langle b_k+b_k^\dag\rangle\neq0$. This values can be calculated by diagonalization of the   
boson part of the obtained Hamiltonian. It can be done by substitution  
$b_k=\beta_k-2f_1\langle a_1^\dag a_1\rangle\lambda_k/\Omega_k$, where $\beta_k$ is the set of 
boson annihilation operators with the same commutation rules as $b_k$ operators, and zero average values.
Hence, we have
\begin{equation}
\langle b_k+b_k^\dag\rangle=\langle \beta_k+\beta_k^\dag\rangle-4f_1\langle a_1^\dag a_1\rangle\lambda_k/\Omega_k=
-4f_1\langle a_1^\dag a_1\rangle\lambda_k/\Omega_k.
\end{equation}   
The non-zero $\langle b_k+b_k^\dag\rangle$ affects the term 
\begin{equation}
f_2[a_1^\dag a_0+a_0^\dag a_1]\sum_k \lambda_k(b_k+b_k^\dag)\rightarrow -4f_1f_2\langle a_1^\dag a_1\rangle\Omega[a_1^\dag a_0+a_0^\dag a_1]=q[a_1^\dag a_0+a_0^\dag a_1],
\end{equation}
where we introduce the short notation $q=-4f_1f_2\langle a_1^\dag a_1\rangle\Omega$.
Here $\langle a_1^\dag a_1\rangle$ should be considered as $\langle a_1^\dag a_1\rangle=n(\omega_1)$.
The back action of the bath modifies the Hamiltonian of the fermions which 
we write as a quadratic form 
\begin{equation}
H_F^{mod}=[\begin{array}{ccc}
           a_0^\dag & a_1^\dag & a_2^\dag 
          \end{array}]
					\left[\begin{array}{ccc}
           0 & q & 0 \\
					 q & \omega_1 & t \\
					 0 & t & \omega_2
          \end{array}\right]
					\left[\begin{array}{ccc}
           a_0 \\
					 a_1  \\
					 a_2 
          \end{array}\right]=\mathbf{a}^\dag \mathcal{H}\mathbf{a}.
\end{equation}
When we introduce the unitary transformation $\mathbf{A}=U \mathbf{a}$ which diagonalizes the Hamiltonian
\begin{equation}
\mathbf{a}^\dag \mathcal{H}\mathbf{a}=\mathbf{a}^\dag U^\dag U\mathcal{H}U^\dag U\mathbf{a}=
\mathbf{A}^\dag \Lambda \mathbf{A}.
\end{equation}
The matrix $\Lambda=U\mathcal{H}U^\dag$ is a diagonal matrix $[\Lambda]_{ij}=\delta_{ij}\lambda_j$,
where $\lambda_j,\,j=0,1,2$ are the eigenenergies of the system.
The general expression for the matrix $U$ is quite complex. We find it approximately up to qubic order 
by small parameters $q$ and $t$. Namely, we consider the unitary matrix $U$ in the form $U=\exp(-S)$, 
and find the anti-unitary matrix $S^\dag=-S$
\begin{equation}
S=\left[
\begin{array}{ccc}
 0 & \frac{q}{\omega _1} & -\frac{qt}{\omega_1\omega_2} \\
 -\frac{q}{\omega_1} & 0 & 0 \\
 \frac{qt}{\omega_1\omega_2} & 0 & 0 \\
\end{array}
\right].
\end{equation}  
Then we calculate $U_{app}$ approximately $\exp(-S)\approx \mathcal{I}- S+S^2/2$,
 where $\mathcal{I}$ is a unit $3\times3$ matrix
\begin{equation}
U_{app}=\left[
\begin{array}{ccc}
 1 & 0 & 0 \\
 0 & 1 & 0 \\
 1 & 0 & 1 \\
\end{array}
\right]-\left[
\begin{array}{ccc}
 0 & \frac{q}{\omega _1} & -\frac{qt}{\omega_1\omega_2} \\
 -\frac{q}{\omega_1} & 0 & 0 \\
 \frac{qt}{\omega_1\omega_2} & 0 & 0 \\
\end{array}
\right]+\frac12 \left[
\begin{array}{ccc}
 -\frac{q^2 t^2}{\omega_1^2 \omega_2^2}-\frac{q^2}{\omega_1^2} & 0 & 0 \\
 0 & -\frac{q^2}{\omega_1^2} & \frac{q^2 t}{\omega_1^2 \omega_2} \\
 0 & \frac{q^2 t}{\omega_1^2 \omega _2} & -\frac{q^2 t^2}{\omega_1^2 \omega_2^2} \\
\end{array}
\right].
\end{equation}
One can check that $U_{app}U^\dag_{app}=\mathcal{I}+O(q^4,q^4t)$. 
Applying the transformation $\mathcal{H}'=U_{app}\mathcal{H}U_{app}^\dag$ we get 
\begin{equation}
\mathcal{H}'=\left[
\begin{array}{ccc}
 -\frac{q^2}{\omega _1} & 0 & 0 \\
 0 & \frac{q^2}{\omega _1}+\omega _1 & t \\
 0 & t & \omega _2 \\
\end{array}
\right]+ O(q^2t,qt^2).
\end{equation}
It doesn't fully diagonalize $\mathcal{H}$, but reduced it to the block-diagonal form. 
Let us introduce the second transformation $W=\exp(-T)$ with the matrix 
\begin{equation}
T=\left[
\begin{array}{ccc}
 0 & 0 & 0 \\
 0 & 0 & -\frac{t}{q^2/\omega_1+\omega _1-\omega _2} \\
 0 & \frac{t}{q^2/\omega_1+\omega _1-\omega _2} & 0 \\
\end{array}
\right].
\end{equation}
Then calculate $W_{app}=\mathcal{I}-T+T^2/2$ and evaluate $\mathcal{H}''=W_{app}\mathcal{H}'W_{app}^\dag$
\begin{equation}
\mathcal{H}''=\left[
\begin{array}{ccc}
 -\frac{q^2}{\omega _1} & 0 & 0 \\
 0 &  \omega_1+\frac{q^2}{\omega _1}+\frac{t^2}{\omega _1-\omega _2} & 0 \\
 0 & 0 & \omega_2-\frac{t^2}{\omega _1-\omega _2} \\
\end{array}
\right]+O(t^3,q^2t^2).
\end{equation}   
We got fully diagonalized matrix with the eigenvalues 
\begin{equation}
\lambda_1=-\frac{q^2}{\omega _1}\approx 0,\quad \lambda_2=\omega _1+\frac{q^2}{\omega _1}+\frac{t^2}{\omega _1-\omega _2}\approx \omega_1,
\quad\lambda_3=\omega_2+\frac{t^2}{\omega _2-\omega _1}\approx \omega_2.
\end{equation}
These eigenvalues have the correct form up to quadratic terms by $q,t$. They can be alternatively 
found in ``Mathematica''. Therefore the composition of $U_{app}$ and $W_{app}$ gives the full unitary 
transformation $U$ 
\begin{equation}
\mathcal{H}''= U\mathcal{H}U^\dag=W_{app}U_{app}\mathcal{H}U_{app}^\dag W_{app}^\dag
\end{equation}
which diagonalizes the Hamiltonian $\mathcal{H}$. Then we have
\begin{equation}
U=W_{app}U_{app}=\left[
\begin{array}{ccc}
 1-\frac{q^2}{2 \omega _1^2} & -\frac{q}{\omega _1} & \frac{q t}{\omega _1 \omega _2} \\
 \frac{q}{\omega _1} & 1-\frac{q^2}{2 \omega _1^2}-\frac{t^2}{2 \left(\omega _1-\omega _2\right)^2} & \frac{t}{\omega _1-\omega _2} \\
 -\frac{q t}{\left(\omega _1-\omega _2\right) \omega _2} & \frac{t}{\omega _2-\omega _1} & 1-\frac{t^2}{2 \left(\omega _1-\omega _2\right)^2} \\
\end{array}
\right]+O(q^2t,qt^2).
\end{equation}
The Hermitian conjugated matrix is
\begin{equation}
U^\dag=U_{app}^\dag W_{app}^\dag=\left[
\begin{array}{ccc}
 1-\frac{q^2}{2 \omega _1^2} & \frac{q}{\omega_1} & -\frac{q t}{\left(\omega_1-\omega_2\right) \omega_2} \\
 -\frac{q}{\omega_1} & 1-\frac{q^2}{2 \omega _1^2}-\frac{t^2}{2 \left(\omega_1-\omega_2\right)^2} & 
\frac{t}{\omega _2-\omega _1} \\
 \frac{q t}{\omega_1 \omega_2} & \frac{t}{\omega_1-\omega_2} & 1-\frac{t^2}{2\left(\omega_1-\omega_2\right)^2} \\
\end{array}
\right]+O(q^2t,qt^2).
\end{equation}   

Then, taking into account the definition $\mathbf{a}=U^\dag\mathbf{A}$ we get the following relations
\begin{align}
a_0=&\Big(1-\frac{q^2}{2\omega_1^2}\Big)A_0+\frac{q}{\omega_1}A_1-\frac{qt}{\left(\omega_1-\omega_2\right)\omega_2}A_2, \\
a_1=&-\frac{q}{\omega_1}A_0+\Big(1-\frac{q^2}{2 \omega _1^2}-\frac{t^2}{2 \left(\omega_1-\omega_2\right)^2}\Big)A_1+
\frac{t}{\omega_2-\omega_1}A_2, \\ 
a_2=&\frac{q t}{\omega_1 \omega_2}A_0+\frac{t}{\omega_1-\omega_2}A_1+
\Big(1-\frac{t^2}{2\left(\omega_1-\omega_2\right)^2}\Big)A_2.
\end{align} 

It  give us the expression for the coherence
\begin{equation}
\langle a_1^\dag a_0+a_0^\dag a_1\rangle=\frac{2q}{\omega_1}
[\langle A_1^\dag A_1\rangle-\langle A_0^\dag A_0\rangle]\approx
-\frac{8f_1f_2\Omega}{\omega_1}n(\omega_1)\Big[n(\omega_1)-n(-q^2/\omega_1)\Big].
\end{equation} 
The expression for another coherence is 
\begin{align}
\langle a_2^\dag a_0+a_0^\dag a_2\rangle=&\frac{2qt}{\omega_1\omega_2}\langle A_0^\dag A_0\rangle+
\frac{2qt}{(\omega_1-\omega_2)}\Big[\frac{\langle A_1^\dag A_1\rangle}{\omega_1}-
\frac{\langle A_2^\dag A_2\rangle}{\omega_2}\Big] = \nonumber \\
=&\frac{2qt}{(\omega_1-\omega_2)\omega_1}\Big[\langle A_1^\dag A_1\rangle-\langle A_0^\dag A_0\rangle\Big]-
\frac{2qt}{(\omega_1-\omega_2)\omega_2}\Big[\langle A_2^\dag A_2\rangle-
\langle A_0^\dag A_0\rangle\Big].
\end{align}

If $\omega_2\gg\omega_1$ then this expression can be simplified
\begin{equation}
\langle a_2^\dag a_0+a_0^\dag a_2\rangle\approx-\frac{t}{\omega_2}\langle a_1^\dag a_0+a_0^\dag a_1\rangle
\end{equation}

In order to compare the obtained result with the one, derived with Green's function technique we need to 
make the substitution $\langle A_0^\dag A_0\rangle=n(-q^2/\omega_1)\rightarrow n(0)=1/2$, since we consider
the linear in $q$ contributions only and $\propto q^2$ terms are beyond of this approximation.   

Then we get the following expressions 
\begin{equation}
\langle a_1^\dag a_0+a_0^\dag a_1\rangle\rightarrow 
-\frac{8f_1f_2\Omega}{\omega_1}n(\omega_1)[n(\omega_1)-1/2]=
\frac{4f_1f_2\Omega}{\omega_1}n(\omega_1)\tanh\Big(\frac{\beta\omega_1}{2}\Big),
\end{equation} 
\begin{equation}
\langle a_2^\dag a_0+a_0^\dag a_2\rangle=-\frac{4f_1f_2\Omega t}{\omega_1-\omega_2}n(\omega_1)
\Big[\frac{\tanh(\beta\omega_2)}{\omega_2}-\frac{\tanh(\beta\omega_1)}{\omega_1}\Big]. 
\end{equation}
Note that these expressions coincide with the Hartree contributions obtained before, see Appendix~\ref{sec:hartree}. 
Indeed, in this chapter we used the same boson-fermion Hamiltonian (see Eq.~\eqref{eq:fermion_hamiltonian}) and 
define the averages as 
\begin{equation}
\langle a_j^\dag a_0+a_0^\dag a_j\rangle=[G_{0j}(0_-)+G_{0j}(0_-)]\approx 
[G^{(0)}_{0j}(0_-)+\Delta G^F_{0j}(0_-)+\Delta G^H_{0j}(0_-)+ \{j\leftrightarrow 0\}],
\end{equation} 
with $j=1,2$, $G_{0j}(\tau)=-\langle\mathrm{T}_\tau a_0(\tau)a_j(0)^\dag\rangle$ full thermal Green's function, that
is presented as the sum of non-interacting Green's function $G^{(0)}_{0j}(\tau)$ and two corrections --
Fock and Hartree, respectively. 
The Hartree corrections are 
\begin{equation}
\Delta G^H_{01}(0_-)=
\frac{2f_1f_2}{\epsilon_1}\Omega n(\epsilon_1)\tanh\Big(\frac{\beta\epsilon_1}{2}\Big),
\end{equation}
\begin{equation}
\Delta G^H_{02}(0_-)=
f_1f_2\sin\theta n(\epsilon_1)\Omega
\Big[\frac{\tanh\Big(\frac{\beta\epsilon_2}{2}\Big)}{\epsilon_2}-
\frac{\tanh\Big(\frac{\beta\epsilon_1}{2}\Big)}{\epsilon_1}\Big].
\end{equation}
Here $\epsilon_2=(\omega_1+\omega_2)/2+\sqrt{(\omega_1-\omega_2)^2/4+t^2}$ and 
$\epsilon_1=(\omega_1+\omega_2)/2-\sqrt{(\omega_1-\omega_2)^2/4+t^2}$ are the largest and middle energies of fermions in the system, $\sin\theta=2t/\sqrt{(\omega_1-\omega_2)^2+4t^2}$. Then, for example, in the limit $t\rightarrow 0$ and  $\omega_2>\omega_1$ these expressions are simplified
\begin{equation}
\Delta G^H_{01}(0_-)=
\frac{2f_1f_2\Omega}{\omega_1} n(\omega_1)\tanh\Big(\frac{\beta\omega_1}{2}\Big),
\end{equation}
\begin{equation}
\Delta G^H_{02}(0_-)=
\frac{2f_1f_2\Omega t}{\omega_1-\omega_2} n(\omega_1)
\Big[\frac{\tanh\Big(\frac{\beta\omega_1}{2}\Big)}{\omega_1}-
\frac{\tanh\Big(\frac{\beta\omega_2}{2}\Big)}{\omega_2}\Big].
\end{equation}
Both expressions are real, so the Hartree contributions to the corresponding coherences just twice larger than the latter expressions. They coincide with the coherences obtained in this section. 


\begin{thebibliography}{99}

\bibitem{Streltsov2017}  Alexander Streltsov, Gerardo Adesso, and Martin B. Plenio. Colloquium: Quantum coherence as a resource. \href{https://doi.org/10.1103/RevModPhys.89.041003}{Rev. Mod. Phys. {\bf 89}, 041003 (2017).}

\bibitem{Schmitt2017} Simon Schmitt, Tuvia Gefen, Felix M. St\"{u}rner, Thomas Unden, Gerhard Wolff, Christoph M{\''u}ller, Jochen Scheuer, Boris Naydenov, Matthew Markham, Sebastien Pezzagna, Jan Meijer, Ilai Schwarz, Martin Plenio, Alex Retzker, Liam P. McGuinness, Fedor Jelezko. Submillihertz magnetic spectroscopy performed with a nanoscale quantum sensor. \href{https://doi.org/10.1126/science.aam5532}{Science {\bf 356}, 6340, 832-837 (2017).}

\bibitem{Zhou2020} Hengyun Zhou, Joonhee Choi, Soonwon Choi, Renate Landig, Alexander M. Douglas, Junichi Isoya, Fedor Jelezko, Shinobu Onoda, Hitoshi Sumiya, Paola Cappellaro, Helena S. Knowles, Hongkun Park, and Mikhail D. Lukin. Quantum Metrology with Strongly Interacting Spin Systems.
 \href{https://doi.org/10.1103/PhysRevX.10.031003}{ Phys. Rev. X {\bf 10}, 031003 (2020).}

\bibitem{Reiserer2016}  Andreas Reiserer, Norbert Kalb, Machiel S. Blok, Koen J. M. van Bemmelen, Tim H. Taminiau, Ronald Hanson, Daniel J. Twitchen, and Matthew Markham. Robust Quantum-Network Memory Using Decoherence-Protected Subspaces of Nuclear Spins.  \href{https://doi.org/10.1103/PhysRevX.6.021040}{ Phys. Rev. X {\bf 6}, 021040 (2016).}

\bibitem{Awschalom2018} 
David D. Awschalom, Ronald Hanson, J\"{o}rg Wrachtrup \& Brian B. Zhou. Quantum technologies with optically interfaced solid-state spins. 
\href{https://doi.org/10.1038/s41566-018-0232-2}{Nature Photonics {\bf 12}, 516–527 (2018).}

\bibitem{Hensgens2017} T. Hensgens, T. Fujita, L. Janssen, Xiao Li, C. J. Van Diepen, C. Reichl, W. Wegscheider, S. Das Sarma \& L. M. K. Vandersypen. Quantum simulation of a Fermi–Hubbard model using a semiconductor quantum dot array. 
\href{https://doi.org/10.1038/nature23022}{Nature {\bf 548}, 70–73 (2017).}

\bibitem{Drost2017} Robert Drost, Teemu Ojanen, Ari Harju \& Peter Liljeroth. \& Liljeroth, P. Topological states in engineered atomic lattices. \href{https://doi.org/10.1038/nphys4080}{Nature Physics {\bf 13}, 668–671 (2017).}

\bibitem{Slot2017} Marlou R. Slot, Thomas S. Gardenier, Peter H. Jacobse, Guido C. P. van Miert, Sander N. Kempkes, Stephan J. M. Zevenhuizen, Cristiane Morais Smith, Daniel Vanmaekelbergh \& Ingmar Swart. Experimental realization and characterization of an electronic Lieb lattice. 
\href{https://doi.org/10.1038/nphys4105}{Nature Physics {\bf 13}, 672-676 (2017).}

\bibitem{Scholes2017} Gregory D. Scholes, Graham R. Fleming, Lin X. Chen, Al{\'a}n Aspuru-Guzik, Andreas Buchleitner, David F. Coker, Gregory S. Engel, Rienk van Grondelle, Akihito Ishizaki, David M. Jonas, Jeff S. Lundeen, James K. McCusker, Shaul Mukamel, Jennifer P. Ogilvie, Alexandra Olaya-Castro, Mark A. Ratner, Frank C. Spano, K. Birgitta Whaley \& Xiaoyang Zhu. Using coherence to enhance function in chemical and biophysical systems.
\href{https://doi.org/10.1038/nature21425}{Nature {\bf 543}, 647–656 (2017).}

\bibitem{Romero2017} Elisabet Romero, Vladimir I. Novoderezhkin \& Rienk van Grondelle.
Quantum design of photosynthesis for bio-inspired solar-energy conversion. 
\href{https://doi.org/10.1038/nature22012}{Nature {\bf 543}, 647–656 (2017).}

\bibitem{Klatzow2019} James Klatzow, Jonas N. Becker, Patrick M. Ledingham, Christian Weinzetl, Krzysztof T. Kaczmarek, Dylan J. Saunders, Joshua Nunn, Ian A. Walmsley, Raam Uzdin, and Eilon Poem. Experimental demonstration of quantum effects in the operation of microscopic heat engines. \href{https://doi.org/10.1103/PhysRevLett.122.110601}{Phys. Rev. Lett. {\bf 122}, 110601 (2019).}

\bibitem{Ono2020} K. Ono, S. N. Shevchenko, T. Mori, S. Moriyama, and Franco Nori. F. Analog of a Quantum Heat Engine Using a Single-Spin Qubit. 
\href{https://doi.org/10.1103/PhysRevLett.125.166802}{Phys. Rev. Lett. {\bf 125}, 166802 (2020).}

\bibitem{Latune2021} Camille L. Latune, Ilya Sinayskiy \& Francesco Petruccione. Roles of quantum coherences in thermal machines.
\href{https://doi.org/10.1140/epjs/s11734-021-00085-1}{Eur. Phys. J. Spec. Top. (2021).} 

\bibitem{Bradley2019}  C. E. Bradley, J. Randall, M. H. Abobeih, R. C. Berrevoets, M. J. Degen, M. A. Bakker, M. Markham, D. J. Twitchen, and T. H. Taminiau. A Ten-Qubit Solid-State Spin Register with Quantum Memory up to One Minute. 
\href{https://doi.org/10.1103/PhysRevX.9.031045}{Phys. Rev. X {\bf 9}, 031045 (2019).}

\bibitem{Stephen2019} C. J. Stephen, B. L. Green, Y. N. D. Lekhai, L. Weng, P. Hill, S. Johnson, A. C. Frangeskou, P. L. Diggle, Y.-C. Chen, M. J. Strain, E. Gu, M. E. Newton, J. M. Smith, P. S. Salter, and G. W. Morley. Deep Three-Dimensional Solid-State Qubit Arrays with Long-Lived Spin Coherence. 
\href{https://doi.org/10.1103/PhysRevApplied.12.064005}{Phys. Rev. Applied {\bf 12}, 064005 (2019).}

\bibitem{DelSanto2020} Flavio Del Santo and Borivoje Daki\'{c}. Coherence Equality and Communication in a Quantum Superposition.
\href{https://doi.org/10.1103/PhysRevLett.124.190501}{Phys. Rev. Lett. {\bf 124}, 190501 (2020).}

\bibitem{DelSanto2018} Flavio Del Santo and Borivoje Daki\'{c}. Two-Way Communication with a Single Quantum Particle.
\href{https://doi.org/10.1103/PhysRevLett.120.060503}{Phys. Rev. Lett. {\bf 120}, 060503 (2018).}

\bibitem{Tan2017} Kok Chuan Tan, Tyler Volkoff, Hyukjoon Kwon, and Hyunseok Jeong. Quantifying the Coherence between Coherent States.
\href{https://doi.org/10.1103/PhysRevLett.119.190405}{Phys. Rev. Lett. {\bf 119}, 190405 (2017).}

\bibitem{Micadei2020_1} Kaonan Micadei, John P. S. Peterson, Alexandre M. Souza, Roberto S. Sarthour, Ivan S. Oliveira, Gabriel T. Landi, Roberto M. Serra, and Eric Lutz. Experimental Validation of Fully Quantum Fluctuation Theorems Using Dynamic Bayesian Networks
 \href{https://doi.org/10.1103/PhysRevLett.127.180603}{Phys.Rev.Lett. {\bf 127}, 180603 (2021).}

\bibitem{Alonso2016} Jose Joaquin Alonso, Eric Lutz, and Alessandro Romito. Thermodynamics of Weakly Measured Quantum Systems.
\href{https://doi.org/10.1103/PhysRevLett.116.080403}{Phys. Rev. Lett. {\bf 116}, 080403 (2016).}

\bibitem{Haase2018} J. F. Haase, A. Smirne, J. Ko\l{}ody\'{n}ski, R. Demkowicz-Dobrza\'{n}ski, and S. F. Huelga, Fundamental limits to frequency estimation: a comprehensive microscopic perspective. 
\href{https://doi.org/10.1088/1367-2630/aab67f}{New J. Phys. {\bf 20}, 053009 (2018).}

\bibitem{Czajkowski2019} Jan Czajkowski, Krzysztof Paw\l{}owski and Rafa\l{} Demkowicz-Dobrza\'{n}ski. Many-body effects in quantum metrology. \href{https://doi.org/10.1088/1367-2630/ab1fc2}{New J. Phys. {\bf 21} 053031 (2019).}

\bibitem{Novo2016} Leonardo Novo, Masoud Mohseni \& Yasser Omar. Disorder-assisted quantum transport in suboptimal decoherence regimes. 
\href{https://doi.org/10.1038/srep18142}{Scientific Reports {\bf 6}, 18142 (2016).}

\bibitem{Micadei2020} Kaonan Micadei, Gabriel T. Landi, and Eric Lutz. Quantum Fluctuation Theorems beyond Two-Point Measurements. \href{https://doi.org/10.1103/PhysRevLett.124.090602}{Phys. Rev. Lett. {\bf 124}, 090602 (2020).}

\bibitem{Diaz2020}  Mar\'{\i}a Garc\'{\i}a D\'{\i}az, Giacomo Guarnieri, Mauro Paternostro. 
Quantum work statistics with initial coherence.
\href{https://doi.org/10.3390/e22111223}{Entropy {\bf 22}(11), 1223 (2020).}

\bibitem{Demkowicz2017} Rafa\l{} Demkowicz-Dobrza\'{n}ski, Jan Czajkowski, and Pavel Sekatski. 
Adaptive Quantum Metrology under General Markovian Noise. 
\href{https://doi.org/10.1103/PhysRevX.7.041009}{Phys. Rev. X {\bf 7}, 041009 (2017).}

\bibitem{Seah2020} Stella Seah, Stefan Nimmrichter, and Valerio Scarani. Maxwell’s Lesser Demon: A Quantum Engine Driven by Pointer Measurements. 
\href{https://doi.org/10.1103/PhysRevLett.124.100603}{Phys. Rev. Lett. {\bf 124}, 100603 (2020).}

\bibitem{Miller2020} Harry J. D. Miller, Giacomo Guarnieri, Mark T. Mitchison, and John Goold.
Quantum fluctuations hinder finite-time information erasure near the Landauer limit. 
\href{https://doi.org/10.1103/PhysRevLett.125.160602}{Phys. Rev. Lett. {\bf 125}, 160602 (2020).}

\bibitem{Francica2020} G. Francica, F. C. Binder, G. Guarnieri, M. T. Mitchison, J. Goold, and F. Plastina. 
Quantum Coherence and Ergotropy. 
\href{https://doi.org/10.1103/PhysRevLett.125.180603}{Phys. Rev. Lett. {\bf 125}, 180603 (2020).}

\bibitem{Seah2019} Stella Seah, Stefan Nimmrichter, Daniel Grimmer, Jader P. Santos, Valerio Scarani, 
and Gabriel T. Landi. Collisional Quantum Thermometry. 
\href{https://doi.org/10.1103/PhysRevLett.123.180602}{Phys. Rev. Lett. {\bf 123}, 180602 (2019).}

\bibitem{Latune2019} C. L. Latune, I. Sinayskiy \& F. Petruccione. 
Quantum coherence, many-body correlations, and non-thermal effects for autonomous thermal machines. 
\href{https://doi.org/10.1038/s41598-019-39300-4}{Scientific Reports {\bf 9}, 3191 (2019).}

\bibitem{Tupkary2021} Devashish Tupkary, Abhishek Dhar, Manas Kulkarni, Archak Purkayastha. 
Fundamental limitations in Lindblad descriptions of systems weakly coupled to baths. 
\href{https://doi.org/10.1103/PhysRevA.105.032208}{Phys. Rev. A {\bf 105}, 032208 (2022)}
 
\bibitem{Zurek2003} Wojciech Hubert Zurek. Decoherence, einselection, and the quantum origins of the classical.  
\href{https://doi.org/10.1103/RevModPhys.75.715}{Rev. Mod. Phys. {\bf 75}, 715 (2003).}

\bibitem{Mohseni2014} Mohseni, M., Omar, Y., Engel, G. S., Plenio, M. {\it Quantum Effects in Biology} 
(Cambridge Univ Press, Cambridge, MA, 2014).

\bibitem{Duan2017} Hong-Guang Duan, Valentyn I. Prokhorenko, Richard J. Cogdell, Khuram Ashraf, Amy L. Stevens, Michael Thorwart, and R. J. Dwayne Miller. Nature does not rely on long-lived electronic quantum coherence for photosynthetic energy transfer. \href{https://doi.org/10.1073/pnas.1702261114}{PNAS {\bf 114} (32) 8493-8498 (2017).}

\bibitem{Guarnieri2018} Giacomo Guarnieri, Michal Kol\'{a}\v{r}, and Radim Filip. 
Steady-State Coherences by Composite System-Bath Interactions. 
\href{https://doi.org/10.1103/PhysRevLett.121.070401}{Phys. Rev. Lett. {\bf 121}, 070401 (2018).}

\bibitem{Guarnieri2020} Giacomo Guarnieri, Daniele Morrone, Barı\c{s} \c{C}akmak, Francesco Plastina, Steve Campbelle.
Non-equilibrium steady-states of memoryless quantum collision models. 
\href{https://doi.org/10.1016/j.physleta.2020.126576}{Phys. Lett. A {\bf 384}, 24, 126576 (2020).}

\bibitem{Reppert2020} Mike Reppert, Deborah Reppert, Leonardo A. Pachon, and Paul Brumer. Equilibrium stationary coherence in the multilevel spin-boson model. 
\href{https://doi.org/10.1103/PhysRevA.102.012211}{Phys. Rev. A {\bf 102}, 012211 (2020).}

\bibitem{Roman2020}  Rom\'{a}n-Ancheyta, R., Kol\'{a}\v{r}, M., Guarnieri, G., Filip, R. 
Enhanced steady-state coherences via repeated system-bath interactions. 
\href{https://doi.org/10.1103/PhysRevA.104.062209}{Phys. Rev. A {\bf 104}, 062209 (2021)}.

\bibitem{Purkayastha2020} Archak Purkayastha, Giacomo Guarnieri, Mark T. Mitchison, Radim Filip \& John Goold.
Tunable phonon-induced steady-state coherence in a double-quantum-dot charge qubit.
\href{https://doi.org/10.1038/s41534-020-0256-6}{npj Quantum Information {\bf 6}, 27 (2020).}

\bibitem{Breuer2007} H.-P. Breuer, F.Petruccione. {\it The Theory of Open Quantum Systems} 
(Oxford University Press, 2007)

\bibitem{Weiss2012} U. Weiss. {\it Quantum Dissipative Systems} (World Scientific Publishing, 2012)

\bibitem{Bateman1953} H. Bateman, A. Erd\'{e}lyi. {\it Higher Transcendental Functions. Vol.2. Bessel Functions, Parabolic Cylinder Functions, Orthogonal Polynomials} (McGraw-Hill, 1953).

\bibitem{Lieb1961} Elliott Lieb,Theodore Schultz, Daniel Mattis. Two soluble models of an antiferromagnetic chain. 
\href{https://doi.org/10.1016/0003-4916(61)90115-4}{Annals of Physics {\bf 16}, 407 (1961).}

\bibitem{Touzard2019} S. Touzard, A. Kou, N. E. Frattini, V. V. Sivak, S. Puri, A. Grimm, L. Frunzio, S. Shankar, 
and M. H. Devoret. Gated Conditional Displacement Readout of Superconducting Qubits. 
\href{https://doi.org/10.1103/PhysRevLett.122.080502}{Phys. Rev. Lett. {\bf 122}, 080502 (2019).}

\bibitem{Ma2021} X. Ma, J. J. Viennot, S. Kotler, J. D. Teufel \& K. W. Lehnert. Non-classical energy squeezing of a macroscopic mechanical oscillator. \href{https://doi.org/10.1038/s41567-020-01102-1}{ 
Nature Physics {\bf 17}, 322-326 (2021).}

\bibitem{Cai2021} M.-L. Cai, Z.-D. Liu, W.-D. Zhao, Y.-K. Wu, Q.-X. Mei, Y. Jiang, L. He, X. Zhang, Z.-C. Zhou 
\& L.-M. Duan. Observation of a quantum phase transition in the quantum Rabi model with a single trapped ion. 
\href{https://doi.org/10.1038/s41467-021-21425-8}{Nature Communications {\bf 12}, 1126 (2021).}

\end{thebibliography}
\end{document}